\documentclass[11pt]{article}

\pdfoutput=1

\usepackage{amsmath,amssymb,amsfonts,amscd,mathrsfs,amsthm}
\usepackage{xcolor}
\definecolor{darkblue}{rgb}{0.1,0.1,.7}
\usepackage[colorlinks, linkcolor=darkblue, citecolor=darkblue, urlcolor=darkblue, linktocpage]{hyperref} 
\usepackage[square, comma, compress,numbers]{natbib}
\usepackage[]{version}
\usepackage[]{graphicx}
\usepackage[]{latexsym}
\usepackage{geometry}
\usepackage[all,cmtip]{xy}
\usepackage[margin=10pt,font=small,labelfont=bf]{caption}
\usepackage{ifthen}
\usepackage{tikz}
\usepackage{array,setspace,mathrsfs,amsfonts,yfonts,dsfont,bbm,colonequals}

\def\wh{\widehat}
\def\beq{\begin{equation}}
\def\eeq{\end{equation}}

\def \a{\alpha}
\def \b{\beta}
\def\to{\rightarrow}
\def\dd{\delta}
\def\DD{\Delta}
\def\La{\Lambda}

\newcommand{\du}[2]{_{#1}^{\phantom{#1}#2}} 
\def\nn{\nonumber}

\def\pd{\partial}

\def\lra{\leftrightarrow}
    
\usepackage{dsfont} 

\def\wt{\widetilde}

\makeatletter
\def\Ddots{\mathinner{\mkern1mu\raise\p@
\vbox{\kern7\p@\hbox{.}}\mkern2mu
\raise4\p@\hbox{.}\mkern2mu\raise7\p@\hbox{.}\mkern1mu}}
\makeatother

\def\bsub{\begin{subequations}}
\def\esub{\end{subequations}}
\newcommand{\ceil}[1]{\left \lceil #1 \right \rceil }
\newcommand{\floor}[1]{\left \lfloor #1 \right \rfloor }

\newcommand{\braket}[3]{\langle #1|#2|#3 \rangle}
\newcommand{\brakket}[2]{\langle #1|#2\rangle}
\newcommand{\ket}[1]{|#1\rangle}
\newcommand{\bra}[1]{\langle #1|}
\newcommand{\expec}[1]{\langle #1 \rangle}
\newcommand{\no}[1]{:\! #1 \!:}
\newcommand{\NO}[1]{{:\!#1\!:}}

\newcommand{\mbb}[1]{{\mathbb #1}}
\newcommand{\mbf}[1]{{\mathbf #1}}
\newcommand{\mca}[1]{{\mathcal #1}}
\newcommand{\msc}[1]{{\mathscr #1}}
\newcommand{\mfr}[1]{{\mathfrak #1}}
\newcommand{\mrm}[1]{{\mathrm #1}}
\newcommand{\reef}[1]{(\ref{#1})}

\def\zb{\bar{z}}
\def\thalf{\tfrac{1}{2}}
\def\half{\frac{1}{2}}
\def\pd{\partial}
\def\a{\alpha}
\def\la{\lambda}
\def\DD{\Delta}
\def\Oo{\mathcal{O}}

\def\sO{\mathrm{O}}

\def\l{\ell} 
\def\eps{\epsilon}

\def\unit{\mathds{1}} 
\def\rnk{r}
\def\ga{\gamma}
\newcommand{\fracpt}[2]{\frac{\partial #1}{\partial #2}}
\newcommand{\sfrac}[2]{\frac{\mathrm{d} #1}{\mathrm{d} #2}}
\def\ldef{\mathrel{\mathop:}=}
\def\rdef{=\mathrel{\mathop:}}

\newcommand{\limu}[1]{\mathrel{\mathop{\sim}\limits_{\scriptstyle{#1}}}}

\newcommand{\dsq}{\Box}

\newcommand{\cO}{\mathcal O}

\newcommand{\be}{\begin{equation}}
\newcommand{\ee}{\end{equation}}
\newcommand{\bea}{\begin{eqnarray}}
\newcommand{\eea}{\end{eqnarray}}

\newcommand{\uud}{\mathrm d}

\numberwithin{equation}{section}
\begin{document}

\vspace*{-.6in} \thispagestyle{empty}
\begin{flushright}
CERN-TH-2016-114\\
YITP-SB-16-16
\end{flushright}
\vspace{1cm} {\Large
\begin{center}
 {\bf The ABC (in any D) of Logarithmic CFT}\\
\end{center}}
\vspace{1cm}
\begin{center}
{\bf Matthijs Hogervorst$^{a}$, Miguel Paulos$^{b}$, Alessandro Vichi$^{b}$ }\\[2cm] 
{
$^{a}$ C.N.\@ Yang Institute for Theoretical Physics, Stony Brook University, USA\\
$^{b}$ Theoretical Physics Department, CERN, Geneva, Switzerland
}
\\
\end{center}
\vspace{14mm}

\begin{abstract}
  Logarithmic conformal field theories have a vast range of applications, from critical percolation to systems with quenched disorder. In this paper we thoroughly examine the structure of these theories based on their symmetry properties. Our analysis is model-independent and holds for any spacetime dimension. Our results include a determination of the general form of correlation functions and conformal block decompositions, clearing the path for future bootstrap applications. Several examples are discussed in detail, including logarithmic generalized free fields, holographic models, self-avoiding random walks and critical percolation. 
\end{abstract}
\vspace{12mm}
\hspace{7mm}May 2016

\newpage

{
\setlength{\parskip}{0.05in}
\tableofcontents
}

\setlength{\parskip}{0.05in}

\section{Introduction}
\label{sec:intro}

Conformal field theories are scale invariant, which seems to require that two-point functions behave as power laws.
But this is not quite true: conformal correlation functions can in fact have logarithms~\cite{Saleur:1991hk,Rozansky:1992td,Gurarie:1993xq}, which contain a scale. Many interesting models turn out to have this property, such as percolation~\cite{Cardy:1999zp}, self-avoiding walks~\cite{Duplantier:1987sh, Saleur:1991hk}, spanning forests \cite{Ivashkevich:1998na}, as well as systems with quenched disorder~\cite{Caux:1995nm,Maassarani:1996jn,Caux:1998sm,Cardy:1999zp,Cardy:2013rqg}. This surprising fact and its consequences will be examined at length in this paper, where we will study {\em logarithmic} Conformal Field Theories (logCFTs) starting from first principles and especially without fixing a particular spacetime dimension.

We should begin by remarking that there is already a vast literature on the subject in the two-dimensional case. In the seminal work~\cite{Gurarie:1993xq}, Gurarie was the first to point out that logarithmic terms in CFT correlation functions are caused by reducible but indecomposable representations of the two-dimensional conformal group.
Subsequent work focused on the constraints coming from conformal symmetry on (chiral) three- and four-point functions, and Operator Product Expansions (OPEs)~\cite{RahimiTabar:1996ub,Flohr:2000mc,Flohr:2001tj,Flohr:2005qq,Rasmussen:2005ta,Nagi:2005sb}.
Infinitely many two-dimensional logCFTs were later constructed by extending the Kac table of the ordinary Virasoro minimal models~\cite{Pearce:2006sz}. Both the representation content and the fusion rules of these ``logarithmic minimal models'' have been studied in detail~\cite{Feigin:2005xs,Feigin:2006iv,Feigin:2006xa,Rasmussen:2007su,Rasmussen:2008xi}. A partially overlapping direction of research has focused on realizing 2$d$ logCFTs as continuum limits of lattice models, see e.g.~\cite{Read:2007qq,Gainutdinov:2012qy,Gainutdinov:2012mr,Gainutdinov:2013tja}.
In spite of these developments, it is fair to say that 2$d$ logCFTs are significantly less understood than their non-logarithmic counterparts. In particular, the computation of non-chiral (also known as bulk or local) correlation functions remains a difficult problem~\cite{Do:2007dn,Vasseur:2011ud,Ridout:2012ew,Fuchs:2013lda,Santachiara:2013gna,Gaberdiel:2007jv,Runkel:2012rp,Fuchs:2016wjr}. 
The references given above can serve as a starting point for the reader. More comprehensive discussions can be found in the review articles~\cite{Flohr:2001zs,Gaberdiel:2001tr,Creutzig:2013hma,Gurarie:2013tma}. A special class of 2$d$ logCFTs, in the form of WZW and sigma models on superspaces, is reviewed in~\cite{Quella:2013oda}. 

Higher-dimensional logCFTs have received much less attention, apart from the determination of constraints on some scalar two- and three-point functions~\cite{Ghezelbash:1997cu}. This state of affairs is unfortunate, since interesting logarithmic theories are certainly not confined to two dimensions. As already mentioned, CFTs coupled to quenched disorder generically flow to logCFTs at long distances. Likewise, the $Q$-state Potts critical point in $2 \leq d<6$ and the $O(N)$ model in $2 \leq d<4$ dimensions become logarithmic in certain limits of their parameters. These logCFTs describe theories with non-local actions, like percolation and polymer statistics. Logarithmic theories are also known to arise as limits of quantum field theories with instantons, like 4$d$ super-Yang-Mills theory~\cite{Frenkel:2006fy,Frenkel:2007ux,Frenkel:2008vz}. An additional reason to be interested in logCFTs comes from holography, see e.g.~\cite{Grumiller2013}. Most of the work in this direction has so far focused on the AdS$_3$/CFT$_2$ correspondence or does not go beyond the level of three point functions.

In this paper we will perform a careful and systematic study of the formal structure of logCFTs in any spacetime dimension. One motivation is the expectation that a broader look at these theories can help us better understand the two-dimensional case. More importantly, we hope that these structural results improve our knowledge about higher-dimensional logarithmic fixed points.
This is especially urgent in the light of the conformal bootstrap~\cite{Rattazzi:2008pe}, which in recent years has proved to be a powerful tool in analyzing CFTs in any dimension. It has been applied in many contexts, for example in computing critical exponents of the 3$d$ Ising and $O(n)$ models to high precision~\cite{ElShowk:2012ht,El-Showk:2014dwa,Kos:2014bka,Kos:2016ysd} but also for understanding structural properties of CFTs analytically~\cite{Komargodski:2012ek,Fitzpatrick:2012yx,Hartman:2015lfa,Hofman:2016awc}. Our work clears the path for any future bootstrap applications to logCFTs.

The outline of this paper is as follows. In section~\ref{sec:confinv} we discuss general consequences of (logarithmic) conformal invariance for $n$-point correlation functions. We start with a discussion of radial quantization in these theories. The Hilbert space contains reducible but indecomposable representations, which means that the dilatation operator cannot be made hermitian. This inevitably leads to the appearance of logarithms in correlation functions. We work out the Ward identities and their solutions, spelling out in detail the general form of two, three and four point functions. Three- and four-point functions must satisfy further constraints from Bose or crossing symmetry.

Section~\ref{sec:CB} is concerned with the derivation of the conformal partial wave and conformal block decompositions of four-point functions. Our main result is to show that conformal blocks of logarithmic primaries in the four point function of logarithmic operators can be determined by computing derivatives of ordinary, non-logarithmic conformal blocks. We show this by solving the Casimir equation {\em \`a la} Dolan and Osborn~\cite{DO2} for a few cases, and then in full generality via radial quantization methods. In order to illustrate the formalism,  we work out a few explicit examples at the end of the section.

In section \ref{sec:ads} we reconsider and extend previous holographic approaches to logarithmic theories. LogCFTs can be modeled holographically by actions containing higher derivatives, which we motivate by coupling bulk theories to bulk disorder. We then provide a thorough discussion of scalar theories, computing all two point functions without recourse to holographic renormalization. We discuss interactions, and show how the resulting structure is consistent with the results of sections~\ref{sec:confinv} and~\ref{sec:CB}. Next we introduce and discuss the holographic version of logarithmic spin-1 multiplets, described by models with higher derivatives of the Maxwell tensor. We finish with some comments on spin-2 models. These holographic toy models should prove useful in future AdS/CMT applications to strongly coupled disordered systems.

In section \ref{sec:examples} we analyze a number of concrete logCFTs. We begin with a 2$d$ example, the triplet model, which is the bosonic sector of the $c=-2$ theory of symplectic fermions. Many results are known for this theory, and we show they are in full agreement with our formalism. Next we consider what we call the logarithmic generalized free field, the logCFT analog of mean field theory. We discuss in detail the four point functions in this model and their conformal block decompositions. Two further examples are considered in a more limited way: the self-avoiding random walks, described by the $O(n)$ model with $n\to 0$, and critical percolation, given by the $Q\to 1$ limit of the Potts model. Both theories have a Lagrangian description in the UV, which allows for computations using the epsilon expansion. We reconsider some existing results in our framework.

We finish this paper with a discussion of several issues and an outlook on future work. Several appendices complement and complete calculations done in the main bulk of the paper.

\section{Consequences of conformal invariance}
\label{sec:confinv}

A CFT is characterized by its symmetry under the action of the conformal group, which in Euclidean signature is $SO(d+1,1)$. Logarithmic CFTs are also invariant under the action of the same group, but what sets these theories apart is that they contain reducible but indecomposable representations, which we call {\em logarithmic multiplets}. In this section, we shall examine the constraints imposed by conformal invariance on correlation functions with insertions of logarithmic operators. For normal CFTs, such constraints and their solutions are well-known, see for instance~\cite{Osborn:1993cr}. In logCFTs the Ward identities satisfied by the correlation functions of the associated operators take an unusual form. Nevertheless, we shall solve them in full generality, with particular attention paid to the two, three and four-point correlation functions.

\subsection{Logarithmic multiplets}
\label{sec:logmul}

We begin by recalling the form of the conformal algebra, for the sake of completeness but also to set our conventions. The algebra $\mathfrak{so}(d+1,1)$ contains as generators $D$ for dilatations, $P^\mu$ for translations, $K^\mu$ for special conformal transformations and $M^{\mu \nu} = -M^{\nu \mu}$ for $d$-dimensional rotations, satisfying non-trivial commutation relations:
\begin{align}
\label{eq:commu}
&i[D,P^\mu]=P^\mu, \qquad i[D,K^\mu]=-K^\mu, \qquad i[P^\mu,K^\nu]= 2 \delta^{\mu\nu} D - M^{\mu\nu}\,,  \nn \\
&i[M^{\mu\nu},X^\rho]= \dd^{\mu \rho} X^\nu - \delta^{\nu \rho} X^\mu \,, \qquad \qquad \text{for } X^\mu = P^\mu, K^\mu\,,\\
  &i [M^{\mu \nu},M^{\rho \sigma}] = \dd^{\mu \rho} M^{\nu \sigma} \pm \text{three terms}. \nn
\end{align} 
Next we consider representations of this algebra. In logarithmic CFTs, states are organized in \emph{logarithmic multiplets} of rank $\rnk \geq 1$. Such a multiplet is built on top of $\rnk$ primary states $\ket{\Oo_a}$, $a = 1,\ldots,\rnk$, obeying the highest-weight condition
\beq
K_\mu \ket{\Oo_a} = 0\,.
\label{eq:prim}
\eeq
The states $\ket{\Oo_a}$ can have arbitrary spin, although we are suppressing $O(d)$ indices for simplicity. A full representation of the conformal algebra consists of the $r$-primary states and their infinite descendants. The latter are obtained by acting an arbitrary number of times with $P_\mu$ on the primary states, exactly like for a standard conformal multiplet with $\rnk=1$.

The generator of dilatations $D$ acts on primary states in the following way:\footnote{Of course, the form of the matrix $\mbf{\Delta}$ is basis-dependent; here we choose a particularly convenient basis.}
\beq
D \ket{\Oo_a}  = -i \mbf{\DD}\du{a}{b} \, \ket{\Oo_b}, \qquad \mbf{\DD} =  \begin{pmatrix}
\DD & 1       & 0       & \cdots  & 0 \\
0       & \DD & 1       & \cdots  & 0 \\
\vdots  & \vdots  & \vdots& \ddots  & \vdots \\
0       & 0       & 0        & \DD & 1       \\
0       & 0       & 0       & 0       & \DD \end{pmatrix}.
\label{eq:Jblock}
\eeq
The Jordan block form of the matrix $\mbf{\DD}$ in~\reef{eq:Jblock} means that for $\rnk > 1$ such representations are indecomposable but reducible.  It is from this simple fact that the entire peculiar structure of logarithmic theories will emerge~\cite{Gurarie:1993xq}. 

On the cylinder $\mathds R\times S^{d-1}$, we normally think of states $\ket{\Oo}$ as energy eigenstates, with $D$ playing the role of the Hamiltonian. In logarithmic CFTs, the states $\ket{\Oo_a}$ are actually \emph{generalized} eigenstates, meaning that they satisfy
\beq
(D+i\DD)^{\rnk-a+1}\ket{\Oo_a}=0\,.
\eeq
Passing to flat space, the states $\ket{\Oo_a}$ correspond to insertions of local operators $\Oo_a$ at the origin. To insert them elsewhere we simply act with the generator of translations. Under rotations and translations, local operators transform as they would in a CFT. However, \reef{eq:Jblock} implies that the action of the $D$ and $K_\mu$ generators is now:
\bsub
\label{eq:Wards}
\begin{align}
  i[D,\Oo_a(x)] &= \left(\mbf{\DD}\du{a}{b}  + \dd_a^b \, x \cdot \fracpt{}{x} \right)\! \Oo_b(x), \label{eq:Dward}\\
  i [K_\mu,\Oo_a(x)] &= - 2 x_\mu \left( \mbf{\DD}\du{a}{b}  + \dd_{a}^{b}\, x \cdot \fracpt{}{x} \right)\! \Oo_b(x) + x^2 \fracpt{}{x^\mu} \Oo_a(x) + 2i x^\la \,\mca{S}_{\la \mu} \cdot \Oo_a(x)\,,
\end{align}
\esub
where $\mca{S}_{\mu \nu}$ is a matrix representation of the $d$-dimensional rotation group, acting on the $O(d)$ indices of $\Oo_a(x)$.  We see that both dilatations and special conformal transformations lead to a mixing between different operators in the multiplet. This mixing is an inevitable consequence of the reducible but indecomposable property of these logarithmic multiplets.   The action of the generators above, together with translations and rotations, determine the Ward identities for correlation functions in the usual way: 
\bea
\langle [G, \Oo_{a_1}^1(x_1) \dotsm \Oo_{a_n}^n(x_n)] \rangle=\sum_{k=1}^n \langle  \Oo_{a_1}^1(x_1) \dotsm [G,\Oo_{a_k}^k(x_k)] \dotsm \Oo_{a_n}^n(x_n)] \rangle=0 \,,
\eea
with $G$ an arbitrary generator of the conformal algebra and the $\Oo_{a}^i(x)$ arbitrary local operators. We see that in general these identities relate correlators of different components of the same multiplet.

There is a formal way of understanding the origin of equations~\reef{eq:Wards}, see e.g.~\cite{RahimiTabar:1996ub}. Let us start with the action of the dilatation generator on a rank-1 primary state $\ket{\Oo}$. 
Formally we have
\bea
\partial_\Delta (i D-\Delta) \ket{\Oo}=0 \qquad \Rightarrow \qquad i D\ket{\partial_\Delta \Oo}=\Delta \ket{\partial_\Delta \Oo}+\ket{\Oo}\,.
\eea
Similarly one deduces
\bea
i D\frac{1}{n!}\ket{\partial_\Delta^n\Oo}=\Delta \frac{1}{n!}\ket{\partial_\Delta^n\Oo}+\frac{1}{(n-1)!}\ket{\partial_\Delta^{n-1}\Oo}.
\eea
It follows that if we make the identification
\bea
\label{eq:formal-id}
\ket{\Oo_a} \; \equiv \; \frac{1}{(r-a)!}\partial_\Delta^{r-a} \ket{\Oo},
\eea
we recover the transformation laws \reef{eq:Jblock} and \reef{eq:Wards} for a rank-$\rnk$ multiplet. This relation is a formal trick that can be useful in solving the Ward identities. Nevertheless, many known logCFTs are limits of one-parameter families of CFTs~\cite{Cardy:1999zp,Cardy:2013rqg} and in those cases the identification~\reef{eq:formal-id} is more than a bookkeeping tool. Indeed, when tuning a parameter $p$ to a special value $p^*$, a logarithmic multiplet can arise when $r$ operators collide to the same scaling dimension. In order to cancel divergences in $(p-p^*)$, one is forced to consider linear combination of operators which become derivatives with respect to the scaling dimension in the $(p-p^*)\rightarrow 0$ limit. Some examples of this phenomenon will be presented in section~\ref{sec:examples}.

In the next subsections, we will use the Ward identities to constrain the form of $n$-point functions of logarithmic operators. Before we do so, we may ask what happens when we consider a finite conformal transformation $x \to x'$ with scale factor
\beq
\label{eq:scalefactor}
\Omega(x) = \vert\!\det(\pd x'^\mu / \pd x^\nu) \vert^{1/d}\,.
\eeq
By exponentiating the action of the generators it is easy to show that, say, a rank-two scalar multiplet of dimension $\DD$ transforms as:
\bsub
\begin{align}
  \Oo'_1(x') &= \frac{1}{\Omega(x)^{\DD}} \left[ \Oo_1(x) - \ln \Omega(x) \, \Oo_2(x) \right]\, , \\
  \Oo'_2(x') &= \frac{1}{\Omega(x)^{\DD}}  \, \Oo_2(x) \,. 
\end{align}
\label{eq:transf}
\esub
Generalizations are straightforward, but here already we see the feature that gives logarithmic CFTs their name, namely the appearance of logarithms. Such logarithms will abound in correlation functions.
As an immediate consequence we notice that, in radial quantization, conjugate states $\bra{\Oo_a}$  have to be defined in an unusual way. In a CFT, such states can be obtained by performing an inversion $I$ which maps $x^\mu \to  x^\mu/ (\mu |x|)^2$, for some scale $\mu$:
\bea
\bra{\Oo}\equiv \lim_{|x|\to 0} \bra{0} I \Oo(x)I=\lim_{|x|\to \infty} (\mu |x|)^{2\Delta} \bra{0} \Oo(x)\,.
\eea
Usually the scale $\mu$ is set to one implicitly. However, the scale is important in a logarithmic theory since now the conjugate states become: 
\begin{subequations}
\bea
\langle \Oo_1|&=&\lim_{|x|\to \infty} (\mu |x|)^{2\Delta} \bra{0}\left[\Oo_1(x) + \ln \mu^2 |x|^2 \,\Oo_2(x)\right]\,,\\
\langle \Oo_2|&=&\lim_{|x|\to \infty} (\mu |x|)^{2\Delta} \bra{0} \Oo_2(x)\,.
\eea
\end{subequations}
We see that, while we may get rid of the overall $\mu$ factor, the scale survives inside the logarithm.

This might seem paradoxical: how can a scale invariant theory contain a scale? To understand how this can be, consider performing a change of basis of states of the form
\bea
\label{eq:changebasis}
\label{eq:fieldred}
\ket{\Oo_a} \to \mbf R_a^{\ b}\ket{\Oo_b}, \qquad \mbf R_a^{\ b}:= \begin{cases}
  R_{b-a} & \text{if} \quad a \leq b \\
  0 & \text{if} \quad a > b \\
\end{cases}\,,
  \eea
  for some fixed coefficients $R_0, \ldots, R_{\rnk-1}$. 
  Since $[\mbf \Delta,\mbf R]=0$, this leaves both the action of the conformal generators and the Ward identities unchanged.\footnote{
    Conversely, it may be shown that any matrix $\mbf{R}$ that commutes with $\mbf{\DD}$ is of the form shown in Eq.~\reef{eq:changebasis}. 
    }
Going back to~\reef{eq:transf}, we see that a change of scale (i.e. $\Omega(x)= \text{const.}$) amounts precisely to performing such a field redefinition. It is then the freedom to change operator basis as in \reef{eq:fieldred} which makes the presence of a scale possible while preserving the full conformal invariance of the theory. We will provide a more thorough discussion of this and related matters in Sec. \ref{sec:scale}, but henceforth we will implicitly work in units where $\mu=1$.

\subsection{Two-point functions}\label{sec:twopt}

Here we will derive the form of the two-point functions of logarithmic operators. Let's consider two scalar multiplets: $\Oo_a$ of dimension $\DD_1$ and rank $\rnk$, and $\wt{\Oo}_b$ of dimension $\DD_2$ and rank $\rnk'$. Without loss of generality, we can assume that $r \leq \rnk'$. By Poincar\'{e} invariance, their two-point function can be written as
\beq
\expec{\Oo_a(x) \wt{\Oo}_b(0)} = \frac{B_{ab}(s)}{|x|^{\DD_1 + \DD_2}}\,,
\eeq
where $s \ldef |x|$ and $B_{ab}(s)$ is a matrix of size $\rnk \times \rnk'$ that we wish to determine. The Ward identities imply that
\bsub
\label{eq:Bsys}
\begin{align}
s \sfrac{}{s} B_{ab} &= - B_{(a+1)b} - B_{a(b+1)}\,,\\
  \left( s \sfrac{}{s} + \Delta_1-\Delta_2 \right)\! B_{ab} &= - 2B_{(a+1)b}\,, \end{align}
\esub
for all $a,b$. Here and in what follows we use the convention that $B_{ab}(s) = 0$ if the labels $a,b$ are unphysical, i.e.\@ if $a > r$ or $b > r'$. Combining both equations, we obtain the useful relation:
\beq
\label{eq:Brec}
B_{a(b+1)} -B_{(a+1)b}=(\Delta_1-\Delta_2) B_{ab}\,.
\eeq
From this it is easy to determine that $\Delta_1=\Delta_2\equiv \Delta$. Indeed, if this wasn't the case we would get immediately $B_{rr'}=0$, and using the same relation successively determines that all other elements would also be zero. Proceeding then with $\Delta_1=\Delta_2$, Eq.~\reef{eq:Brec} now implies that the matrix element $B_{ab}$ only depends on $a+b$ and thus we set
\beq
B_{ab}(s) \rdef \beta_{a+b-1}(s).
\eeq
In the new variables, Eqs.~\reef{eq:Bsys} implies
\beq
s \sfrac{}{s} \beta_n(s) = -2 \beta_{n+1}(s)\,. \label{eq:betasys}
\eeq
Moreover, Eq.~\reef{eq:Brec} implies that $\beta_n = 0$ if $n > \rnk$. Consequently, the system of differential equations~\reef{eq:betasys} can be solved in terms of $\rnk$ undetermined constants $k_{1},\ldots,k_\rnk$:
\beq
\beta_n(s) = \sum_{m=0}^{\rnk - n} k_{n+m}\, \frac{(-1)^m}{m!} (\ln x^2)^m , \qquad n = 1,\ldots,\rnk \,.
\eeq
Summarizing, the Jordan block form of the representation~\reef{eq:Jblock} forces various logarithmic terms to be present in the $\expec{\Oo_a \wt{\Oo}_b}$ correlation function.

There are several important simplifications possible at this stage. First, we remark that after a suitable change of basis, all two-point functions of operators in different multiplets can be made to vanish. Since the proof of this statement is slightly technical, we refer to Appendix~\ref{sec:twopts} for details. Next, we remark that for the correlator $\expec{\Oo_a \Oo_b}$ of two identical multiplets, we can always assume that $k_\rnk \neq 0$. If this is not the case, the bottom component $\Oo_\rnk$ of the multiplet completely decouples from the theory. But if $k_\rnk \neq 0$, there exists a field redefinition which allows us to set $k_1 = \ldots = k_{\rnk-1} = 0$. Indeed, the $r$ undetermined constants precisely match the number of free parameters in the field redefinition matrix $\mathbf R$ in \reef{eq:fieldred}, and we may use this freedom to set such parameters to zero.

In conclusion, the two-point functions of a logarithmic multiplet $\Oo_a$ of dimension $\DD$ can always be brought to the canonical form
\beq
\boxed{
\expec{\Oo_a(x) \Oo_b(0)} = \frac{k_\Oo}{|x|^{2\DD}} \times  \begin{cases}
   \dfrac{(-1)^n}{n!} \left( \ln x^2 \right)^n  & \text{if} \quad n \equiv \rnk + 1 - a - b \geq 0  \\
  0 & \text{if} \quad n < 0
\end{cases}
}
\label{eq:2ptscal}
\eeq
for some constant $k_\Oo \neq 0$.  In particular, if $\Oo_a(x)$ is of rank $\rnk = 2$, we have
\beq
\label{eq:2ptk}
\expec{\Oo_a(x) \Oo_b(0)} = \frac{k_\Oo}{|x|^{2\DD}} \begin{pmatrix} - \ln x^2 & 1 \\ 1 & 0 \end{pmatrix}_{ab}\,,
\eeq
which is a standard result in $d=2$ dimensions~\cite{Gurarie:1993xq}.

One particular consequence of these results is that unitarity is broken. Reflection positivity would require the two point functions $\expec{\Oo_a(x) \Oo_a(-x)} $ to be positive, for all $x$ and (hermitian) fields $\Oo_a$. However it is evident from~\reef{eq:2ptscal} that this is not possible unless all multiplets have rank $r=1$ and $k_\Oo \geq 0$. The same conclusion can be drawn by inspecting the  matrix of inner products $\expec{\Oo_a |\Oo_b}$. As shown in appendix~\ref{sec:gram}, this matrix always has negative eigenvalues.\footnote{The number of negative eigenvalues is $\floor{r/2}$ if $k_\Oo>0$, otherwise it's $\ceil{r/2}$.}
As an important consequence, the unitarity bounds~\cite{Minwalla:1997ka} on operator dimensions that apply to ordinary CFTs do not hold for logCFTs. The sign of the overall normalization $k_\Oo$ is unimportant for this conclusion. Of course, all of these statements follow essentially from the fact that the dilatation generator is not hermitian, which means in particular that time translations on the cylinder are not implemented by a unitary operator.

The generalization to traceless symmetric tensors of spin $\l > 0$ is straightforward.\footnote{By spin $\l$ we mean the 
traceless-symmetric representation of the rotation group $O(d)$, given by a single row Young tableau with $\l$ boxes.}  As in the case of ordinary CFTs, the resulting correlation function features the inversion tensor~\cite{Osborn:1993cr}
\beq
I_{\mu \nu}(x) \ldef \dd_{\mu \nu} - 2 \frac{x_\mu x_\nu}{x^2}.
\eeq
To simplify correlators of spinning operators $\Oo_{\mu_1 \dotsm \mu_\l}(x)$, we use a coordinate-free notation~\cite{Costa:2011mg}:
\beq
\Oo^{(\l)}(x;z) \ldef \Oo_{\mu_1 \dotsm \mu_\l}(x) \, z^{\mu_1} \dotsm z^{\mu_\l} \,,
\eeq
where $z^\mu$ is an auxiliary vector satisfying $z \cdot z = 0$. With this notation, the two-point functions of a logarithmic spin-$\l$ multiplet can be brought into the following form: 
\beq
\expec{\Oo^{(\l)}_a(x;z) \Oo^{(\l)}_b(0;z')} = \frac{k_\Oo}{|x|^{2\DD}}\!\left( I_{\mu \nu}(x)z^\mu z'^\nu \right)^\l \times  \begin{cases}
   \dfrac{(-1)^n}{n!} \left( \ln x^2 \right)^n  & \text{if} \quad n \equiv \rnk + 1 - a - b \geq 0  \\
  0 & \text{if} \quad n < 0
\end{cases}
\,,
\eeq
again for some undetermined constant $k_\Oo \neq 0$.

\subsection{Three-point functions}\label{sec:three-pt}

We will now study constraints on three-point functions in a similar fashion to the previous section, restricting our analysis to scalar-scalar-spin $\l$ correlators for simplicity. Let us first consider a normal CFT with two scalar primaries $\phi,\chi$ with scaling dimensions $\DD_\phi$, $\DD_\chi$ and a spin-$\l$ primary $\Oo^{(\l)}$ of dimension $\DD_\Oo$. Conformal invariance forces their three-point function to take the following form:
\beq
\expec{\phi(x_1)  \chi(x_2) \Oo^{(\l)}(x_3;z)} = \la^{\phi \chi \Oo} \, \mbf{P}_{\DD_\phi  \DD_\chi  \DD_\Oo}(x_i) \left(X \cdot z\right)^\l\,,
\label{eq:3ptspin}
\eeq
where $\la^{\phi \chi \Oo}$ is an OPE coefficient, 
\beq
\label{eq:Polyakov}
\mbf{P}_{\DD_1 \DD_2 \DD_3}(x_i) = \frac{1}{|x_{12}|^{\DD_1 + \DD_2 - \DD_3}|x_{13}|^{\DD_1 + \DD_3 - \DD_2}|x_{23}|^{\DD_2 + \DD_3 - \DD_1}}\,, \quad x_{ij} \ldef x_i - x_j\,,
\eeq
and
\beq
\label{eq:capitalX}
X^\mu =\frac{|x_{13}||x_{23}|}{|x_{12}|} \left( \frac{x_{13}^\mu}{x_{13}^2} - \frac{x_{23}^\mu}{x_{23}^2} \right).
\eeq

We want to generalize this to the case where all operators are part of logarithmic multiplets, where $\phi_a,\chi_b$ have rank $\rnk_1,\rnk_2$ and $\displaystyle{\Oo_c^{(\l)}}$ has rank $\rnk_3$.  This logarithmic three-point function takes the form: 
\beq
\expec{\phi_a(x_1)  \chi_b(x_2) \Oo_c^{(\l)}(x_3;z)} = K_{abc}(x_i)\,  \mbf{P}_{\DD_\phi  \DD_\chi  \DD_\Oo}(x_j) \left(X \cdot z\right)^\l
\label{eq:3ptlog}.
\eeq
We can obtain constraints on the functions $K_{abc}(x_i)$ using the $D$ and $K_\mu$ Ward identities. To simplify the resulting expressions, let's introduce the variables
\beq
\label{eq:taudef1}
\tau_1 \ldef \ln \frac{|x_{23}|}{|x_{12}||x_{13}|}\,,
\quad \tau_2 \ldef \ln \frac{|x_{13}|}{|x_{12}||x_{23}|}\,,
\quad \tau_3 \ldef \ln \frac{|x_{12}|}{|x_{13}||x_{23}|}\,,
\eeq
or equivalently
\bea
\label{eq:taudef2}
\tau_i=\partial_{\Delta_i} \ln[\mbf{P}_{\DD_1 \DD_2 \DD_3}(x_i)].
\eea
The Ward identities look now extremely simple:
\beq
\label{eq:3ptWard}
\fracpt{}{\tau_1} K_{abc} = K_{(a+1)bc}\,, \quad
\fracpt{}{\tau_2} K_{abc} = K_{a(b+1)c}\,, \quad
\fracpt{}{\tau_3} K_{abc} = K_{ab(c+1)}\,.
\eeq
We defer the proof of Eq.~\reef{eq:3ptWard} to section~\ref{sec:npoint}. Again, we use the convention that ${K_{abc}(\tau_i) = 0}$ if any of the labels $a,b,c$ is unphysical. 
The most general solution to Eqs.~\reef{eq:3ptWard} depends on $\rnk_1 \rnk_2 \rnk_3 $ coefficients $\la^{\phi \chi \Oo}_{ijk}$ as follows:
\beq
\label{eq:3ptgensol}
K_{abc}(\tau_i) = \sum_{k=0}^{\rnk_1 - a}\sum_{l=0}^{\rnk_2 - b}\sum_{m=0}^{\rnk_3 - c} \la^{\phi \chi \Oo}_{(a+k)(b+l)(c+m)} \frac{\tau_1^k}{k!}\frac{\tau_2^l}{l!}\frac{\tau_3^m}{m!}\,.
\eeq
Conformal invariance does not constrain the different OPE coefficients $\la^{\phi \chi \Oo}_{ijk}$. However, when two or more of the fields are identical, additional constraints will come from Bose symmetry.  Below, we will spell out these constraints for the case where the two scalars belong to rank-two multiplets.

\subsubsection{Examples}\label{sec:3ptranktwo}

As the simplest example of the formulae above, let us consider the case of two rank-1 scalars $\phi$, $\chi$ of dimension $\DD_\phi$, $\DD_\chi$ and one  rank-$\rnk$ scalar field $\Oo_p$ of dimension $\DD_\Oo$. 
In this case, the three-point function reads \beq
\label{eq:3ptsimpgen}
 \expec{\phi(x_1) \chi(x_2) \Oo_p(x_3)}=\mbf{P}_{\DD_\phi \DD_\chi \DD_\Oo}(x_j)\, K_p(\tau_3)\, , \quad K_p(\tau_3) = \sum_{n=0}^{\rnk - p} \la^{\phi \chi \Oo}_{p+n} \, \frac{\tau_3^n}{n!}\,,
 \eeq
 where $\la^{\phi \chi \Oo}_{1 \dotsm \rnk}$ are the relevant OPE coefficients. 
In more detail, for $\rnk=2$  we find
\begin{subequations}
\label{eq:3ptrnk2}
\bea
\expec{\phi(x_1) \chi(x_2) \Oo_1(x_3)}&=&\mbf{P}_{\DD_\phi \DD_\chi \DD_\Oo}(x_j) \left( \lambda^{\phi \chi \Oo}_2 \tau_3+\lambda^{\phi \chi \Oo}_1 \right) \,,\\
\expec{\phi(x_1) \chi(x_2) \Oo_2(x_3)}&=&\mbf{P}_{\DD_\phi \DD_\chi \DD_\Oo}(x_j) \, \lambda^{\phi \chi \Oo}_2.
\eea
\end{subequations}

For a more complicated example, consider the three-point function of a rank-two scalar primary $\phi_a$ and a rank-$\rnk$ primary of spin $\l$:
\beq
\expec{\phi_a(x_1) \phi_b(x_2) \Oo_c^{(\l)}(x_3;z)} = K_{abc}(\tau_i) \, \mbf{P}_{\DD_\phi \DD_\phi \DD_\Oo}(x_j) \, (X \cdot z)^{\l}\,,
\eeq
where $a,b = 1,2$ and $c = 1,\ldots,\rnk$. As a starting point we consider the general solution~\reef{eq:3ptgensol}. However, since there are two insertions of the same multiplet, 
we have to take Bose symmetry into account, which requires 
\beq
\label{eq:rank2bose}
K_{abc}(\tau_1,\tau_2,\tau_3) = (-1)^\l \, K_{bac}(\tau_2,\tau_1,\tau_3).
\eeq
In particular, $K_{11c}(\tau_i)$ and $K_{22c}(\tau_i)$ will be even (resp.\@ odd) under the exchange $\tau_1 \lra \tau_2$ if $\l$ is even (resp.\@ odd). Consequently, we will treat the cases where $\l$ is even and odd separately.

First, we consider the case of odd $\l$. Concretely, Eq.~\reef{eq:rank2bose} implies that the coefficients $\la^{\phi \phi \Oo}_{ijk}$ obey
\beq
\la^{\phi \phi \Oo}_{11k} = \la^{\phi \phi \Oo}_{22k} = 0\,, \qquad \la^{\phi \phi \Oo}_{12k} = - \la^{\phi \phi \Oo}_{21k}\,, \qquad 1 \leq k \leq \rnk.
\eeq
Consequently, there are only $\rnk$ undetermined OPE coefficients. After defining the functions
\beq
\La_c(\tau_3) \ldef \sum_{n=0}^{\rnk - c} \la^{\phi \phi \Oo}_{12(c+n)} \, \frac{\tau_3^n}{n!}\,, \qquad 1 \leq c \leq \rnk \,,
\eeq
it is possible to write the functions $K_{abc}$ in the following compact form:
\bsub
\begin{align}
 & K_{11c} = (\tau_2 - \tau_1)\, \La^{}_{c}(\tau_3) \,, \\
 & K_{12c} = - K_{21c} = \La^{}_{c}(\tau_3)\,, \\
 & K_{22c} = 0\,.
\end{align}
\esub
Notice that the correlator $\expec{\phi_1 \phi_1 \Oo^{(\l)}}$ is generally nonzero, despite the fact that $\l$ is odd in this case. Although this might seem paradoxical, an explanation comes by inspecting the $\phi_1 \times \phi_1$ OPE. As shown in appendix \ref{appendix:OPE}, this expansion does not contain the primary operator $\Oo^{(\l)}$, but it does include contributions  from its descendants, which have a different spin and consequently a different parity under Bose symmetry.

Second, we consider the case of even $\l$. Here Bose symmetry only requires that
\beq
\la^{\phi \phi \Oo}_{12k} = \la^{\phi \phi \Oo}_{21k}\,, \qquad 1 \leq k \leq \rnk \,,
\eeq
so there are $3\rnk$ undetermined OPE coefficients. Introducing the quantities
\beq
\begin{pmatrix}
  \; \La^{1}_c(\tau_3) \; \\
  \; \La^{2}_c(\tau_3) \; \\
  \; \La^{3}_c(\tau_3) \;
\end{pmatrix}
\ldef \sum_{n=0}^{\rnk - c}
\begin{pmatrix}
  \la^{\phi \phi \Oo}_{11(c+n)} \\
  \la^{\phi \phi \Oo}_{12(c+n)} \\
  \la^{\phi \phi \Oo}_{22(c+n)}
\end{pmatrix}
\frac{\tau_3^n}{n!}\,, \qquad 1 \leq c \leq \rnk \,,
\eeq
the functions $K_{abc}$ can be written in the compact form:
\bsub
\begin{align}
  K_{11c} &= \La_c^{1}(\tau_3) + ( \tau_1 + \tau_2)\, \La_c^{2}(\tau_3) + \tau_1 \tau_2 \, \La_c^{3}(\tau_3)\,,\\
  K_{12c} &= \La^{2}_{c}(\tau_3) + \tau_1  \, \La^{3}_c(\tau_3) \,,\\
  K_{21c} &= \La^{2}_{c}(\tau_3) + \tau_2  \, \La^{3}_c(\tau_3)  \,,\\
  K_{22c} &= \La^{3}_c(\tau_3)\,.
\end{align}
\esub
Finally, as a special case of the above, consider the three-point function of $\phi_a$ itself:
\beq
\expec{\phi_a(x_1) \phi_b(x_2) \phi_c(x_3)} = K_{abc}(\tau_i) \, \mbf{P}_{\DD_\phi \DD_\phi \DD_\phi}(x_j).
\eeq
In this case, Bose symmetry is even more constraining, and the most general solution to the Ward identities will only depend on four coefficients $\la^{\phi \phi \phi}_i$, $i=1, \ldots, 4$:
\bsub
\label{eq:3ptphi}
\begin{align}
  K_{111} &= \la^{\phi \phi \phi}_1 +  \la^{\phi \phi \phi}_2 \sum_i \tau_i  + \la^{\phi \phi \phi }_3 \sum_{i < j} \tau_i \tau_j  + \la^{\phi \phi \phi }_4 \, \tau_1 \tau_2 \tau_3\,,\\
  K_{112} &= \la^{\phi \phi \phi }_2 + \la^{\phi \phi \phi }_3 \left(\tau_1 + \tau_2 \right) + \la^{\phi \phi \phi }_4\,  \tau_1 \tau_2\,, \\ 
  K_{122} &= \la^{\phi \phi \phi }_3 + \la^{\phi \phi \phi }_4\, \tau_1\,, \\ 
  K_{222} &= \la^{\phi \phi \phi }_4\,.
\end{align}
\esub
All other $K_{abc}$ (e.g.\@ $K_{121}$ and $K_{221}$) are related to the above solutions by cyclic permutations of the $\tau_i$.

\subsubsection{Conserved currents}\label{sec:conserved}

So far, we have considered constraints coming from conformal symmetry alone on three-point functions. Some additional constraints apply to \emph{conserved currents}, which are spin-$\l$ operators $J_{\mu_1 \dotsm \mu_\l}$ whose correlators are conserved at non-coincident points:
\beq
\fracpt{}{y^{\mu_1}} \expec{\Oo_1(x_1) \dotsm \Oo_n(x_n) J_{\mu_1 \dotsm \mu_\l}(y)} = 0 \qquad y \neq x_1,\ldots,x_n \,,
\eeq
for arbitrary insertions of $\Oo_1,\ldots,\Oo_n$. Current conservation puts a constraint on the dimension $\DD_J$ of $J$:
\beq
J_{\mu_1 \dotsm \mu_\l}(x) \;\; \text{conserved} \; \quad \rightarrow \quad \DD_J = \l + d-2\,.
\eeq
This is a consequence of conformal invariance and holds both for ordinary and logarithmic CFTs. We will see that in logarithmic CFTs current conservation forces various three-point functions to vanish.

For definiteness, we consider the case where $J$ itself is a rank-one tensor operator --- i.e.\@ $J$ has no logarithmic partners. Furthermore, we will specialize to the three-point function $\expec{\phi_a \phi_b J}$ where $\phi_a$ is a rank-two scalar of dimension $\DD_\phi$. The strategy to derive these constraints is the following. The correlator $\expec{\phi_a \phi_b J}$ can be written as
\beq
\expec{\phi_a(x_1) \phi_b(x_2) J_{\mu_1 \dotsm \mu_\l}(x_3)} = K_{ab}(\tau) \, \mbf{P}_{\DD_\phi \DD_\phi \DD_J}(x) \left[ X_{\mu_1} \dotsm X_{\mu_\l} - \text{traces} \right]\,,
\eeq
for some matrix $K_{ab}$ determined in Sec.~\ref{sec:3ptranktwo}. 
Then it is shown in~\cite{Osborn:1993cr} that
\beq
\fracpt{}{x_3^{\mu_1}}\mbf{P}_{\DD_\phi \DD_\phi \DD_J}(x) \left[X_{\mu_1} \dotsm X_{\mu_\l} - \text{traces} \right] = 0 \qquad [\DD_J = \l + d-2]\,,
\eeq
at $x_3 \neq x_1,x_2$. We must then have also
\beq
\fracpt{}{x_3^{\mu}} \expec{\phi_a(x_1) \phi_b(x_2) J_{\mu \dotsm \mu_\l}(x_3)} =0 \quad \lra \quad
\fracpt{}{x_3^\mu}K_{ab}(\tau_i) = 0\,.
\label{eq:Kcons}
\eeq
This is the equation that we will use to get concrete constraints on OPE coefficients. In what follows, we will consider $\l$ odd and even separately.

First, for odd $\l$, there is only one OPE coefficient, namely $\la^{\phi \phi J}_{12}$. We have
\beq
K_{11} = \la_{12}^{\phi \phi J}\, (\tau_2 - \tau_1)\,, \quad
K_{12} = -K_{21} = \la_{12}^{\phi \phi J}\,, \quad
K_{22} = 0\,. \qquad [\text{odd } \l]
\eeq
The only constraint comes from applying Eq.~\reef{eq:Kcons} to $K_{11}$, which requires that $\la^{\phi \phi J}_{12} = 0$, i.e.\@ $J$ does not couple to $\phi_a \times \phi_b$ at all. A different way to arrive at this conclusion comes from the OPE $\phi_1 \times \phi_1 \sim J$. Consider for definiteness the case $\l=1$, where for arbitrary $\DD_J$ we have
\beq
\label{eq:currentOPE}
\phi_1(x)\phi_1(0) \sim  \frac{\la_{12}^{\phi \phi J}}{|x|^{2\DD_\phi - \DD_J + 1}} \left[ - x^\mu x^\nu \pd_\nu J_\mu(0) - \frac{1}{(\DD_J + 1)(\DD_J - d +1)} x^2 \pd^\mu J_\mu(0) + \sO(x^3)  \right]\,.
\eeq
In the limit $\DD_J \to \l + d-2 = d-1$ the second term in the OPE blows up, hence requiring that the OPE remains finite forces $\la_{12}^{\phi \phi J} = 0$.

For even $\l$ there are three OPE coefficients $\la_{11}^{\phi \phi J}$, $\la_{12}^{\phi \phi J} = \la_{21}^{\phi \phi  J}$ and $\la_{22}^{ \phi \phi J}$, and the three-point functions are
\beq
K_{11} = \la_{11}^{\phi \phi J} + \la_{12}^{\phi \phi J} \, (\tau_1 + \tau_2) + \la_{22}^{\phi \phi J} \, \tau_1 \tau_2\,, \quad
K_{12} = \la_{12}^{\phi \phi J} + \la_{22}^{\phi \phi J} \, \tau_1\,, \quad
K_{22} = \la_{22}^{\phi \phi J}\,. \qquad [\text{even } \l]
  \eeq
  Applying Eq.~\reef{eq:Kcons} to $K_{22}$ does not give any constraints. However, applying it $K_{11}$ and $K_{12}$ shows that $\la_{22}^{\phi \phi J}$ must vanish. We find no additional constraints on the coefficients $\la_{11}^{\phi \phi J}$ and $\la_{12}^{\phi \phi J}$. As above, this argument is buttressed by analyzing the $\phi_1 \times \phi_2 \sim J$ OPE, taking $\l = 2$ for definiteness. For arbitrary $\DD_J$ we have
\begin{multline}
  \phi_1(x)\phi_2(0)  \sim  \frac{\la_{22}^{\phi \phi J}}{|x|^{2\DD_\phi - \DD_J + 2}} \Big[-  \frac{ \ln x^2}{4}  x^\mu x^\nu J_{\mu \nu}(0) + \left( -\frac{\ln x^2}{8} + \frac{1}{4(\DD_J+2)} \right) x^\mu x^\nu (x \cdot \pd) J_{\mu \nu}(0)  \\
    + \frac{1}{2(\DD_J + 2)(\DD_J - d)} x^2 x^\mu \pd^\nu J_{\mu \nu}(0) + \sO(x^4)\Big] \; + \; \frac{\la_{12}^{\phi \phi J}}{|x|^{2\DD_\phi - \DD_J + 2}} \times (\text{finite})\,.
\end{multline}
If $\la_{22}^{\phi \phi J} \neq 0$ the term proportional to $x^2 x^\mu \pd^\nu J_{\mu \nu}$  blows up in the limit $\DD_J \to d$. A similar argument applies to the $\phi_1 \times \phi_1$ OPE.

It may be interesting in future work to generalize this argument and to find constraints on OPE coefficients for conserved currents of rank $\rnk$ in more general three-point functions.

\subsection{Four-point functions}\label{sec:fourpt}

Let us now turn to the constraints of conformal symmetry on scalar four-point functions. We first discuss the non-logarithmic case, considering four different rank-1 scalar primaries $\phi^i$ of dimension $\DD_i$. We recall that their four-point function can always be written as
\beq
\label{eq:4ptnorm}
\expec{\phi^1(x_1)\phi^2(x_2)\phi^3(x_3)\phi^4(x_4)} = F(u,v)\, \mbf{P}_{\DD_1 \DD_2 \DD_3 \DD_4}(x_i)\,,
\eeq
where $F(u,v)$ is a function depending on two independent cross ratios $u,v$
\beq
u = \frac{x_{12}^2 x_{34}^2}{x_{13}^2 x_{24}^2}\,, \qquad v = \frac{x_{14}^2 x_{23}^2}{x_{13}^2 x_{24}^2}\,,
\eeq
and $\mbf{P}$ is a scale factor:
\beq
{\mbf{P}}_{\DD_1 \DD_2 \DD_3 \DD_4}(x_i) = \prod_{i < j} \frac{1}{|x_{ij}|^{\DD_i + \DD_j - \Sigma/3}}\,, \qquad \Sigma \ldef \sum_{i=1}^4 \DD_i\,.
\eeq

We want to generalize Eq.~\reef{eq:4ptnorm} to the case of four logarithmic scalars $\phi^i_a$ of {rank $\rnk_i$}. In this case, we write
\beq
\expec{\phi^1_a(x_1)\phi^2_b(x_2)\phi^3_c(x_3)\phi^4_d(x_4)} = F_{abcd}(u,v,x_i)\, \mbf{P}_{\DD_1 \DD_2 \DD_3 \DD_4}(x_j),
\eeq
and we wish to determine the constraints on the tensor $F_{abcd}$ imposed by conformal invariance. It will be useful to introduce four new variables $\zeta_i$:
\beq
\label{eq:zetadef1}
\zeta_1 \ldef \frac{1}{3} \ln \frac{|x_{23}||x_{24}||x_{34}|}{|x_{12}|^2|x_{13}|^2|x_{14}|^2}, \quad \zeta_{2},\zeta_3,\zeta_4 = \text{cyclic permutations of } \zeta_1\,,
\eeq
or equivalently
\bea
\label{eq:zetadef2}
\label{eq:zetas}
\zeta_i=\partial_{\Delta_i} \ln [\mbf{P}_{\DD_1 \DD_2 \DD_3 \DD_4}(x_j)].
\eea
In terms of the $\zeta_i$ variables, the Ward identities take a particularly simple form:
\begin{align}
&\fracpt{}{\zeta_1} F_{abcd} = F_{(a+1)bcd}\,, \quad
\fracpt{}{\zeta_2} F_{abcd} = F_{a(b+1)cd}\,,\nonumber\\
&\fracpt{}{\zeta_3} F_{abcd} = F_{ab(c+1)d}\,,\quad
\fracpt{}{\zeta_4} F_{abcd} = F_{abc(d+1)}\,.
\label{eq:4ptWard}
\end{align}
This is proved in Sec.~\ref{sec:npoint}. Notice that the Ward identities do not restrict the dependence of $F_{abcd}$ on $u$ and $v$, since the latter are conformally invariant. An immediate consequence of Eq.~\reef{eq:4ptWard} is that the functions $F_{abcd}$ are polynomials in the $\zeta_i$, the degree of which will depend on the different ranks $\rnk_i$.

It is possible to write down a completely general solution to Eqs.~\reef{eq:4ptWard} similar to Eq.~\reef{eq:3ptgensol} for the three-point case. However, we will only work out the details for two specific cases. First we consider a four-point function with two insertions of a rank-two scalar and two insertions of a rank-one scalar. Second, we consider the case of four identical rank-two scalar primaries.

\subsubsection{Example: two logarithmic and two normal insertions}\label{sec:warmup}

As an exercise, we will show how to solve the Ward identities~\reef{eq:4ptWard} for the correlation function
\beq
\label{eq:FdefOnly2}
\expec{\phi_a(x_1) \phi_b(x_2) \chi(x_3) \chi(x_4)} = F_{ab}(u,v,\zeta_i) \, \mbf{P}_{\DD_\phi \DD_\phi \DD_\chi \DD_\chi}(x_i)\,,
\eeq
where $\phi_a$ is a rank-two scalar primary of dimension $\DD_\phi$ and $\chi$ is a normal scalar primary of dimension $\DD_\chi$ (i.e. a rank-1 operator).
An immediate consequence of~\reef{eq:4ptWard} is that the $F_{ab}$ will not depend on $\zeta_3$ and $\zeta_4$.

Next, we stress that the $F_{ab}$ must be consistent with Bose symmetry. The exchange of $x_3 \lra x_4$ acts on the $(u,v,\zeta_i)$ variables as follows:
\beq
x_3 \lra x_4 \quad : \quad u \to u/v\,, \quad v \to 1/v\,, \quad \zeta_1,\zeta_2 \to \zeta_1,\zeta_2\,, \quad \zeta_3 \lra \zeta_4\,.
\eeq
This implies the following crossing symmetry relation:
\beq
\label{eq:cross1}
F_{ab}(u,v,\zeta_1,\zeta_2) = F_{ab}(u/v,1/v,\zeta_1,\zeta_2)\,.
\eeq
Likewise, exchanging $x_1$ and $x_2$ acts on the variables as follows:
\beq
x_1 \lra x_2 \quad : \quad u \to u/v\,, \quad v \to 1/v\,, \quad \zeta_1 \lra \zeta_2\,,\quad \zeta_3,\zeta_4 \to \zeta_3,\zeta_4\,.
\eeq
This implies a second crossing relation, namely
\beq
\label{eq:cross2}
F_{ab}(u,v,\zeta_1,\zeta_2) = F_{ba}(u/v,1/v,\zeta_2,\zeta_1)\,.
\eeq

At this point, we will compute the functions $F_{ab}$ one by one. We will start with $F_{22}$, which depends only on $u$ and $v$. We can therefore write 
$F_{22} = \mca{F}_3(u,v)$
for some function $\mca{F}_3(u,v)$.
Taking into account the crossing symmetry relation~\reef{eq:cross1}, the latter must satisfy
\beq
\mca{F}_3(u,v) = \mca{F}_3(u/v,1/v)\,.
\eeq
Next, we consider $F_{12}$, which according to Eq.~\reef{eq:4ptWard} obeys
\beq
\fracpt{}{\zeta_1} F_{12} = F_{22} = \mca{F}_3(u,v), \quad \fracpt{}{\zeta_2} F_{12} = 0\,.
\eeq
Taking again Eq.~\reef{eq:cross1} into account, we conclude that
\beq
\label{eq:F12}
F_{12} = \mca{F}_2(u,v) + \zeta_1 \,\mca{F}_3(u,v)\,, \quad \mca{F}_2(u,v) = \mca{F}_2(u/v,1/v)\,,
\eeq
for some function $\mca{F}_2(u,v)$. $F_{21}$ can be obtained from Eq.~\reef{eq:F12} using Eq.~\reef{eq:cross2}:
\beq
F_{21} = \mca{F}_2(u,v) + \zeta_2\, \mca{F}_3(u,v).
\eeq
Finally, the function $F_{11}$ is given by
\beq
\label{eq:F11}
F_{11} = \mca{F}_1(u,v) + \left(\zeta_1 + \zeta_2 \right) \mca{F}_2(u,v) + \zeta_1 \zeta_2 \, \mca{F}_3(u,v)\,,
\eeq
for some function $\mca{F}_1(u,v)$ which obeys
\beq
\mca{F}_1(u,v) = \mca{F}_1(u/v,1/v).
\eeq
In conclusion, the correlation function $\expec{\phi_a \phi_b \chi \chi}$ is fixed by conformal symmetry up to three functions $\mca{F}_i(u,v)$, all of which satisfy a crossing symmetry identity:
\beq
\label{eq:crossgen}
\mca{F}_i(u,v) = \mca{F}_i(u/v,1/v), \qquad i=1,2,3\,.
\eeq

\subsubsection{Example: four insertions of a rank-two operator}\label{sec:ranktwo4pt}

Let's now turn to the four-point function of a single rank-two field $\phi_a$ of dimension $\DD_\phi$:
\beq
\expec{\phi_a(x_1) \phi_b(x_2) \phi_c(x_3) \phi_d(x_4)} = F_{abcd}(u,v,\zeta_i) \, \mbf{P}_{\DD_\phi \DD_\phi \DD_\phi \DD_\phi}(x_j)\,.
\label{eq:Fdef}
\eeq
As in the previous example, the solution for $F_{abcd}$ must be consistent with Bose symmetry. Notice that all permutations of four points $x_i$ can be obtained by combining the exchange $x_1 \lra x_2$ with the cyclic permutation $(x_1,x_2,x_3,x_4) \to (x_2,x_3,x_4,x_1)$. The latter acts on the $(u,v,\zeta_i)$ variables as
\beq
(x_1,x_2,x_3,x_4) \to (x_2,x_3,x_4,x_1) \quad : \quad u \lra v\,, \quad (\zeta_1,\zeta_2,\zeta_3,\zeta_4) \to (\zeta_2,\zeta_3,\zeta_4,\zeta_1)\,.
\eeq
This means that the functions $F_{abcd}$ must satisfy the following crossing identities:
\bsub
\label{eq:cross3}
\begin{align}
  F_{abcd}(u,v,\zeta_1,\zeta_2,\zeta_3,\zeta_4) &= F_{bacd}(u/v,1/v,\zeta_2,\zeta_1,\zeta_3,\zeta_4) \\
  &= F_{bcda}(v,u, \zeta_2,\zeta_3,\zeta_4,\zeta_1).
\end{align}
\esub
Equivalent forms can be obtained by combining these identities in different ways.

The computation of the $F_{abcd}$ follows the same steps as the computation from Sec.~\ref{sec:warmup}, and we will not spell out the details. In the end the correlators $\expec{\phi_a \phi_b \phi_c \phi_d}$ involve five functions\footnote{Note that the functions $\mca{F}_i$ are not the same as those defined in the previous section. We will refer to the proper definition wherever a confusion will appear.} $\mca{F}_{1},\ldots, \mca{F}_5$ as follows:
\bsub
\label{eq:4ptmaster}
\begin{align}
   F_{1111} &= \mca{F}_1(u,v) + \sum_i \zeta_i \, \mca{F}_2(u,v) + \left(\zeta_1 \zeta_2 + \zeta_3 \zeta_4 \right) \mca{F}_3(u,v) \nn \\
   &\qquad + \left(\zeta_1 \zeta_3 + \zeta_2 \zeta_4 \right) \mca{F}_3(1/u,v/u) + \left(\zeta_1 \zeta_4 + \zeta_2 \zeta_3 \right) \mca{F}_3(v,u) \nn \\ 
   & \qquad + \sum_{i<j<k} \zeta_i \zeta_j \zeta_k \, \mca{F}_4(u,v) + \zeta_1 \zeta_2 \zeta_3 \zeta_4 \, \mca{F}_5(u,v)\,, \\
   F_{1112} &= \mca{F}_2(u,v) + \zeta_1 \, \mca{F}_3(v,u) + \zeta_2 \, \mca{F}_3(1/u,v/u) + \zeta_3 \, \mca{F}_3(u,v)  \nn \\
   & \qquad + \left(\zeta_1 \zeta_2 + \zeta_1 \zeta_3 + \zeta_2 \zeta_3\right) \mca{F}_4(u,v) + \zeta_1 \zeta_2 \zeta_3 \, \mca{F}_5(u,v) \,,\\
   F_{1122} &= \mca{F}_3(u,v) + \left(\zeta_1 + \zeta_2 \right) \mca{F}_4(u,v) + \zeta_1 \zeta_2 \, \mca{F}_5(u,v) \,,\\
  F_{1222} &= \mca{F}_4(u,v) + \zeta_1 \,\mca{F}_5(u,v) \,,\\
  F_{2222} &= \mca{F}_5(u,v).
\end{align}
\esub
The conformally invariant functions $\mca{F}_i(u,v)$ have the following crossing properties:
\bsub
\label{eq:cross4}
\begin{align}
  \mca{F}_i(u,v) &= \mca{F}_i(u/v,1/v) = \mca{F}_i(v,u) \quad \text{if} \quad i=1,2,4,5;\\
  \mca{F}_3(u,v) &= \mca{F}_3(u/v,1/v).
\end{align}
\esub
All other four-point functions (like $F_{2122}$) can be obtained from~\reef{eq:4ptmaster} by using Eqs.~\reef{eq:cross3} and~\reef{eq:cross4} repeatedly.

\subsection{$n$-point functions}
\label{sec:npoint}
The discussion of three- and four-point functions can be easily extended to scalar $n$-point functions.\footnote{The generalization to spinning $n$-point functions is straightforward but not important for this work, see e.g.~\cite{Costa:2011mg} for an introduction.} This will allow us to derive the solution of the Ward identities once and for all. Consider therefore the correlation function of $n$ scalars $\Oo^i_{a}$ of rank $\rnk_i$ and dimension $\DD_i$. It can be written in the following form:
\bea
\label{eq:nptcor}
\langle \Oo^1_{a_1}(x_1) \dotsm \Oo^n_{a_n}(x_n) \rangle= \mbf P_{\Delta_1 \dotsm \Delta_n}(x_i) F_{a_1 \dotsm a_n}(x_i)\,,
\eea
where we have extracted a scale factor $\mbf{P}$:
\beq
\label{eq:Pndef}
\mbf{P}_{\DD_1 \dotsm \DD_n}(x_i) \ldef \prod_{i < j} \frac{1}{|x_{ij}|^{\kappa_1(\DD_i + \DD_j) -  \kappa_2\Sigma_n}}\,, \qquad \Sigma_n \ldef \sum_{i=1}^n \DD_i
\eeq
writing $\kappa_1 = 2/(n-2)$ and $\kappa_2 = 2/[(n-1)(n-2)]$. The function $\mbf{P}$ is defined such that under any conformal transformation $x \to x'$ with scale factor $\Omega(x)$ --- cf.\@ Eq.~\reef{eq:scalefactor} --- it transforms as
\beq
\label{eq:Ptrafo}
\mbf{P}_{\DD_1 \dotsm \DD_n}(x'_1,\ldots,x'_n) = \prod_{i=1}^n \frac{1}{\Omega(x_i)^{\DD_i}} \, \mbf{P}_{\DD_1 \dotsm \DD_n}(x_1,\ldots,x_n)\,.
\eeq
To prove this, the identity $|x' - y'|^2 = \Omega(x) \Omega(y) |x-y|^2$ may be used. 
The function $F_{a_1 \dotsm a_n}(x_i)$ can depend on $n(n-1)/2$ Lorentz scalars, out of which $n(n-3)/2$ are cross ratios $u_i$ and the other $n$ variables are Poincar\'{e} but not conformally invariant.
We will parametrize these remaining $n$ variables as follows:
\begin{align}
  \zeta_i^{(n)} &\ldef \partial_{\Delta_i} \ln[\mbf P_{\Delta_1 \dotsm \Delta_n}(x_j)] \\
  &= \frac{2}{(n-1)(n-2)} \, \ln\!\left( \frac{ \prod_{k < l} |x_{kl}| }{ \prod_{j \neq i} |x_{ij}|^{n-1} } \right)\,, \nn
\end{align}
for $i=1,\ldots,n$. 
By construction, the $\zeta_i^{(n)}$ transform as
\beq
\label{eq:zetatrans}
\zeta_i^{(n)}(x') =  \zeta_i^{(n)}(x) - \ln \Omega(x_i)\,.
\eeq
Clearly, $\zeta_i^{(n)}$ reduces to $\tau_i$ for $n=3$ and $\zeta_i$ for $n=4$.

To derive the Ward identities, consider performing an infinitesimal transformation $x \to x'$ with scale factor ${\Omega(x) = 1 + \eps \a(x) + \sO(\eps^2)}$, where $\eps$ is a small parameter. Under such a transformation we have
\begin{align}
  \delta \Oo_a(x) &\ldef \Oo'_a(x') - \Oo_a(x) \nn \\
  &= -\eps\,\alpha(x) \left[\Delta\,  \Oo_a(x)+\Oo_{a+1}(x)\right] + \sO(\eps^2)\,.
\end{align}
Consequently the correlator~\reef{eq:nptcor} transforms as
\begin{multline}
  \dd \expec{\Oo_{a_1}^1(x_1) \dotsm \Oo_{a_n}^n(x_n)} = -\eps \sum_{i=1}^n \a(x_i) \Big[ \DD_i \, \expec{\Oo_{a_1}^1(x_1) \dotsm \Oo_{a_n}^n(x_n)} \\
    + \, \expec{\Oo_{a_1}^1(x_1) \dotsm \Oo_{a_i+1}^i(x_i) \dotsm \Oo_{a_n}^n(x_n)}  \Big] \, + \, \sO(\eps^2)\,.
\end{multline}
At the same time, using Eqs.~\reef{eq:Ptrafo} and~\reef{eq:zetatrans} it may be shown that
\beq
\dd \big( \mbf{P}_{\Delta_1 \dotsm \Delta_n}(x) F_{a_1 \dotsm a_n}(u,\zeta)\big) = -\eps \sum_{i=1}^n \a(x_i)\!\left[ \DD_i \, F_{a_1 \dotsm a_n} + \fracpt{}{\zeta_i^{(n)}} F_{a_1 \dotsm a_n} \right]\!\mbf{P}_{\DD_1 \dotsm \DD_n}(x) + \sO(\eps^2)\,.
\eeq
Equating these two expressions we conclude
\bea
\label{eq:nptWard}
\frac{\partial}{\partial \zeta_i^{(n)}} F_{a_1 \dotsm a_i \dotsm a_n}(u,\zeta) =F_{a_1 \dotsm(a_i+1) \dotsm a_n}(u,\zeta) \qquad i=1,\ldots,n\,.
\eea
The Ward identities for three-point~\reef{eq:3ptWard} and four-point~\reef{eq:4ptWard} functions are a special case of Eq.~\reef{eq:nptWard}. Notice that the dependence on the cross ratios $u_i$ is not constrained. 
The general solution to Eq.~\reef{eq:nptWard} is
\bea
\label{eq:nptSol}
F_{a_1 \dotsm a_n}(u_i,\zeta_i^{(n)}) =\sum_{k_1=0}^{r_1-a_1}\dotsm \sum_{k_n=0}^{r_n-a_n} \mathcal F_{(a_1+k_1) \dotsm (a_n+k_n)}(u_i)\, \prod_{i=1}^n \frac{\big[ \zeta_i^{(n)}\big]^{k_i}}{k_i!} 
\eea
which is given in terms of $r_1\times r_2\times \dotsm \times r_n$ functions $\mathcal F_{a_1 \dotsm a_n}(u_i)$ that depend only on the cross ratios. As before, if any of the inserted operators are identical, Bose symmetry will impose additional constraints on the $\mca{F}_{a_1 \dotsm a_n}(u_i)$.

\subsection{Scale dependence}\label{sec:scale}

We will now formalize the brief statements of Sec.~\ref{sec:logmul} on the scale dependence of logarithmic correlation functions. Let us recall what the issue is. On the one hand dimensional analysis requires that logarithms have dimensionless arguments, and hence we must introduce a scale $\mu$. On the other, in a conformal theory we expect that all scales should drop out. The way that these two statements are reconciled in a logCFT is that the operator basis itself becomes scale dependent. Indeed, there is an ambiguity in the choice of operator basis, since all Ward identities in the theory are preserved under the transformation
\bea
\Oo_a(x) \to \mbf R_a^{\ b} \, \Oo_b(x), \qquad \mbf R_a^{\ b}:= \begin{cases}
  R_{b-a} & \text{if} \quad a \leq b \\
  0 & \text{if} \quad a > b \\
\end{cases}\,.
\eea
This ambiguity precisely cancels the ambiguity in the choice of scale $\mu$. To see this, consider writing a Callan-Symanzik type equation for correlation functions,
\bea
\mu \frac{d}{d \mu} \langle \Oo_{a_1}(x_1) \ldots \Oo_{a_n}(x_n)\rangle=0
\eea
Let us write $\Oo_a^\mu:=\mbf R_{a}^{\ b}(\mu) \Oo_b$, with $\mbf R_a^{\ b}(\mu_0)=\delta_{a}^{b}$. Then we can write the equation in the form
\bea
\left(\mu \fracpt{}{\mu} \dd_{a_1}^{b_1}\dotsm \dd_{a_n}^{b_n} +\sum_{k=1}^n \dd_{a_1}^{b_1}\dotsm \gamma_{a_k}^{\ b_k}\dotsm  \dd_{a_n}^{b_n} \right) \langle \Oo^\mu_{b_1}(x_1) \ldots \Oo^\mu_{b_n}(x_n)\rangle=0. \label{eq:callan}
\eea
where we have defined
\bea
\gamma_a^{\ b}\ldef \left(\mbf R^{-1} \partial_\mu \mbf R \right)_a^{\ b}.
\eea

In order for such an equation to be true, it must necessarily reduce to the Ward identities. Since $\mu\, \partial_\mu$ implements a change of scale it is perhaps not surprising that the only possibility is to equate the above with the dilatation Ward identity, by setting
\beq
\label{eq:gammadef}
\gamma_a^{\ b} = \mbf{\DD}\du{a}{b} - \DD \, \dd_a^{b} = \begin{pmatrix}
0 & 1       & 0       & \cdots  & 0 \\
0       & 0 & 1       & \cdots  & 0 \\
\vdots  & \vdots  & \vdots& \ddots  & \vdots \\
0       & 0       & 0        & 0 & 1       \\
0       & 0       & 0       & 0       & 0 \end{pmatrix}.
\eeq
In this way we see indeed that the scale dependence is cancelled by a shift in the operator basis as determined by $\gamma_a^{\ b}$. 

A different interpretation of the Callan-Symanzik equation~\reef{eq:callan} is that it provides us with a prescription to compare correlation functions at different scales, once some initial conditions are given.  Suppose that we choose a field redefinition that brings the two-point function $\expec{\Oo_a \Oo_b}_{\mu }$ in the canonical form~\reef{eq:2ptscal} at the initial scale $\mu_0$. If we now change the scale to some new value $\mu'$, the correlator $\expec{\Oo_a \Oo_b}_{\mu'}$ will no longer be in the canonical form, since the coefficients $k_1, \ldots, k_{\rnk-1}$ may be nonzero. We could choose to cancel this change by redefining our operator basis as before, or instead we could simply say that the coefficients are now scale dependent. Concretely, the coefficients $k_j$ would transform as follows under a scale transformation:
\beq
\mu \fracpt{}{\mu} k_i(\mu) = - 2 \gamma\du{i}{j}  k_j(\mu) \label{eq:runk}
\eeq
A similar reasoning applies to three- and four-point functions. Consider for instance the scale dependence of the three-point function $\expec{\phi \chi \Oo_a}$, where $\phi,\chi$ are non-logarithmic scalars and $\Oo_a$ is a logarithmic scalar of rank $\rnk$.  The general form of this three-point function was computed in Sec.~\ref{sec:3ptranktwo}, and it was found that it depends on $\rnk$ coefficients $\la^{\phi \chi \Oo}_1,\ldots,\la^{\phi \chi \Oo}_\rnk$.  The functions $K_a$ obey:
\beq
\left(\mu \fracpt{}{\mu} \dd_a^{\ a'} + \gamma\du{a}{a'} \right)\! K_{a'}(\tau_3;\mu) = 0
\eeq
Evidently, this equation is satisfied if we make the $\tau_i$ variables dimensionless, i.e.\@ by replacing $\tau_i \to \tau_i - \ln \mu$. Equivalently, we can keep the $\tau_i$ with a fixed scale $\mu_0$, provided that the coefficients $\la_a$ transform under scale changes as
\beq
\label{eq:lamudep}
\mu \fracpt{}{\mu} \la^{\phi \chi \Oo}_i = - \gamma\du{i}{j}  \la^{\phi \chi \Oo}_j\,.
\eeq
The same reasoning applies to more general logarithmic three-point functions. Equations \reef{eq:runk} and \reef{eq:lamudep} now determine the running of all correlation functions.

\section{Conformal block decompositions}
\label{sec:CB}

In the previous section we have studied the general constraints of conformal symmetry on correlators of logarithmic multiplets. In particular, we have seen that four-point correlation functions are determined up to a set of seemingly arbitrary functions of cross-ratios. However, this analysis neglects the existence of the operator product expansion. Due to the state-operator correspondence, the OPE implies that such functions are sums of contributions from in principle all possible operators in the theory, organized into {\em conformal blocks}. Each conformal block captures the contribution of an entire multiplet to the four-point function, which includes not only primary states but also descendants. Our goal in this section is to understand the conformal block decompositions of the four point functions of logarithmic multiplets.

We will do this in several steps. We begin by reviewing how such decompositions are derived in CFTs. Next, we generalize the fact that conformal blocks are eigenfunctions of the Casimir operator to the logarithmic case. This allows us to work out the conformal block decomposition in a couple of examples. The upshot is that the logarithmic blocks are essentially derivatives of ordinary ones. In principle this approach could be extended to all possible four-point functions, but we shall not pursue this here. Instead we move on to a more general analysis based on radial quantization techniques. More concretely, we will achieve two separate goals. Firstly, a precise understanding of the decomposition in terms of contributions which we will call logarithmic conformal blocks. In particular we will derive precise (but complicated) expressions for these contributions in terms of radial quantization matrix elements. Secondly, we shall show that these expressions are related in a simple way to those corresponding to non-logarithmic conformal blocks. This means that we will not have to compute the logarithmic conformal blocks explicitly, but rather we will determine them in terms of certain derivatives of ordinary blocks.

\subsection{Conformal blocks in CFTs}\label{sec:blocksCFT}

Let us start by considering the four-point function
\beq
\label{eq:4pt0}
\expec{\phi^1(x_1) \phi^2(x_2) \phi^3(x_3) \phi^4(x_4)}=F(u,v)\, \mbf{P}_{\DD_1 \DD_2 \DD_3 \DD_4}(x_i)\,,
\eeq
of four different scalar primaries $\phi^i$ of dimension $\DD_i$, see Eq.~\reef{eq:4ptnorm}. Thanks to the state-operator correspondence every CFT has a radial quantization Hilbert space, spanned by primary operators $\ket{\Oo}$ and their descendants $P_{\mu_1} \dotsm P_{\mu_n} \ket{\Oo}$. We will use the shorthand notation $\ket{\Oo;\a} \equiv P_{\a} \ket{\Oo}$ to describe such states, interpreting $\a$ as a multi-index. Again, we ignore any $O(d)$ indices belonging to the state $\ket{\Oo}$.  These descendant states can be organized into states with a well-defined spin and scaling dimension. 
In such a basis, the different descendants are orthogonal:
\beq
\label{eq:normgab}
\brakket{\Oo;\a}{\Oo;\b} = k_\Oo \, \dd_{\a\b} \, g_{\a}(\DD_\Oo;\l_\Oo)\,,
\eeq
where $k_\Oo$ is the normalization of the $\expec{\Oo \Oo}$ two-point function, $\DD_\Oo$ its scaling dimension and $\l_\Oo$ its spin. The norms $g_\a(\DD_\Oo;\l_\Oo)$ depend only on the quantum numbers of the operator $\Oo$ and can be computed using the conformal algebra~\cite{Pappadopulo:2012jk}. 
Assuming that the four-point function~\reef{eq:4pt0} is radially ordered, we can then insert a complete set of states, which yields
\beq
\label{eq:dec1}
\expec{\phi^1 \phi^2 \phi^3 \phi^4} \; = \sum_{\text{primaries } \Oo} k_\Oo^{-1} \sum_{\a} \braket{0}{\phi^1(x_1) \phi^2(x_2)}{\Oo;\a} \, g_\a(\DD_\Oo;\l_\Oo)^{-1} \, \braket{\Oo;\a}{\phi^3(x_3) \phi^4(x_4)}{0}.
 \eeq
The radial quantization matrix element $\braket{0}{\phi^1(x_1) \phi^2(x_2)}{\Oo;\a}$ is proportional to an OPE coefficient $\la^{12\Oo}$, but otherwise fixed by conformal symmetry. The same holds, mutatis mutandis, for the matrix element $\braket{\Oo;\a}{\phi^3(x_3) \phi^4(x_4)}{0}$. We will therefore write
\bsub
\label{eq:Mme}
 \begin{align}
 \braket{0}{\phi^1(x_1) \phi^2(x_2)}{\Oo;\a} &\rdef \la^{12\Oo} \, \mca{M}[1,2,\a,\DD_\Oo,\l_\Oo]\,,\\
 \braket{\Oo;\a}{\phi^3(x_3) \phi^4(x_4)}{0} &\rdef \la^{34\Oo} \, \mca{M}'[3,4,\a, \DD_\Oo, \l_\Oo].
 \end{align}
 \esub
 The matrix elements $\mca{M}[1,2,\a,\DD,\l]$ and $\mca{M}'[3,4,\a,\DD,\l]$ can in principle be computed using the conformal algebra, but their precise form is not important here.  In conclusion, we can rewrite Eq.~\reef{eq:dec1} as
 \bea
&&   \expec{\phi^1(x_1) \phi^2(x_2) \phi^3(x_3) \phi^4(x_4)}  = \sum_\Oo k_\Oo^{-1} \la^{12 \Oo} \la^{34\Oo} \, W^{(\l_\Oo)}_{\DD_\Oo}(x_i;\DD_j)\,, \\
 &&   W^{(\l)}_{\DD}(x_i;\DD_j) \ldef \sum_\a  \mca{M}[1,2,\a,\DD,\l]\, g_{\a}(\DD;\l)^{-1} \mca{M}'[3,4,\a, \DD, \l]\,.\label{eq:cpwnormal}
 \eea
It follows that the \emph{conformal partial waves} $W_\DD^{(\l)}(x_i;\DD_j)$ are universal objects in CFT. They depend only on the positions $x_i$ and scaling dimensions $\DD_{i}$ of the external operators $\phi^i(x_i)$ and the scaling dimension $\DD$ and spin $\l$ of the exchanged operators $\Oo$. 

The partial waves have the same conformal properties as the four-point functions, and hence we can strip off a scale factor
\beq
\label{eq:CBdef}
 W_\DD^{(\l)}(x_i;\DD_j) = G_\DD^{(\l)}(u,v;\DD_j) \, \mbf{P}_{\DD_1 \DD_2 \DD_3 \DD_4}(x_i).
 \eeq
 leaving behind a \emph{conformal block} $G_\DD^{(\l)}(u,v;\DD_i)$ which depends only on $u$ and $v$ and the external dimensions $\DD_{1\dotsm 4}$. It follows that the function $F(u,v)$ can be decomposed in terms of the conformal blocks:
 \beq
 \label{eq:CBdecnorm}
 F(u,v) = \sum_\Oo k_\Oo^{-1} \la^{12 \Oo} \la^{34 \Oo} \, G_{\DD_\Oo}^{(\l_\Oo)}(u,v;\DD_i)\,.
 \eeq
We stress that the conformal blocks $G_{\DD}^{(\l)}(u,v;\DD_i)$ defined in Eq.~\reef{eq:CBdef}  are normalized in an unconventional way. Often one defines conformal blocks $\widehat{G}_{\DD}^{(\l)}$ as follows:
\beq
\label{eq:normalBlock}
 G_{\DD}^{(\l)}(u,v;\DD_i) \rdef u^{-(\DD_1 + \DD_2 + \DD_3 + \DD_4)/6}\,  v^{(-\DD_1 + 2\DD_2 + 2\DD_3 - \DD_4)/6}\; \widehat{G}_{\DD}^{(\l)}(u,v;\rho_1,\rho_2)\,.
 \eeq
 The latter depend on the external dimensions only through the linear combinations ${\rho_1 \equiv (\DD_2 - \DD_1)/2}$ and ${\rho_2 \equiv (\DD_3 - \DD_4)/2}$.\footnote{The quantities $\rho_1$ and $\rho_2$ are usually denoted as $a$ and $b$ in the CFT literature.} They can be characterized as solutions to a second-order differential equation~\cite{DO2} with the following asymptotic behavior: 
 \beq
 \label{eq:CBasymp}
 \widehat{G}_\DD^{(\l)}(u,v;\rho_1,\rho_2) \; \limu{u \to 0,v \to 1} \; (-1)^\l \frac{\l!\, \Gamma\!\left(d/2-1\right)}{2^\l \, \Gamma\!\left(\l + d/2-1\right)} u^{\DD/2} \, C_\l^{\left(d/2-1\right)}\!\left(\frac{1-v}{2\sqrt{u}} \right) \,,
 \eeq
 where the $C_j^{(a)}$ are Gegenbauer polynomials. There are systematic methods to compute the scalar conformal blocks $\widehat{G}_\DD^{(\l)}(u,v;\rho_1,\rho_2)$ for arbitrary $d$; we refer the reader to~\cite{DO2,DO1,DO3,Hogervorst:2013sma,Kos:2013tga} as a point of entry in the relevant literature.

\subsection{Exchange of logarithmic operators}\label{sec:logExchange}

Before discussing the conformal block decomposition of a four-point function of logarithmic operators, we will consider the exchange of a logarithmic rank-$\rnk$ operator to the four-point function~\reef{eq:4pt0} of four normal scalars. 
The discussion will revolve around the Casimir equation, a second-order PDE that the conformal block must obey. Let us start recalling the general framework~\cite{DO2}. We start from the quadratic Casimir of $\mfr{so}(d+1,1)$, namely
 \bea
L^2 \ldef \half L_{AB} L^{AB} =  \frac12 M_{\mu\nu}M^{\mu\nu}-D^2 + \frac12\left(P^\mu K_\nu+K^\mu P_\nu\right)\,,
\eea
where we denote the generators of $\mfr{so}(d+1,1)$ as $L_{AB} = - L_{BA}$. These generators act on local operators as first-order differential operators: $[L_{AB},\phi^i(x)] = \mfr{L}_{AB}^{(i)}  \phi^i(x)$. A short calculation then leads to the following identity:
\beq
\mca{L}^2  \braket{0}{\phi^1(x_1)\phi^2(x_2)}{\Psi} = \braket{0}{\phi^1(x_1)\phi^2(x_2) L^2}{\Psi}\,,
\qquad
\mca{L}^2 \ldef \frac{1}{2} \left( \mfr{L}_{AB}^{(1)} +  \mfr{L}_{AB}^{(2)}  \right)^2\,,
\eeq
for any arbitrary state $\ket{\Psi}$ in the Hilbert space of the CFT. 
Acting on a state $\ket{\Oo;\a}$, the Casimir evaluates to
\beq
\label{eq:CasOnState}
L^2 \ket{\Oo;\a} = \mca{C}_2(\DD_\Oo,\l_\Oo) \ket{\Oo;\a}, \quad \mca{C}_2(\DD,\l)  \ldef \DD(\DD-d) + \l(\l+d-2)\,.
\eeq
We conclude that the partial wave $W_\DD^{(\l)}$ obeys the following second-order differential equation:
\beq\label{eq:casWave}
\mca{L}^2 \cdot W_\DD^{(\l)}(x_i;\DD_j) =  \mca{C}_2(\DD,\l) \, W_\DD^{(\l)}(x_i;\DD_j)\,.
\eeq
By extracting the scale factor $\mbf{P}$, this equation descends to a second-order PDE for the conformal block $G_\DD^{(\l)}$:
\beq
\label{eq:casNormal}
\mca{D}_{u,v} \cdot G_\DD^{(\l)}(u,v;\DD_i) = \mca{C}_2(\DD,\l) \, G_\DD^{(\l)}(u,v;\DD_i)\,.
\eeq
This is the promised Casimir differential equation for $G_\DD^{(\l)}$. The precise form for the differential operator $\mca{D}_{u,v}$ is shown in~\cite{DO2}.\footnote{To be precise, the Casimir operator from that reference applies to the function $\widehat{G}_\DD^{(\l)}$ --- it is straightforward to obtain $\mca{D}_{u,v}$ from their expression using~\reef{eq:normalBlock}.}

We will now consider a more general case, namely one where the exchanged operator $\Oo_p$ is part of a logarithmic multiplet of rank $\rnk > 1$, hence $p$ takes values $p=1,\ldots,\rnk$. We will denote the dimension and spin of $\Oo_p$ as $\DD_\Oo$ resp.\@ $\l$. In that case, the Gram matrix
\beq
\mbf{G}_{p\a;q\b} \ldef \brakket{\Oo_p;\a}{\Oo_q;\b}
\eeq
will be non-trivial, due to  mixing between the different primaries $\Oo_p$ within the multiplet. We will require the inverse of the Gram matrix, $\mbf{G}^{p\a;q;\b}$, which is given by
\beq
\mbf{G}^{p\a;q;\b} = k_\Oo^{-1} \dd_{\a\b} \, V^{pq}(\pd_{\DD_\Oo}) \cdot g^{-1}_\a(\DD_\Oo,\l_\Oo), \quad
\label{eq:Vdef}
V^{pq}(\pd) =   \begin{cases}
    \pd^n /n!   & \text{if} \quad n \equiv p + q - \rnk -1 \geq 0  \\
  0 & \text{if} \quad n < 0
\end{cases}\,,
\eeq
provided that the two-point function $\expec{\Oo_p \Oo_q}$ is of canonical form. 
This computation is outlined in appendix~\ref{sec:gram}. By definition, the contribution of $\Oo_p$ to the four-point function~\reef{eq:4pt0} is given by 
\beq
\label{eq:unitRes}
\mathcal W^{(r)}_\Oo := \sum_{p,q = 1}^{\rnk} \sum_{\a,\b} \braket{0}{\phi^1(x_1)\phi^2(x_2)}{\Oo_q;\a} \mbf{G}^{p\a;q\b} \braket{\Oo_q;\b}{\phi^3(x_3) \phi^4(x_4)}{0} \,.  
\eeq
We will now show that the above partial wave obeys a differential equation similar to~\reef{eq:casWave}.
To set up this differential equation, we first remark that the Casimir $L^2$ does not act simply on $\ket{\Oo_p}$ or any descendant $\ket{\Oo_p;\a}$. The reason is that the matrix $\mbf{\DD}$ is not diagonal. As a matter of fact, we have
\beq
\label{eq:L2minCas}
\big( L^2 - \mca{C}_2(\DD_\Oo,\l_\Oo) \big) \ket{\Oo_p;\a} = \sum_{q = 1}^\rnk  \left[\dd_{q,p+1} \pd_{\DD_\Oo}  + \dd_{q,p+2} \half \pd_{\DD_\Oo}^2 \right] \mca{C}_2(\DD_\Oo,\l_\Oo) \ket{\Oo_{q};\a}\,.
\eeq
However, by acting $\rnk$ times with the operator in the LHS of~\reef{eq:L2minCas}, we obtain
\beq
\big( L^2 - \mca{C}_2(\DD_\Oo,\l_\Oo) \big)^\rnk \ket{\Oo_p;\a} = 0\,.
\eeq
It is easy to see that this observation leads to a higher-order Casimir equation for the above partial wave. For convenience, let's strip off a trivial scaling factor from the partial wave~\reef{eq:unitRes}:
\beq
W^{(r)}_\Oo \; \rdef \; \mbf{P}_{\DD_1 \dotsm \DD_4} \, G_\Oo(u,v)
 \eeq
Then applying the Casimir trick $\rnk$ times yields
\beq
\label{eq:logCasEq}
\big( \mca{D}_{u,v} - \mca{C}_2(\DD_\Oo,\l_\Oo)  \big)^\rnk \cdot G_\Oo(u,v) = 0\,.
\eeq
Indeed, this is a PDE of order $2\rnk$ that generalizes the normal Casimir equation~\reef{eq:casNormal}.

Finally, we will find the appropriate solution for Eq.~\reef{eq:logCasEq}. As a starting point, notice that a general solution is given by
\beq
\label{eq:sol00}
G_{\Oo} = \sum_{n=0}^{\rnk-1} a_n \, \frac{\pd^n}{\pd \DD_\Oo^n} G_{\DD_\Oo}^{(\l_\Oo)}(u,v;\DD_1,\DD_2,\DD_3,\DD_4)
\eeq
where $G_{\DD}^{(\l)}$ are the standard conformal blocks from the previous section. The only thing left to do is to compute the relative coefficients $a_n$ appearing in this sum. We do so by analyzing the limit of the $\phi^i$ four-point function where $x_3 \to x_4$. To be precise, we first compute the leading $\phi^3(x_3) \phi^4(x_4) \sim \Oo_p(x_4)$ OPE, which depends on $\rnk$ OPE coefficients $\la^{34\Oo}_i$ with $i=1,\ldots,\rnk$. Next, we plug this OPE into the four-point function. This gives
\begin{multline}
  \label{eq:asymp}
  u^{(\DD_1 + \DD_2 + \DD_3 +\DD_4)/6} \, G_{\Oo}(u,v)  \limu{x_3 \to x_4} k_\Oo^{-1}   \frac{(-\half)^\l \l! \, \Gamma(d/2-1)}{\Gamma(d/2-1+\l)} \, u^{\DD/2} \, C^{(d/2-1)}_\l\left(\frac{1-v}{2\sqrt{u}}\right) \\
  \times \; \sum_{p,q=1}^\rnk \la^{12\Oo}_p \la^{34\Oo}_q \; \begin{cases}
    \ln(\sqrt{u})^k / k! & \text{if} \quad k \equiv p + q - \rnk -1 \geq 0  \\
   0 & \text{if} \quad k < 0
   \end{cases} \nn
\end{multline}
where $\la^{12\Oo}_p$ are the OPE coefficients appearing in the three-point function $\expec{\phi^1 \phi^2 \Oo_p}$. By comparing this to the conformal block asymptotics~\reef{eq:CBasymp} we read off the $a_n$:
\beq
\label{eq:sol01}
a_n = \frac{k_\Oo^{-1}}{n!} \sum_{p+q=n+\rnk+1}^{} \la^{12\Oo}_p \la^{34\Oo}_{q}\,.
\eeq
This concludes the computation of logarithmic conformal blocks for the case of four-point functions of rank-one scalars.

For practical purposes, we will print the results for ranks $\rnk=2,3$. If the exchanged operator $\Oo$ has rank two, then the total contribution to the four-point function~\reef{eq:4pt0} is
\beq
\label{eq:rank2}
G_\Oo =  k_\Oo^{-1} \left[ \la^{12\Oo}_1 \la^{34\Oo}_2 + \la^{12\Oo}_2 \la^{34\Oo}_1 + \la^{12\Oo}_2 \la^{34\Oo}_2 \fracpt{}{\DD_\Oo} \right]\!G_{\DD_\Oo}^{(\l_\Oo)}(u,v;\DD_1,\DD_2,\DD_3,\DD_4)\,, \qquad [\rnk = 2].
\eeq
For rank $\rnk=3$ we have instead
\begin{multline}
\label{eq:rank3}
  G_\Oo = k_\Oo^{-1} \Big[  \la^{12\Oo}_1 \la^{34\Oo}_3  + \la^{12\Oo}_2 \la^{34\Oo}_2 + \la^{12\Oo}_3 \la^{34\Oo}_1  \\
    + \left( \la^{12\Oo}_2 \la^{34\Oo}_3 +  \la^{12\Oo}_3 \la^{34\Oo}_2  \right)\!\fracpt{}{\DD_\Oo} + \frac{1}{2!}  \la^{12\Oo}_3 \la^{34\Oo}_3 \frac{\pd^2}{\pd \DD_\Oo^2} \Big]G_{\DD_\Oo}^{(\l_\Oo)}(u,v;\DD_1,\DD_2,\DD_3,\DD_4)\,, \qquad [\rnk = 3]\,.
\end{multline}
In general, if the exchanged operator is of rank $\rnk$, the conformal block decomposition will contain derivatives up to order $\rnk - 1$ of the conformal block.

\subsection{Example: one external logarithmic operator}\label{sec:logExternal}

In this section, we will consider a problem that is orthogonal to the case considered above: we have in mind the four-point function
\beq
\label{eq:chi3phi}
\expec{\chi(x_1)\chi(x_2) \chi(x_3) \phi_a(x_4)} = \mbf{P}_{\DD_\chi \DD_\chi \DD_\chi \DD_\phi}(x_i) F_a(\tau_4)
\eeq
where $\chi$ is a rank-one scalar primary and $\phi_a$ has rank two. Following the logic of Sec.~\ref{sec:fourpt}, the functions $F_a$ can be decomposed as follows:
\beq
F_1 = \mca{F}_1(u,v) + \tau_4 \, \mca{F}_2(u,v)\,, \quad F_2 = \mca{F}_2(u,v)
\eeq
in terms of two conformally invariant functions $\mca{F}_{1,2}(u,v)$.\footnote{
  Although the $\mca{F}_i$ obey various crossing identities, they won't play a role in this section. 
}
Consequently, we need to consider two different conformal block decompositions, treating $\mca{F}_1$ and $\mca{F}_2$ separately. For simplicity, we will consider the exchange of a rank-one operator $\Oo$ of dimension $\DD_\Oo$ and spin $\l$. We will denote the contribution of $\Oo$ to the four-point functions~\reef{eq:chi3phi} as
\beq
\begin{pmatrix}\mca{F}_1(u,v) \\ \mca{F}_2(u,v) \end{pmatrix} = \begin{pmatrix} G_{\Oo,1}(u,v) \\ G_{\Oo,2}(u,v) \end{pmatrix} \; + \; \text{other multiplets}\,.
\eeq
A priori, the functions $G_{\Oo,i}(u,v)$ can depend on three OPE coefficients:  $\la^{\chi \chi \Oo} \propto \expec{\chi \chi \Oo}$ and $\la^{\chi \phi \Oo}_{i} \propto \expec{\chi \phi_i \Oo}$ for $i=1,2$.

We will now use the conformal Casimir to construct the partial waves $G_{\Oo,i}$. Let us first consider the four-point function $\expec{\chi \chi \chi \phi_2}$. The Casimir trick applies as before, and we obtain:
\beq
\label{eq:firstGuy}
\big(\mca{L}^2 - \mca{C}_2(\DD_\Oo,\l_\Oo) \big) \left[ \mbf{P}_{\DD_\chi\DD_\chi\DD_\chi\DD_\phi}(x) \, G_{\Oo,2}(u,v) \right] = 0\,.
\eeq
This is the differential equation we encountered above, and we know its solution:  $G_{\Oo,2}$ must be proportional to a normal conformal block. Taking into account the asymptotics of this four-point function in the limit $x_3 \to x_4$, we conclude that
\beq
\label{eq:G2cb}
G_{\Oo,2}(u,v) = \la^{\chi \chi \Oo} \la^{\chi \phi \Oo}_2 \, G_{\DD_\Oo}^{(\l_\Oo)}(u,v;\DD_\chi,\DD_\chi,\DD_\chi,\DD_\phi)\,.
\eeq
Next, we consider the $\expec{\chi \chi \chi \phi_1}$ four-point function. This time, we obtain
\beq
\label{eq:secondGuy}
\big(\mca{L}^2 - \mca{C}_2(\DD_\Oo,\l_\Oo) \big) \left[ \mbf{P} \, G_{\Oo,1}(u,v) \right] =
-\big(\mca{L}^2 - \mca{C}_2(\DD_\Oo,\l_\Oo) \big) \left[ \tau_4 \, \mbf{P} \,  G_{\Oo,2}(u,v) \right],
\quad \mbf{P} \ldef \mbf{P}_{\DD_\chi\DD_\chi\DD_\chi\DD_\phi}(x)\,.
\eeq
This has the form of an inhomogeneous differential equation, the RHS playing the role of a source term. We can rewrite the RHS using the identity
\beq
-\big(\mca{L}^2 - \mca{C}_2(\DD_\Oo,\l_\Oo) \big) \left[ \tau_4 \, \mbf{P} \,  G_{\Oo,2}(u,v) \right]
= \big(\mca{L}^2 - \mca{C}_2(\DD_\Oo,\l_\Oo) \big) \left[ \mbf{P} \,  \fracpt{}{\DD_\phi} G_{\Oo,2}(u,v) \right]
\eeq
which follows from deriving~\reef{eq:firstGuy} with respect to $\DD_\phi$. Therefore the general solution to~\reef{eq:secondGuy} is
\beq
\label{eq:G1cb}
G_{\Oo,1}(u,v) = \gamma \, G_{\DD_\Oo}^{(\l_\Oo)}(u,v;\DD_\chi,\DD_\chi,\DD_\chi,\DD_\phi) +
\la^{\chi \chi \Oo} \la^{\chi \phi \Oo}_2 \, \fracpt{}{\DD_\phi} G_{\DD_\Oo}^{(\l_\Oo)}(u,v;\DD_\chi,\DD_\chi,\DD_\chi,\DD_\phi)
\eeq
for some constant $\gamma$. The latter can also be fixed by considering the $x_3 \to x_4$ limit of the four-point function:
\beq
\gamma = \la^{\chi \chi \Oo} \la^{\chi \phi \Oo}_1\,.
\eeq

\subsection{Conformal partial waves: general case} 
\label{sec:general-cb}
We will now derive general formulae for the conformal partial waves in the logarithmic case using radial quantization methods. We consider the four-point function
\beq
\label{eq:4ptlog}
\expec{\phi^1_a(x_1)\phi^2_b(x_2)\phi^3_c(x_3)\phi^4_d(x_4)}
\eeq
of logarithmic operators $\phi^i_a$ of rank $\rnk_i$. We want to mimic what we did in Sec.~\ref{sec:blocksCFT} for ordinary CFTs and insert a complete set of states in this expression. 
Inserting the identity operator as in~\reef{eq:unitRes} in the correlator~\reef{eq:4ptlog} means that we must evaluate radial quantization matrix elements
\beq
\braket{0}{\phi^1_a(x_1) \phi^2_b(x_2)}{\Oo_p;\a} \quad \text{and} \quad  \braket{\Oo_q;\a}{\phi^3_c(x_3) \phi^4_d(x_4)}{0}.
\eeq
We will show that these are related to the universal matrix elements $\mca{M}$ and $\mca{M}'$ from Eq.~\reef{eq:Mme} in a simple way. The argument is the following. Recall that the three-point function $\expec{\phi^1_a \phi^2_b \Oo_p}$ can be expressed as
\beq
\label{eq:3ptnw}
\expec{\phi^1_a(x_1) \phi^2_b(x_2) \Oo_p^{(\l_\Oo)}(x_3;z)} = K_{abp}(\tau_1,\tau_2,\tau_3)\, \mbf{P}_{\DD_1 \DD_2 \DD_\Oo}(x_i) \left(X \cdot z\right)^{\l_\Oo}\,.
\eeq
The functions $K_{abp}(\tau_i)$ can be expressed in terms of a finite number of OPE coefficients $\la^{12\Oo}_{ijk}$, as discussed in Sec.~\ref{sec:three-pt}. 
Next, using Eq.~\reef{eq:taudef2} we can replace the $\tau_i$ variables in Eq.~\reef{eq:3ptnw} by partial derivatives with respect to $\DD_1,\DD_2$ and $\DD_\Oo$. The same holds at the level of the matrix elements, so we conclude that
\beq
\braket{0}{\phi^1_a(x_1) \phi^2_b(x_2)}{\Oo_p;\a} = K_{abp}\!\left(\fracpt{}{\DD_1},\fracpt{}{\DD_2},\fracpt{}{\DD_\Oo}\right) \mca{M}[1,2,\a,\DD_\Oo,\l_\Oo]\,,
\eeq
where the matrix element $\mca{M}$ is precisely the one from Eq.~\reef{eq:Mme}.

Likewise the three-point functions $\expec{\phi^3_c \phi^4_d \Oo_q}$ are described by a set of functions $K'_{cdq}(\tau_i)$, which encode a number of OPE coefficients $\la^{34\Oo}_{ijk}$. 
Therefore we can also relate the matrix elements $\braket{\Oo_q;\a}{\phi^3_c(x_3) \phi^4_d(x_4)}{0}$ to the matrix elements $\mca{M}'$ from Eq.~\reef{eq:Mme}:
\beq
\braket{\Oo_q;\a}{\phi^3_c(x_3) \phi^4_d(x_4)}{0} = K'_{cdq}\!\left(\fracpt{}{\DD_3},\fracpt{}{\DD_4},\fracpt{}{\DD_\Oo}\right) \mca{M}'[3,4,\a,\DD_\Oo,\l_\Oo].
\eeq
Bringing everything together, it follows that the contribution of the operator $\Oo_p$ to the four-point function~\reef{eq:4ptlog} is given by 
\begin{align}
  \label{eq:logPW}
  \expec{\phi^1_a\phi^2_b\phi^3_c\phi^4_d}  \sim & \,
  k_\Oo^{-1} \sum_{pq} \sum_\a \left[ K_{abp}\!\left(\fracpt{}{\DD_1},\fracpt{}{\DD_2},\fracpt{}{\DD_\Oo}\right) \mca{M}[1,2,\a,\DD_\Oo,\l_\Oo] \right] \\
  & \quad \times \left[ V^{pq}\!\left( \fracpt{}{\DD_\Oo} \right)  g_\a(\DD_\Oo,\l_\Oo)^{-1} \right]  \left[ K'_{cdq}\!\left(\fracpt{}{\DD_3},\fracpt{}{\DD_4},\fracpt{}{\DD_\Oo}\right) \mca{M}'[3,4,\a,\DD_\Oo,\l_\Oo] \right]. \nn
\end{align}
Formula~\reef{eq:logPW} is not very enlightening, but as we will now show it can in fact be rewritten in a more useful way, in terms of a differential operator acting on a normal partial wave $W_{\DD}^{(\l)}(x_i,\DD_j)$. 

To see this, let's first assume that the exchanged multiplet $\Oo$ is non-logarithmic, i.e.\@ of unit rank. Then it's possible to commute the sum over descendants $\a$ with the operators $K_{ab}(\pd_{\DD_1},\pd_{\DD_2})$ and $K'_{cd}(\pd_{\DD_3},\pd_{\DD_4})$. By doing so, it immediately follows that the partial wave becomes
\beq
  \expec{\phi^1_a\phi^2_b\phi^3_c\phi^4_d}  \sim k_\Oo^{-1}\,  K_{ab}\!\left(\fracpt{}{\DD_1},\fracpt{}{\DD_2}\right) K'_{cd}\!\left(\fracpt{}{\DD_3},\fracpt{}{\DD_4}\right) \cdot W_{\DD_\Oo}^{(\l_\Oo)}(x_i;\DD_j) \,,\qquad [\rnk_\Oo = 1]
\eeq
which is of the desired form, namely a differential operator acting on an ordinary partial wave, cf. Eq.~\reef{eq:cpwnormal}.
Suppose now that the exchanged operator $\Oo_p$ is logarithmic, i.e.\@ of rank $\rnk_\Oo > 1$. In this case, the sum over descendants $\a$ can no longer be commuted with the differential operators $K_{abp}$, $V^{pq}$ and $K'_{cdq}$, since both the matrix elements $\mca{M}, \mca{M}'$ and the norm $g_\a$ depend on the dimension $\DD_\Oo$. It is still possible to reorganize the expression~\reef{eq:logPW}, but this requires using the precise form of the operators $K_{abp}$ and $K'_{cdq}$ as fixed by conformal invariance. This computation is outlined in Appendix~\ref{sec:gram}. In the end the partial wave~\reef{eq:logPW} can be expressed in terms of the OPE coefficients $\la^{12\Oo}_{ijk}$ and $\la^{34\Oo}_{ijk}$ as:
\begin{align}
  \label{eq:fullPW}
  \hspace{-5mm} \expec{\phi^1_a\phi^2_b\phi^3_c\phi^4_d}  \; &\sim \; \mca{D}^{abcd}_\Oo \cdot W_{\DD_\Oo}^{(\l_\Oo)}(x_i;\DD_j),\\
  \mca{D}^{abcd}_\Oo &\ldef k_\Oo^{-1} 
  \sum_{l_1=0}^{\rnk_1-a}   \sum_{l_2=0}^{\rnk_2-b}\sum_{l_3=0}^{\rnk_3-c} \sum_{l_4 = 0}^{\rnk_4-d}
  \sum_{p,q=1}^{\rnk_\Oo}  \la^{12\Oo}_{(a+l_1)(b+l_2)p} \, \la^{34\Oo}_{(c+l_3)(d+l_4)q}   \nn \prod_{i=1}^4 \frac{1}{l_i!} \frac{\pd^{l_i}}{\pd \DD_i^{l_i}} \; V^{pq}\!\left(\fracpt{}{\DD_\Oo}\right). 
  \end{align}
This ``master formula'' describes the most general partial wave for a scalar four-point function in a logarithmic CFT.

\subsection{From partial waves to conformal blocks}

We will now obtain the conformal blocks from the conformal partial waves. Recall from Sec.~\ref{sec:fourpt} that a logarithmic four-point function can be written as
 \beq
 \expec{\phi^1_a(x_1)\phi^2_b(x_2)\phi^3_c(x_3)\phi^3_d(x_4)} = F_{abcd}(u,v,\zeta_i) \, \mbf{P}_{\DD_1 \DD_2 \DD_3 \DD_4}(x_j).
 \eeq
Let's isolate the part of $F_{abcd}$ that is independent of the $\zeta_i$ variables:
\beq
\label{eq:hdef}
 F_{abcd}(u,v,\zeta_i) = h_{abcd}(u,v) + \sO(\zeta_i).
 \eeq
 In what follows, we will develop a conformal block decomposition for the function $h_{abcd}(u,v)$.
The crucial idea will be to use Eq.~\reef{eq:zetadef2}: it implies that if we act with the differential operator $\mca{D}^{abcd}_\Oo$ from Eq.~\reef{eq:fullPW} on a partial wave, we obtain
 \beq
 \mca{D}^{abcd}_\Oo \cdot W_{\DD_\Oo}^{(\l_\Oo)}(x_i;\DD_j) = \left[ \mca{D}^{abcd}_\Oo \cdot G_{\DD_\Oo}^{(\l_\Oo)}(u,v;\DD_i)  \right]  \mbf{P}_{\DD_1 \DD_2 \DD_3 \DD_4}(x_j)  + \sO(\zeta_i).
 \eeq
 Since this is true for all values of $a,b,c,d$, it follows that the functions $h_{abcd}(u,v)$ admit the following conformal block decomposition: 
\beq
\label{eq:hcbdef}
\boxed{
  h_{abcd}(u,v) = \sum_\Oo \mca{D}^{abcd}_\Oo \cdot G_{\DD_\Oo}^{(\l_\Oo)}(u,v;\DD_i)\,.
  }
 \eeq
 This decomposition is very similar to the ordinary one~\reef{eq:CBdecnorm}, the main difference being that here every term is a linear combination of a normal conformal block \emph{and} its partial derivatives.
 
At this point, we can compare the master formula~\reef{eq:fullPW} to results obtained previously using the Casimir equation. First we consider the exchange of a rank-$\rnk$ operator $\Oo_p$  exchanged in a four-point function of non-logarithmic scalars $\phi^i$, as in Sec.~\ref{sec:logExchange}. Recall that the three-point function $\expec{\phi^1 \phi^2 \Oo_p}$ was characterized by $\rnk$ OPE coefficients $\la^{12\Oo}_p$, and likewise the three-point function  $\expec{\phi^3 \phi^4 \Oo_p}$ was characterized by $\rnk$ OPE coefficients $\la^{34\Oo}_q$. 
According to the formula~\reef{eq:fullPW}, the complete partial wave associated to $\Oo$ is then given by 
\beq
k_\Oo^{-1} \sum_{p,q = 1}^{\rnk} \la^{12\Oo}_p \la^{34\Oo}_q \, V^{pq}\left(\fracpt{}{\DD_\Oo} \right) G_{\DD_\Oo}^{(\l_\Oo)}(u,v;\DD_1,\DD_2,\DD_3,\DD_4)\,.
\eeq
It is easy to see that this expression is equivalent to Eqs.~\reef{eq:sol00} and~\reef{eq:sol01} that we derived before. Second, we consider the exchange of a non-logarithmic operator $\Oo$ to the four-point function $\expec{\chi \chi \chi \phi_a}$, where $\chi$ is of rank one and $\phi_a$ has rank two. This was considered before in Sec.~\ref{sec:logExternal}. Again, it is easy easy to see that formula~\reef{eq:fullPW} is equivalent to the previous results~\reef{eq:G2cb} and~\reef{eq:G2cb}.

\subsection{Rank-two four-point function}\label{sec:gooddiffop}

Let's now turn to the four-point function of a single rank-two operator $\phi_a$. In Sec.~\ref{sec:ranktwo4pt}, it was shown that the correlators $\expec{\phi_a \phi_b \phi_c \phi_d}$ are encoded by five different functions $\mca{F}_i(u,v)$ of $u$ and $v$, satisfying a number of crossing relations~\reef{eq:cross4}. Here we will develop a conformal block decomposition for each of the $\mca{F}_i$ and consider the consequences of Bose symmetry.

Consider the contribution of a rank-$\rnk_\Oo$ exchanged operator $\Oo_p$ of spin $\l$. Recall from Sec.~\ref{sec:3ptranktwo} that depending on whether $\l_\Oo$ is even or odd, Bose symmetry imposes various constraints on the OPE coefficients $\la^{\phi \phi \Oo}_{ijk}$. If $\l_\Oo$ is even, let's denote 
\beq
\la^{\phi \phi \Oo}_{11p} \equiv a^\Oo_p\,, \quad \la^{\phi \phi \Oo}_{12p} = \la^{\phi \phi \Oo}_{21p} \equiv b^\Oo_{p}\,, \quad \la^{\phi \phi \Oo}_{22p} \equiv c^\Oo_p\,, \qquad [\l_\Oo \text{ even}]
\eeq
and if $\l_\Oo$ is odd
\beq
\la^{\phi \phi \Oo}_{11p} =0\,, \quad \la^{\phi \phi \Oo}_{12p} = -\la^{\phi \phi \Oo}_{21p} \equiv b^\Oo_{p}\,, \quad \la^{\phi \phi \Oo}_{22p} = 0\,, \qquad [\l_\Oo \text{ odd}].
\eeq
For simplicity, we will set $k_\Oo$ to one, which can always be accomplished by a redefinition of the OPE coefficients.

Consider first the case where $\l_\Oo$ is even. Then the relevant differential operators are given by
\bsub
\label{eq:evendiffop}
\begin{align}
  \mca{D}_\Oo^{1111} &=  \sum_{p,q=1}^{\rnk_\Oo}\,\big[ a^\Oo_p + b^\Oo_p\, (\pd_{\DD_1} + \pd_{\DD_2}) + c^\Oo_p \, \pd_{\DD_1}\pd_{\DD_2} \big] \nn \\
  & \hspace{40mm} \times \big[ a^\Oo_q + b^\Oo_q\, (\pd_{\DD_3} + \pd_{\DD_4}) + c^\Oo_q \, \pd_{\DD_3}\pd_{\DD_4} \big] V^{pq}(\pd_{\DD_\Oo}) \,, \label{eq:f1}\\
   \mca{D}_\Oo^{1112} &= \sum_{p,q=1}^{\rnk_\Oo} \big[ a^\Oo_p + b^\Oo_p \, (\pd_{\DD_1} + \pd_{\DD_2}) + c^\Oo_p \, \pd_{\DD_1}\pd_{\DD_2} \big] \big[b^\Oo_q + c^\Oo_q\, \pd_{\DD_3} \big] V^{pq}(\pd_{\DD_\Oo}) \,, \label{eq:f2} \\
   \mca{D}_\Oo^{1122} &=   \sum_{p,q=1}^{\rnk_\Oo} \big[ a^\Oo_p + b^\Oo_p \, (\pd_{\DD_1} + \pd_{\DD_2}) + c^\Oo_p \, \pd_{\DD_1}\pd_{\DD_2} \big] c^\Oo_q\, V^{pq}(\pd_{\DD_\Oo})\,, \label{eq:f3} \\
   \mca{D}_\Oo^{1221} &=   \sum_{p,q=1}^{\rnk_\Oo} \big[ b^\Oo_p + c^\Oo_p \, \pd_{\DD_1} \big]\big[ b^\Oo_q + c^\Oo_q \, \pd_{\DD_4} \big]  V^{pq}(\pd_{\DD_\Oo})\,, \label{eq:f3alt} \\
 \mca{D}_\Oo^{1222} &=  \sum_{p,q=1}^{\rnk_\Oo} \big[ b^\Oo_p + c^\Oo_p \, \pd_{\DD_1} \big] c^\Oo_q\, V^{pq}(\pd_{\DD_\Oo}) \,, \quad 
  \mca{D}_\Oo^{2222} =  \sum_{p,q=1}^{\rnk_\Oo} c^\Oo_p c^\Oo_q \, V^{pq}(\pd_{\DD_\Oo})\,. \label{eq:d2222}
\end{align}
\esub
When acting on a conformal block, Eqs.~\reef{eq:f1},~\reef{eq:f2} and~\reef{eq:f3} can be rewritten using the identity~\reef{eq:del12}.

Similarly, for odd $\l_\Oo$ we have
\bsub
\label{eq:odddiffop}
\begin{align}
  \mca{D}_\Oo^{1111} &=  \sum_{p,q=1}^{\rnk_\Oo}\, b^\Oo_p b^\Oo_q \, (\pd_{\DD_2} - \pd_{\DD_1})(\pd_{\DD_4} - \pd_{\DD_3})   V^{pq}(\pd_{\DD_\Oo}) \,,\label{eq:f1n}\\
  \mca{D}_\Oo^{1112} &=  \sum_{p,q=1}^{\rnk_\Oo} b^\Oo_p b^\Oo_q \, (\pd_{\DD_2} - \pd_{\DD_1})   V^{pq}(\pd_{\DD_\Oo})\,, \label{eq:f2n} \\
  \mca{D}_\Oo^{1221} &= - \sum_{p,q=1}^{\rnk_\Oo}  b^\Oo_p b^\Oo_q \, V^{pq}(\pd_{\DD_\Oo}) \label{eq:f3nalt} \,,\\
  \mca{D}_\Oo^{1122} &= \mca{D}_\Oo^{1222} = \mca{D}_\Oo^{2222} = 0\,.
\end{align}
\esub

The full conformal block decompositions for the functions $\mca{F}_i$ are obtained by summing over both even- and odd-spin operators. We obtain:
\bea
\mca{F}_1(u,v) =\sum_\Oo \mca{D}_\Oo^{1111} \cdot G_{\DD_\Oo}^{(\l_\Oo)}(u,v;\DD_i) \,,
\qquad  & \mca{F}_2(u,v) = \, \displaystyle{\sum_\Oo} \mca{D}_\Oo^{1112} \cdot G_{\DD_\Oo}^{(\l_\Oo)}(u,v;\DD_i)\,, \\
\mca{F}_3(u,v) = \sum_\Oo \mca{D}_\Oo^{1122} \cdot G_{\DD_\Oo}^{(\l_\Oo)}(u,v;\DD_i)\,,
\qquad &  \mca{F}_3(v,u) = \, \displaystyle{\sum_\Oo}  \mca{D}_\Oo^{1221} \cdot G_{\DD_\Oo}^{(\l_\Oo)}(u,v;\DD_i)\,,\\
 \mca{F}_4(u,v) = \sum_\Oo \mca{D}_\Oo^{1222} \cdot G_{\DD_\Oo}^{(\l_\Oo)}(u,v;\DD_i)\,,
\qquad & \mca{F}_5(u,v) = \, \displaystyle{\sum_\Oo}  \mca{D}_\Oo^{2222} \cdot G_{\DD_\Oo}^{(\l_\Oo)}(u,v;\DD_i)\,.
\eea
The conformal blocks are to be evaluated at $\DD_1 = \DD_2 = \DD_3 = \DD_4 \equiv \DD_\phi$.

We  conclude this section with some remarks about crossing equations and the conformal bootstrap. It may readily be checked that the crossing relation $\mca{F}_i(u,v) = \mca{F}_i(u/v,1/v)$ is trivially satisfied for all $\mca{F}_i$. This follows from the conformal block identity Eq.~\reef{eq:uvRels}.
The crossing identity $\mca{F}_i(u,v) = \mca{F}_i(v,u)$ however is non-trivial, so e.g.\@ for $\mca{F}_1(u,v)$ we must have
\beq
\label{eq:bs1}
  \sum_\Oo  \mca{D}_\Oo^{1111} \cdot  \left[ G_{\DD_\Oo}^{(\l_\Oo)}(u,v;\DD_i) - G_{\DD_\Oo}^{(\l_\Oo)}(v,u;\DD_i) \right] = 0\,.
\eeq
The same bootstrap equation holds for $\mca{F}_2$, $\mca{F}_4$ and $\mca{F}_5$, with $\mca{D}_\Oo^{1111}$ replaced by $\mca{D}_\Oo^{1112}$, $\mca{D}_\Oo^{1222}$ and $\mca{D}_\Oo^{2222}$. 
Only for the function $\mca{F}_3$, we have two inequivalent conformal block decompositions~\cite{Kos:2014bka}, namely
\beq
\label{eq:bs2}
  \sum_\Oo \mca{D}_\Oo^{1122}  \cdot  G_{\DD_\Oo}^{(\l_\Oo)}(u,v;\DD_i) = \sum_\Oo \mca{D}_\Oo^{1221} \cdot G_{\DD_\Oo}^{(\l_\Oo)}(v,u;\DD_i)\,.
\eeq
Summarizing, imposing that the $\phi_a$ four-point function is crossing symmetric leads to five distinct bootstrap equations.


\section{Holographic logCFT}
\label{sec:ads}

In this section, we will review and derive new results on logCFTs using the AdS/CFT correspondence. Holographic duals to logCFTs were first considered for scalar operators in \cite{Ghezelbash1999} and \cite{Kogan1999c}; a thorough discussion is given in~\cite{Bergshoeff2012a}. For the interesting spin-2 case there have been several investigations, mostly in 3$d$ bulk theories, with a review of this and related topics in ref.~\cite{Grumiller2013}. 

In all known cases, the holographic duals of logCFTs involve higher derivative equations of motion. Before we begin a detailed analysis of some examples, let us see how such actions can be motivated. The idea is to consider a scalar field propagating in (Euclidean) AdS$_{d+1}$, coupled linearly to a random potential $V(x)$:
\bea
S=-\int \uud^{d+1}x \sqrt{g}\bigg[\frac 12\phi (\dsq-m^2) \phi-V \phi \bigg], \qquad \overline{V(x)V(y)}= \frac{v^2}{\sqrt{g}}\, \delta(x-y),
\eea
Here $g$ is the determinant of the AdS metric, and the bar represents disorder averaging.
To obtain the disorder averaged correlators of $\phi$ we use the replica trick~\cite{Cardy:1996xt}. 
Introducing $N$ copies of $\phi$ and performing the disorder average the action becomes
\bea
S_N=-\int \uud^{d+1}x \sqrt{g}\left[\frac 12\sum_{a=1}^N \phi_a (\dsq-m^2) \phi_a- \frac{v^2}2\left(\sum_{a=1}^N\phi_a\right)^2\right]\,.
\eea
Since this is a free theory, all correlation functions reduce at most to products of two point functions. Furthermore, there are only two kinds of two point functions: $\langle \phi_a \phi_a\rangle$ and $\langle \phi_a \phi_b\rangle$ with $a\neq b$, which in the $N\to 0$ limit become $\overline{\langle \phi \phi\rangle}$ and $\overline{\langle\phi\rangle\langle \phi\rangle}$ respectively. Hence, for the purpose of computing correlation functions we can write the action in terms of two fields, $\phi_1$ and $N-1$ copies of what we will call $\phi_2$. Then in the $N\to 0$ limit we have
\bea
S_{N=0}=-\int \uud^{d+1}x \sqrt{g}\left[\frac 12\phi_+ (\dsq-m^2) \phi_- -\frac{v^2}2\left(\phi_-\right)^2\right]\,,
\eea
with $\phi_\pm=\phi_1\pm \phi_2$. We are now free to integrate out field $\phi_-$ using its equation of motion to finally get
\bea
S=-\frac{1}{4v^2}\,\int \uud^{d+1}x \sqrt{g}\left[\frac {1}{2}\phi_+ (\dsq-m^2)^2 \phi_+ \right].
\eea
The action is indeed higher derivative, and as we shall see shortly it describes in fact a boundary logCFT of rank-2. If the reader finds this derivation troublesome, notice that the same result could have been obtained by computing the averaged correlators directly, since this is a free theory.

It is possible to obtain higher ranks in a similar fashion but we must allow for a more general, non-gaussian, potential.
It is also straightforward to obtain a similar result for spin-1 and linearized spin-2. Going to non-linear level at spin-2 seems more difficult but it seems quite likely that the action \reef{eq:nmg} shown below may be derived along similar lines.

\subsection{Scalars}
\label{sec:holoscalar}
Here we shall consider scalar theories in some detail. We begin by showing how to reconstruct all two point functions in a straightforward way which does not require holographic renormalization \cite{Haro2001,Skenderis2002}. The procedure is general and may be applied to higher spins. We also compute some sample three- and four-point functions, finding perfect agreement with the general results of Sec.~\ref{sec:three-pt} and Sec.~\ref{sec:CB}.

\subsubsection{Two point functions}
Our starting point is the action
\bea
S = -\int d^{d+1} x \sqrt{g} \left[\frac 12 \phi(\dsq-m^2)^{\rnk}\phi \right]. \label{eq:scalaraction}
\eea
which generalizes the $r=2$ case deduced at the beginning of this section. The metric is given by:
\bea
\uud s^2 =g_{\mu\nu}\uud x^\mu \uud x^{\nu}=\frac{\uud z^2+\uud x_i \uud x^i}{z^2}.
\eea
Provided that $\rnk \geq 2$ the action is higher-derivative and the theory is non-unitary. One way to see this is to think of the equation of motion satisfied by the field $\phi$ as a limiting case of having $\rnk$ distinct masses. The propagator then necessarily includes modes with negative norm. The fact that all masses are taken to be identical leads to degeneracies, which are resolved by the appearance of logarithms. Indeed, solving the equation of motion close to the AdS boundary at $z=0$ gives
\bea
\phi (z,x) &\limu{z \to 0} & z^\Delta \sum_{i=1}^{\rnk} \tilde \phi_i^{(0)}(x) \frac{1}{(i-1)!}\ln^{i-1}(z)+z^{d-\Delta} \sum_{i=1}^{\rnk} \phi_i^{(0)}(x) \frac{(-1)^{i-1}}{(i-1)!}\ln^{i-1}(z)+ \, \ldots \label{eq:phiexp}
\eea
with $\Delta$ the largest root of the equation $m^2=\Delta(\Delta-d)$. The $\phi_i^{(0)}(x)$ are fixed boundary values or sources, and the $\tilde \phi_i^{(0)}(x)$ will shortly be related to expectation values of some operators $\langle \mathcal O_i\rangle$. Acting with the the dilatation generator $z\partial_z+x^\mu \partial_\mu$ on $\phi$ induces the map:
\begin{align}
\tilde \phi_i^{(0)} \, &\to \, (x^\mu \partial_\mu+\Delta)  \tilde \phi_i^{(0)} \, + \, \tilde\phi_{i+1}^{(0)}\,, \label{eq:ops} \\
\phi_i^{(0)} \, &\to \,  (x^\mu \partial_\mu+d-\Delta)\phi_i^{(0)}  - \, \phi_{i+1}^{(0)}\,.\label{eq:sources}
\end{align}
This is consistent with the transformation law \reef{eq:Dward} for logarithmic multiplets. One can check the same is true under the action of the other generators, so that $\tilde{\phi}_i^{(0)}$ and $\phi_i^{(0)}$ transform as logarithmic primaries of dimension $\DD$ resp.\@ $d-\DD$.
The normalizations in~\reef{eq:phiexp} were chosen to obtain these particularly simple transformations, which are in agreement with our conventions. 

The source terms in the expansion of the field $\phi$ can be chosen to couple to the logarithmic multiplet in the following way:
\bea
Z_{\mbox{\tiny bulk}}[\phi_i^{(0)}]=\langle \exp\left(\kappa \int \uud^d x\, \sum_{i=1}^\rnk \mathcal O_{\rnk-i+1}(x) \,\phi_i^{(0)}(x) \right) \rangle_{\mbox{\tiny logCFT}}\,. \label{eq:z}
\eea
This form is invariant under conformal transformations, provided that operators and sources transform appropriately as in  \reef{eq:ops} and \reef{eq:sources}. Since~\reef{eq:z} defines a generating functional for logCFT correlation functions, insertions of $\cO_i$ in correlators can be obtained by switching on sources. Delta function sources are particularly useful, and may be obtained by the use of bulk-to-boundary propagators. To construct propagators appropriate for general rank $\rnk$, we start from the ordinary bulk-to-boundary propagator $K_\DD(z,x)$ satisfying the $\rnk=1$ equation of motion:
\bea
K_\Delta(z,x)=N_\Delta\, \frac{z^{\Delta}}{(z^2+x^2)^\Delta} \; \limu{z \to 0} \;  z^{d-\Delta}\,\delta(x)+z^\Delta \frac{N_\Delta}{|x|^{2\Delta}}+\ldots
\eea
where $N_\Delta=\Gamma(\Delta)/[\pi^{d/2} \Gamma(\Delta-d/2) ]$ was chosen so that $K_\DD$ has a pure delta function source at the boundary. It follows that to obtain a source for the log components one can use:
\bea
\frac{1}{(j-1)!}\partial_\Delta^{j-1} K_\Delta(z,x)\quad \Rightarrow \quad \phi_k^{(0)}(x)=\delta_{jk} \, \delta(x).
\eea
The next step is to identify the map between asymptotic values $\tilde \phi^{(0)}_i$ and expectation values $\langle \cO_j \rangle$. The simplest way to do this is to consider the bulk-to-bulk propagator~\cite{Klebanov1999}. The idea is that a boundary operator insertion can be obtained by taking a bulk point to the boundary. Since the normalization of the bulk-to-bulk propagator is fixed by the action, this will give us the desired relation. The advantage of this method is that it will avoid us the somewhat cumbersome procedure of holographic renormalization. The propagator satisfies the equation of motion:
\beq
\label{eq:eom}
\left(\dsq - m^2\right)^{\rnk} G^{[\rnk]}_{\Delta}(x_1,z;x_2,y)=\frac{\delta^{d}(x_1-x_2) \delta(z-y)}{\sqrt{g}}\,.
\eeq
Such propagators $G_\DD^{[\rnk]}(x_1,z;x_2,y)$ are not to be confused with the conformal blocks $G_{\DD}^{(\l)}(u,v)$ from Sec.~\ref{sec:CB}. To solve this equation, consider first the known rank-1 bulk-to-bulk propagator, 
\beq
G^{[1]}_{\Delta}(\xi) =\frac{\Gamma(\Delta)}{2 \pi^{\frac d2}\Gamma(1+\Delta-d/2)}\, \frac{1}{\xi^\Delta}\, _2F_1\!\left(\Delta,\frac{1+2\Delta-d}2,1+2\Delta-d,-\frac{4}{\xi}\right)\,,
\eeq
with $\xi \ldef [(z-y)^2+(x_1-x_2)^2]/(zy)$. To compute the propagator for rank $\rnk \geq 2$, we notice the following identity
\bea
(\Box-m^2)^k \partial_{m^2}^i G^{[1]}_\Delta &=& \frac{i!}{(i-k)!} \partial_{m^2}^{i-k}G^{[1]}_\Delta \qquad \,, [i \geq k \geq 0]
\eea
which follows from the $\rnk=1$ equation of motion.  
A solution to the rank-$\rnk$ equation of motion~\reef{eq:eom} is therefore furnished by
\bea
G_\Delta^{[r]}(\xi) =\frac{\partial^{r-1}_{m^2}G^{[1]}_\Delta(\xi)}{(r-1)!}.\, 
\eea
For instance,  for ranks two and three we have:
\bea
G_\Delta^{[2]} &=&\frac{\partial_\Delta G^{[1]}_\Delta}{2\Delta-d}, \qquad
G_\Delta^{[3]} =\frac{\partial_\Delta^2-2\partial_\Delta}{2(2\Delta-d)^2}\,G^{[1]}_\Delta. \label{eq:bb2}
\eea
To determine the map between asymptotic values and operators, let us consider first the rank-1 case. Taking one of the points to the boundary one obtains
\bea
G^{[1]}_{\Delta}(z,x_1;y,x_2) \; \limu{z \to 0} \; z^\Delta \frac{K_\Delta(y,x_1-x_2)}{2\Delta-d}\,+\ldots
\eea
Since $K_\Delta$ corresponds to an insertion of $\kappa \mathcal O$, we determine from the above that
\bea
\tilde \phi^{(0)}(x)=\frac{\kappa \langle \cO(x) \rangle}{2\Delta-d}.
\eea
One may then compute the two point function by reading off the expectation value as a function of the source, by simply expanding the bulk-to-boundary propagator
\bea
K_\Delta \limu{z\to0} z^\Delta \frac{N_\Delta}{|x|^{2\Delta}}&=&z^\Delta\frac{\kappa \langle \cO(x)\rangle_{\phi^{(0)}=\delta(x)}}{2\Delta-d}=z^\Delta\frac{\kappa^2\langle \cO(x)\cO(0)\rangle}{2\Delta-d}\nonumber \\
&\Rightarrow& \langle \cO(x) \cO(0)\rangle= \left(\frac{N_\Delta (2\Delta-d)}{\kappa^2}\right)\, \frac{1}{|x|^{2\Delta}}\,.
\eea
Now let us consider the rank-2 case. Repeating the logic, with the propagators $K_\Delta, \partial_\Delta K$ 
corresponding to insertions of $\kappa \cO_2,\kappa \cO_1$ respectively, we find the dictionary
\bea
\tilde \phi_1^{(0)}&=&\frac{\kappa \langle \cO_1\rangle}{(2\Delta-d)^2}-2 \frac{\kappa \langle \cO_2\rangle}{(2\Delta-d)^3}, \qquad \tilde \phi_2^{(0)}=\frac{\kappa \langle \cO_2\rangle}{(2\Delta-d)^2}.
\eea
To compute the two point functions we switch on sources and work to linear order. Expanding $K_\Delta$
we learn
\bea
\frac{\kappa^2 \langle \cO_1(x) \cO_2(0)\rangle}{(2\Delta-d)^2}&=&\frac{N_\Delta}{|x|^{2\Delta}}, \qquad \langle \cO_2(x) \cO_2(0)\rangle=0
\eea
and from $\partial_\Delta K$ we also get
\bea
\frac{\kappa^2 \langle \cO_1(x)\cO_1(0)\rangle}{(2\Delta-d)^2}&=&\frac{N_\Delta}{|x|^{2\Delta}}\left(-\ln x^2+\frac{\partial_\Delta N_\Delta}{N_\Delta}+\frac{2}{2\Delta-d}\right).
\eea
Altogether the two point functions are consistent with those of a rank-2 logCFT. 

We would like to make a choice such that the two point functions take the canonical form \reef{eq:2ptscal}. For this we can use the freedom to do field redefinitions of the form $\cO_i \to \cO_i+\sum_{j>i} \alpha_j \cO_{j}$ (cf. Eq. \reef{eq:fieldred}). In practice we perform such a redefinition on the sources, by writing
\bea
\phi(x,z)  \; \limu{z \to 0} \; z^\Delta \tilde \phi_1^{(0)}(x) + z^\Delta \ln(z)\tilde \phi_2^{(0)}(x) + z^\Delta \big(\phi_1^{(0)}(x)+\alpha \phi_2^{(0)}(x)\big)+z^\Delta \ln(z)\tilde \phi_2^{(0)}(x)+\ldots
\eea
while keeping the form of the couplings \reef{eq:z} fixed.
With this choice, an insertion of $\kappa \cO_1$ now corresponds to a bulk-to-boundary propagator of the form $\partial_\Delta K_\Delta+\alpha K_\Delta$. Similarly, the identification between subleading terms and operators also gets corrected,
\bea
\tilde \phi_1^{(0)}&=&\frac{\kappa \langle \cO_1\rangle}{(2\Delta-d)^2}-2 \frac{\kappa \langle \cO_2\rangle}{(2\Delta-d)^3}-\alpha\frac{\kappa \langle \cO_2\rangle}{(2\Delta-d)^2}  
\eea
and so does the $\langle \cO_1\cO_1\rangle$ two point function
\bea
\frac{\kappa^2 \langle \cO_1(x)\cO_1(0)\rangle}{(2\Delta-d)^2}=\ldots + 2\frac{\alpha N_\Delta}{|x|^{2\Delta}}\,,
\eea
where the dots stand for the previous result. Hence, choosing
\bea
\kappa=(2\Delta-d)\sqrt{N_\Delta}, \qquad \alpha=-\frac{1}{(2\Delta-d)}-\frac {\partial_\Delta N_\Delta}{2\,N_\Delta} \label{eq:normal}
\eea
we finally obtain
\bea
\langle \cO_1(x)\cO_1(0)\rangle=-\frac{\ln x^2}{|x|^{2\Delta}}, \qquad 
\langle \cO_1(x)\cO_2(0)\rangle=\frac 1{|x|^{2\Delta}}, \qquad \langle \cO_2(x)\cO_2(0)\rangle=0.
\eea

It should be clear that this entire procedure can be easily generalized to any choice of rank. Let us merely outline the most salient features of such a calculation. The general rank-$r$ bulk-to-bulk propagator satisfies
\bea
G_\Delta^{[r]}(z,x_1; y,x_2) \; \limu{z \to 0} \frac{z^\Delta}{(2\Delta-d)^{r}}  
\sum_{k=0}^{r-1}\left[\binom{r-1}{k} \ln^{k} z (\partial_\Delta^{r-1-k} K_\Delta+\ldots)\right]\,,
\eea
where the dots stand for terms with less derivatives of $K_\Delta$. Switching on a source $\phi^{(0)}_i$ turns on an operator $\kappa \cO_{r-i+1}$ and corresponds to a bulk-to-boundary propagator $\frac{1}{(i-1)!}\partial_\Delta^{i-1} K_\Delta$, from which one identifies
\bea
\tilde \phi_i^{(0)}=(r-1)!\frac{\kappa \langle \cO_i\rangle}{(2\Delta-d)^r}+ \sum_{j>i} a^i_j \langle\cO_j\rangle\,,
\eea
for some coefficients $a_j^i$.
The two point functions can be determined by expanding bulk-to-boundary propagators:%
\bea
\frac{1}{(i-1)!}\partial_\Delta^{i-1} K_\Delta \to z^\Delta N_\Delta \sum_{k=1}^{i}\left[\ln^{k-1}(z)\frac{(-1)^{i-1-k} \ln^{i-k} x^2}{(i-k)!(k-1)!}\,   +\ldots\right],
\eea
where now the dots stand for smaller powers of logarithms. From this expression we can read off the $\tilde \phi_i^{(0)}$ and deduce
\bea
\frac{(r-1)!}{(2\Delta-d)^r N_\Delta } \kappa^2\langle \cO_i(x) \cO_{j}\rangle=\, \frac{(-1)^{m}}{m!} \frac{\ln^{m}(x^2)}{|x|^{2\Delta}}+\mbox{\ldots}, \qquad m=r+1-i-j\geq 0 \label{eq:twopt}\eea
and zero otherwise. This is in agreement with our general results \reef{eq:2ptscal} and also those of \cite{Bergshoeff2012a}.

\subsubsection{Interactions}

Let us now introduce interactions. As a simple but illuminating example, consider a cubic interaction between rank-2 and rank-1 multiplets:%
\bea
S=-\frac 12\int \uud^{d+1}x \sqrt{g} \left[\psi(\dsq-m_0^2) \psi+\phi(\dsq-m^2)^2 \phi +g_{\mbox{\tiny N}} \phi \psi\psi\right].\label{eq:intaction}
\eea
The field $\psi$ couples to a single boundary operator $\mathcal Q$ with dimension $\Delta_0$. It is straightforward to compute the three point functions using Witten diagrams. One uses the bulk-to-boundary propagators $K_\Delta, K_{\Delta_0}$ for insertions of $\cO_2$ and $\mathcal Q$ respectively, and $\partial_\Delta K_\Delta+\alpha K_\Delta$ for inserting $\cO_1$, with $\alpha$ as in \reef{eq:normal}. In particular,
\begin{subequations}\label{eq:ads3pt}
\bea
\kappa\langle \mathcal Q(x_1) \mathcal Q(x_2)\cO_2(x_3) \rangle&=&  g_{\mbox{\tiny N}}\, c(\Delta)\,\mbf P_{ \Delta_0 \Delta_0\Delta}(x_i), \\
\kappa^2\langle \mathcal Q(x_1)\cO_2(x_2) \mathcal \cO_2(x_3) \rangle&=&  g_{\mbox{\tiny N}}\,d(\Delta)\,\mbf P_{\Delta_0 \Delta \Delta}(x_i),\\
\kappa^3\langle \cO_2(x_1) \mathcal \cO_2(x_2) \mathcal \cO_2(x_3)\rangle&=& g_{\mbox{\tiny N}}\, e(\Delta)\,\mbf P_{\Delta \Delta \Delta}(x_i)\,
\eea
\end{subequations}
with $c,d,e$ some non-zero constants depending among others on $\Delta$. From these equations we may obtain the three-point function with logarithmic operator insertions, for instance:
\bea
\kappa \langle \mathcal Q(x_1) \mathcal Q(x_2)\cO_1(x_3) \rangle&=&(\partial_\Delta+\alpha)(\kappa \langle \mathcal Q \mathcal Q\cO_2\rangle)\nonumber \\
&=& g_{\mbox{\tiny N}} \,\mbf P_{\Delta,\Delta_0,\Delta_0}\,\left[c(\Delta) \tau_3+\partial_\Delta c(\Delta)+\alpha\, c(\Delta)\right] \label{eq:LNN}.
\eea
This is in perfect agreement with \reef{eq:3ptrnk2}, with OPE coefficients:
\bea
\label{eq:opernk2}
\lambda_2^{\mathcal Q\mathcal Q \mathcal O}=g_{\mbox{\tiny N}}\frac{ c(\Delta)}{\sqrt{N_\Delta} (2\Delta-d)}, \qquad \lambda_1^{\mathcal Q\mathcal Q \mathcal O}=g_{\mbox{\tiny N}}\frac{\partial_\Delta c(\Delta)+\alpha\, c(\Delta)}{\sqrt{N_\Delta} (2\Delta-d)},
\eea
where we have used \reef{eq:normal}. Other three point functions involving $\cO_1$ may be obtained in a similar fashion.

As a cross-check we would like to see that the same couplings are reproduced by the conformal block expansion of the four-point function of $\mathcal Q$. An exchange diagram for the logarithmic multiplet will involve the rank-2 bulk-to-bulk propagator. Since this is in turn given by derivatives of the ordinary propagator, we can deduce the conformal block decomposition of the rank-2 (and indeed rank-$r$) starting from the one for rank-1. Consider then contribution to the four-point function of $\mathcal Q$ where a rank-1 multiplet is exchanged. That is, we keep the form of the action \reef{eq:intaction} fixed, but simply switch the rank of the field $\phi$ from two to one. Such a contribution will contain the conformal block for the exchange of the field $\phi$, %
\bea
\prod_{i<j} |x_{ij}|^{2\DD_0/3} \, \langle \mathcal Q \mathcal Q \mathcal Q \mathcal Q\rangle_{r=1} \supset (\lambda^{\mathcal Q\mathcal Q\mathcal O})^2 \, G_{\Delta}^{(0)}(u,v)
\eea
with $G_\Delta^{(0)}(u,v)$ the conformal block (not to be confused with the bulk-to-bulk propagator). The OPE coefficient here is $\lambda_{\mathcal Q\mathcal Q\mathcal O}^2=\frac{c(\Delta)^2}{\hat \kappa^2}$ where we have put a hat on $\kappa$ to emphasize that such a constant depends on the rank. In the rank-1 case in particular, one has $\hat \kappa^2=(2\Delta-d) N_\Delta$. With this result we are ready to compute the rank-2 case. Since the bulk-to-bulk propagator satisfies \reef{eq:bb2}, we have
\bea
\prod_{i<j} |x_{ij}|^{2\DD_0/3}  \langle \mathcal Q \mathcal Q\mathcal Q\mathcal Q\rangle_{r=2} &\supset& \frac{1}{2\Delta-d} \fracpt{}{\DD} \left[\frac{c(\Delta)^2}{N_\Delta (2\Delta-d)} G_\Delta^{(0)}(u,v)\right]
\nonumber \\
&=& 2\lambda_1^{\mathcal Q\mathcal Q \mathcal O}\lambda_2^{\mathcal Q\mathcal Q \mathcal O}\, G_\Delta^{(0)}(u,v)+(\lambda_2^{\mathcal Q\mathcal Q \mathcal O})^2\, \partial_\Delta G_\Delta^{(0)}(u,v)\,,
\eea
where $\lambda_{1,2}^{\mathcal Q\mathcal Q \mathcal O}$ are the coefficients determined in Eq.~\reef{eq:opernk2}.
This expression nicely matches with the form of conformal block expansion that was deduced in Sec.~\ref{sec:logExchange}, in particular in Eq.~\reef{eq:rank2}.

A final point to notice is that we have introduced a single parameter, $g_{\mbox{\tiny N}}$, whereas we expect there to be two independent OPE coefficients. The answer to this apparent puzzle is that, unlike in the rank-1 case, there are now new cubic interactions involving derivatives that cannot be removed by field redefinitions. For instance, we could include an interaction
\bea
g_{\mbox{\tiny L}} (\dsq-m^2)\phi\psi^2.
\eea
Since the operator $\dsq-m^2$ annihilates the ordinary bulk-to-boundary propagator $K_\Delta$, such a term can only correct a cubic interaction sourced by at least one logarithmic insertion. Furthermore, the action of the operator removes the logarithm from the propagator, and hence this interaction can only correct the non-logarithmic piece of the correlator \reef{eq:LNN}, that is, the coefficient $\lambda_1^{\mathcal Q\mathcal Q \mathcal O}$. This is exactly as it should be for consistency with the Ward identities, which would be violated if we could do the reverse, namely modifying the coefficient of the logarithmic piece without altering the normal operator three-point function.

%
\subsection{Higher spins}
\subsubsection{Spin-1}
Here we shall for the first time construct holographic actions which describe spin-1 logarithmic multiplets. Consider first the ordinary bulk action of a spin-1 field $V_\mu$:
\bea
\label{eq:spin1action}
S =\int d^{d+1}x \sqrt{g} \,\left(\frac 14\,F_{\mu \nu} F^{\mu \nu}+\frac 12 m^2 V_\mu V^\mu\right), \qquad F_{\mu\nu}=\partial_\mu V_\nu-\partial_\nu V_\mu\,.
\eea
In this case we have
\bea
m^2=\Delta(\Delta-d)+(d-1)\,,
\eea
which vanishes when $\Delta=d-1$, {\em i.e.} when $V_\mu$ is a gauge field and couples to a boundary conserved current. Integrating by parts we get
\bea
S=-\frac 12\int d^{d+1}x \sqrt{g} \left( V^\mu D^{(1)}_{\mu \nu}V^\nu\right)\,,
\eea
with
\bea
D^{(1)}_{\mu\nu} \ldef  (\dsq-m^2) g_{\mu \nu}-\nabla_\mu \nabla_\nu \,. 
\eea
As an ansatz for a higher-rank action, we can define higher-order differential operators $D^{(\rnk)}$ in a natural way,
\bea
D^{(1)}_{\mu\nu} \; \to \; D^{(r)}_{\mu\nu} \ldef D^{(1)\ \rho}_{\ \ \mu}\,D^{(r-1)}_{\rho \nu} 
\eea
and replace $D^{(1)}_{\mu \nu}$ in~\reef{eq:spin1action} by $D^{(\rnk)}_{\mu \nu}$.
To see that the new action describes a logarithmic multiplet, it is sufficient to consider the bulk-to-bulk propagator $G_{\DD,\mu \nu}^{(\rnk)}$. Using
\bea
[D_{\mu \nu}^{(1)},\partial_\Delta]=(2\Delta-d)\, g_{\mu \nu}\,,
\eea
we see that the general propagator will be again given by combinations of derivatives with respect to the dimension of the rank-1 propagator $G_{\DD,\mu \nu}^{(1)}$. For instance for rank-2 it is easy to check
\bea
D^{(2)}_{\mu \nu} G_{\Delta, \nu\rho}^{(2)}=\frac{\delta^{d}(x_1-x_2) \delta(z-y)}{\sqrt{g}} \; \Rightarrow \; G_{\Delta, \mu\nu}^{(2)}=\frac{\partial_\Delta G_{\Delta, \mu\nu}^{(1)}}{2\Delta-d}\,.
\eea
Such derivatives immediately lead to logarithms in the approach to the boundary as in the scalar case.

Let us focus on the interesting case of $\Delta=d-1$. The rank-2 action can now be written
\bea
S_{2} &=& \frac 14 \int d^{d + 1} x \sqrt{g} (\nabla_a F^{a b} \nabla_c
F^c_b) = - \frac 14\int F \wedge \star d \delta F, \qquad \delta=\star d \star,
\eea
with $\star$ the usual Hodge-dual operator. The generalization to the rank $r$ case is straightforward, namely
\bea
S_r &=&-\frac 14 \int F \wedge \star (d \delta)^{(r-1)} F.
\eea
We can also write an equivalent action by introduction of auxiliary fields. The action is then
\bea
S=\int \uud^{d+1}x\sqrt{g} \sum_{i=1}^r \left(\frac 14 F^{(i)}\wedge \star F^{(r-i+1)}-\frac 12 A^{(i+1)}\wedge \star A^{(r-i+1)}\right)\,,
\eea
with fields $A_\mu^{(i)}$, $i=1,\ldots r$ (zero otherwise). The equations of motion are 
\bea
d \star F^{(i)}=\star A^{(i+1)}\,.
\eea
In particular notice that gauge invariance of $A^{(1)}$ is unbroken. This will be the field that will couple to the conserved current operator in the logCFT. Indeed the boundary values of the fields $A^{(i)}$ are identified as sources for boundary operators of the form $\int A^{(i)} J^{(r-i+1)}$  similarly to the scalar case, with $J^{(r)}$ being identified as the conserved current sitting at the bottom of a logarithmic multiplet with spin-1 and dimension $\Delta=d-1$.

\subsubsection{Spin-2}
Before we finish this section, let us just make a few comments on the spin-2 case. It would be straightforward to extend the preceding examples to spin-2 by ``squaring'' the Fierz-Pauli equation of motion for a massive spin-2 field. However it far more interesting to consider the massless case, and in particular a fully non-linear action. The correspondence between logarithmic CFTs with $c=0$ and higher derivative gravity theories is nicely reviewed in \cite{Grumiller2013}, with a focus on the three-dimensional bulk theories. Here we merely point out a simple example, that of ``new massive gravity'' in $d+1$ dimensions \cite{Bergshoeff2009b,Bergshoeff2009c,Lu2011,Deser2011}. The action is
\bea
S=\frac{M_P^{d-1}}2 \int \uud^{d+1} x\sqrt{g}\bigg(R+2\Lambda+\lambda \left(S^{\mu\nu}S_{\mu\nu}-S^2\right)\bigg) \label{eq:nmg}\,,
\eea
with $\Lambda$ the cosmological constant and $S$ the Schouten tensor
\bea
S_{\mu\nu}\equiv \frac{1}{d-1} \bigg( R_{\mu\nu}-\frac{1}d\, g_{\mu \nu} R\bigg).
\eea
In general the action propagates two sets of spin-2 modes, one massless and one massive. By tuning the coupling $\lambda$ the massive mode can be made massless. This happens precisely at the point where the coefficient $c$ measuring the two-point function of the boundary stress-tensor becomes zero. Simultaneously, there are logarithmic modes which appear, lifting the degeneracy between the two sets of spin-2 massless modes. We conclude that such a theory describes a $c=0$ logarithmic CFT, i.e. one where the stress-tensor has a logarithmic partner. 

It is clear that this argument generalizes. Theories with higher derivative terms (say $2n$ derivatives) in the action will propagate several massive spin-2 modes, and it is always possible to tune coefficients such that their masses become zero. At this point logarithms appear, and we expect to be able to obtain a rank-$r$ logarithmic theory. However, we should point out that what singles out theories such as the one above is that they have several other nice properties. They can be derived as particular non-unitary limits \cite{Paulos2012} of the bimetric theory of Hassan and Rosen \cite{Hassan2012b}, which guarantees that they are free from the Boulware-Deser instability. They also satisfy a version of the holographic $c$-theorem \cite{Myers2011}. To our knowledge, no simple candidate satisfying these properties exists for a rank-3 logarithmic CFT in general dimension. Finally it would be extremely interesting to find out whether the action above can be derived by a bulk disorder averaging, as we did for a scalar theory at the beginning of this section.



\section{Examples of logCFTs}
\label{sec:examples}

So far, we have developed a formalism to analyze correlation functions in logarithmic CFTs. In particular, we know the constraints of conformal invariance and Bose symmetry on $n$-point functions of scalar fields. We will now check how our formalism applies to several models, most of which are familiar from the literature.


\subsection{Triplet model}\label{sec:triplet}

Our first case study is the triplet model, which is essentially the theory of a pair of symplectic fermions in two dimensions. The triplet theory has been intensively studied in a series of papers by Ga\-berdiel and Kausch~\cite{Kausch:1995py,Gaberdiel:1996np,Gaberdiel:1998ps,Kausch:2000fu}. In particular, various correlation functions have been computed using purely algebraic methods, and in this section we will compare these existing results to our formalism.

The triplet model is the Virasoro minimal model $\mca{M}(1,2)$ of central charge $c=-2$ extended by a triplet of fields $W^0, W^{\pm}$ of Virasoro weight $h = 3$. The modes of the $W^a$ combine with the Virasoro generators $L_n$ to form an extended chiral algebra~\cite{Adamovic:2007er}. In particular, the zero modes of the $W^{0,\pm}$ generate a global $SL(2)$ symmetry --- not to be confused of the $SL(2,\mbb{C})$ global conformal group.  The model has four primary states with respect to the chiral algebra: two $SL(2)$ singlet representations at $h=0$ and $h=-1/8$, and two $SL(2)$ doublets of weight $h=1$ and $h=3/8$. The chiral theory can be uniquely extended to a non-chiral (i.e.\@ local) CFT. The $h=-1/8$ representation becomes a rank-one scalar primary $\mu(x)$ of dimension $\DD = -1/4$. Second, the $h=3/8$ representation gives rise to a scalar $\nu^{\a \bar{\a}}(x)$ of dimension $\DD = 3/4$, transforming in the adjoint representation of $SL(2)$. Finally, the $h=0$ and $h=1$ multiplets combine into a larger representation $\mca{R}$ with a rather complicated structure. We will only be interested in the lowest level of $\mca{R}$, which contains a logarithmic doublet $\omega_a = (\omega,\Omega)$ of dimension $\DD = 0$. The operator $\Omega$ functions as a unit operator, as its OPE with any other operator is simply $\Omega \times \Oo = \Oo$. However, conformal invariance requires that $\expec{\Omega}$ vanishes. After a suitable rescaling, its logarithmic partner $\omega(x)$ has a vev $\expec{\omega} = 1$.

In the rest of this section, we will analyze correlation functions involving the $\omega_a(x)$ doublet and $\mu(x)$.
We will think of these operators as primaries of the global conformal group $SL(2,\mbb{C})$ and forget about the underlying chiral algebra from now on.

First, let's consider $n$-point functions of the rank-two scalar $\omega_a = (\omega,\Omega)$. We will use the convention that its two-point function is in the canonical form:
\beq
\expec{\omega_a(x)\omega_b(0)} = \begin{pmatrix} -\ln x^2  & 1 \\ 1 & 0 \end{pmatrix}\,,
\eeq
which is consistent with $\expec{\omega} = 1$. The three-point functions are given by
\begin{align}
\label{eq:omega3pt}
&\expec{\Omega \Omega \Omega} = 0\,, \qquad
\expec{\omega \Omega \Omega} = 1\,, \qquad
\expec{\omega(x_1) \omega(x_2) \Omega(x_3)} = \tau_1 + \tau_2\,,\\
&\expec{\omega(x_1) \omega(x_2) \omega(x_3)} = \tau_1 \tau_2 + \tau_1 \tau_3 + \tau_2 \tau_3\,,
\end{align}
see Eq.~(14) of~\cite{Kausch:2000fu}.\footnote{In the conventions of~\cite{Kausch:2000fu} we set $\La = 1$, $\Oo = 2$, $\mca{Z} = 0$, and moreover we redefine $\omega \to \omega/2$.} Clearly this is consistent with our formula~\reef{eq:3ptphi}, with
\beq
\la^{\omega \omega \omega}_1 = \la^{\omega \omega \omega}_2 = \la^{\omega \omega \omega}_4 = 0\,, \quad
\la^{\omega \omega \omega}_3 = 1\,.
\eeq
The four-point functions $\expec{\omega_a \omega_b \omega_c \omega_d}$ have appeared for instance in Eq.~(14) of~\cite{Kausch:2000fu}.
We have checked that they are consistent with~\reef{eq:4ptmaster}, with
$\mca{F}_i \rdef \mca{F}^{\omega}_i$ given by
\bsub
\label{eq:triplet4pt}
\begin{align}
  \mca{F}^{\omega}_1(u,v) &= \frac{1}{54} \ln(u/v^2) \ln(u^2/v) \ln(uv)\,, \\
  \mca{F}^{\omega}_2(u,v) &= -\frac{1}{12} \left[ (\ln u)^2 - (\ln u)(\ln v) + (\ln v)^2 \right] \,,\\
  \mca{F}^{\omega}_3(u,v) &= \frac{1}{6} \ln\!\left(v/u^2\right)\,, \\
  \mca{F}^{\omega}_4(u,v) &= 1\,, \qquad \mca{F}^{\omega}_5  = 0\,.
\end{align}
\esub
It is easy to verify that these $\mca{F}^{\omega}_i$ obey the required crossing relations~\reef{eq:cross4}.

Next, let us consider four-point functions involving the twist field $\mu$ of dimension $\DD = -1/4$. We can read off the mixed four-point functions $\expec{\omega_a \omega_b \mu \mu}$ from Appendix B.1 of~\cite{Kausch:2000fu} and find that they are consistent with the formulas from Sec.~\ref{sec:warmup}. Concretely, the $\mca{F}_i(u,v)$ are: 
\bsub
\label{eq:mixed4pt}
\begin{align}
  \mca{F}_1(u,v) &= \left[  \frac{1}{36} \ln^2(v/u^2) + \frac{2}{3}\ln 2 \ln(v/u^2)+4(\ln 2)^2-\frac{1}{4} \left[\mca{H}(u,v)\right]^2 \right] \, (u^2/v)^{1/24}\,, \\
  \mca{F}_2(u,v) &= -\left[\frac{1}{6} \ln(v/u^2) + 2\ln 2 \right] (u^2/v)^{1/24} \,,\\
  \mca{F}_3(u,v) &= (u^2/v)^{1/24}\,.
\end{align}
\esub
Here we introduce the shorthand notation
\begin{align}
  \mca{H}(z,\zb) &\ldef-4\ln 2 + \ln(z\zb) - \sqrt{1-z}M^{(1)}(z) - \sqrt{1-\zb}M^{(1)}(\zb) \,,\\
  M^{(1)}(x) &\ldef  \fracpt{}{b} \, {}_2F_1\!\left(1,1/2-b;1-b;x \right) \big|_{b = 0} \nn \\
  &\phantom{:}= \sum_{n=0}^{\infty} \frac{(1/2)_n}{n!} \, x^n \left[ \psi(1+n) - \psi(\thalf+n) - 2 \ln 2\right] = -\frac{x}{2} + \mrm{o}(x^2)\,,
\end{align}
where we have expressed $\mca{H}$ in terms of the familiar $z,\zb$ coordinates:
\beq
u=z\zb\,,  \qquad v=(1-z)(1-\zb)\,.
\eeq

Again, we want to check that the four-point functions~\reef{eq:mixed4pt} are consistent with crossing symmetry. For $\mca{F}_2$ and $\mca{F}_3$ this is easy, because any function of $u^2/v$ is automatically invariant under $u \to u/v$, $v \to 1/v$. In the $z,\zb$ coordinates, this exchange is equivalent to mapping $(z,\zb) \to (z',\zb')$ with $z' = z/(z-1)$ and $\zb' = \zb/(\zb-1)$. Then we remark that
\beq
\sqrt{1-x'}M^{(1)}(x') = \sqrt{1-x}M^{(1)}(x) - \ln(1-x)\,, \qquad x' \ldef x/(x-1)
\eeq
as follows from the hypergeometric identities. This implies that
\beq
\mca{H}(z',\zb') = \mca{H}(z,\zb)\,,
\eeq
which proves that $\mca{F}_1$ is crossing symmetric too.

The last correlator in the triplet model we consider will be the four-point function of $\mu$ alone, see e.g.\@ Appendix B.3 of~\cite{Kausch:2000fu}:
\begin{align}
 & \expec{\mu(x_1) \mu(x_2) \mu(x_3) \mu(x_4)} = \prod_{i < j} |x_{ij}|^{1/6} \, \mca{F}^\mu(u,v)\,,\\
 & \mca{F}^\mu(u,v) =  \tfrac{1}{2} (uv)^{1/6} \left[K(z)K(\zb)\left(  8 \ln 2 - \ln z\zb \right) + K(z)M^{(2)}(\zb) + K(\zb)M^{(2)}(z) \right],
\end{align}
where we introduced the the shorthand notations 
\begin{align}
  K(x) &\ldef {}_2F_1\!\left(1/2,1/2;1;x\right),\\
    M^{(2)}(x) &\ldef \fracpt{}{b} \, {}_2F_1\!\left(1/2-b,1/2-b;1-2b;x\right)\big|_{b = 0}\nn \\
  &= 2 \sum_{n=0}^{\infty} \frac{(1/2)_n^2}{(n!)^2} \, x^n \left[ \psi(1+n) - \psi(\thalf+n) - 2 \ln 2\right]\,.
\end{align}
We will first verify that this expression transforms correctly under $(u,v) \to (v,u)$, or equivalently $(z,\zb) \to (1-z,1-\zb)$. According to a hypergeometric identity we have
\beq
\pi K(1-x) = \left( 4 \ln 2 - \ln x \right) K(x) + M^{(2)}(x)
\eeq
which in turn implies that
\beq
\mca{F}^\mu(u,v) = \pi (uv)^{1/6}  \left[ K(z)K(1-\zb) + K(\zb)K(1-z)\right]\,.
\eeq
This makes the crossing symmetry $\mca{F}^\mu(u,v) = \mca{F}^\mu(v,u)$ manifest. To prove the second crossing symmetry $\mca{F}^\mu(u,v) = \mca{F}^\mu(u/v,1/v)$, it is helpful to use that with $x' = x/(x-1)$
\begin{align}
K(x') &= (1-x)^{1/2} K(x)\,, \\
M^{(2)}(x') &= (1-x)^{1/2} \left[  M^{(2)}(x) - \ln(1-x)K(x) \right]\,.
\end{align}

\subsection{Generalized free field theory}\label{sec:GFF}

Let us now turn to an example of a higher-dimensional logCFT: the logarithmic counterpart of the familiar \emph{generalized free field} (GFF), which is defined in any dimension $d$~\cite{ElShowk:2011ag}. This is a theory of a rank-two multiplet $\phi_a$ of dimension $\DD_\phi$ with a canonical\footnote{Of course, it's possible to redefine $\phi_1 \to  \phi_1 + \mca{Z} \phi_2$ for some arbitrary $\mca{Z}$, but this only changes the discussion in a superficial way.} two-point function:
\beq
\label{eq:GFF2pt}
\expec{\phi_a(x) \phi_b(0)} = \frac{1}{|x|^{2\DD_\phi}} \begin{pmatrix} - \ln x^2 & 1 \\  1 & 0 \end{pmatrix}_{ab}\,.
\eeq
All other correlators of $\phi_a$ and composite operators are defined starting from~\reef{eq:GFF2pt} via Wick's theorem. We will refer to this theory as a ``logarithmic GFF''.
Similarly one could construct a logarithmic CFT starting from a higher-rank multiplet or with a different field content (free fermions, for instance).

The logarithmic GFF has an equivalent definition as the limit of an ordinary (but non-unitary) CFT. Take two GFFs $\chi_1$, $\chi_2$ with two-point functions
\beq
\expec{\chi_1(x) \chi_1(0)} = \frac{1}{|x|^{2(\DD_\phi+\eps)}}, \quad \expec{\chi_2(x)\chi_2(0)} = \frac{-1}{|x|^{2\DD_\phi}}, \quad\expec{\chi_1 \chi_2} = 0.
\eeq
Then in the limit $\eps \to 0$, the fields
\beq
\label{eq:phiaschi}
\phi_1 \ldef \frac{1}{\sqrt{\eps}} \left( \chi_1 - \mu^\eps \, \chi_2 \right), \quad \phi_2 \ldef \sqrt{\eps}\, \chi_2
\eeq
transform as a logarithmic multiplet with two-point functions as in Eq.~\reef{eq:GFF2pt}. Here $\mu$ is an arbitrary scale that we will set to unity. The advantage of this limit construction is that it allows to do computations in a non-logarithmic GFF, for which many exact results are known.

\subsubsection{Four-point functions}

We will mostly be interested in the $\phi_a$ four-point function and its conformal block decomposition. As a starting point, it will be useful to understand the operator content of the $\phi_a \times \phi_b$ OPE. This can be found from  the limit construction: the $\chi_i \times \chi_j$ OPE is known, and consists of the unit operator $\unit$ as well as a tower of ``double-trace'' primaries:
\beq
\chi_i \times \chi_j \; \sim \; \pm \dd_{ij} \unit \; + \; \sum_{\l,n} \Oo_{ij}^{(\l,n)}
\eeq
which are schematically given by
\beq
\Oo_{ij}^{(\l,n)} \; \sim \; \chi_i \, \pd_{\mu_1} \dotsm \pd_{\mu_\l} (\pd^2)^n \, \chi_j + \ldots\,.
\eeq
The operator $\Oo_{ij}^{(\l,n)}$ has scaling dimension $[\chi_i] + [\chi_j] + \l + 2n$ and {spin $\l$}. Under the exchange of the fields $\chi_i \lra \chi_j$, it transforms as
\beq
\Oo_{ij}^{(\l,n)} = (-1)^\l \, \Oo_{ji}^{(\l,n)}\,.
\eeq
Hence for even $\l$ and fixed $n$ there are three different double-trace operators, whereas for odd $\l$ there is only one. 
We expect the even-spin primaries to form rank-three multiplets in the logarithmic GFF. For the simplest case, the $n=0$ scalar, this logarithmic triplet is defined by 
\beq
\label{eq:scalar triplet}
\mca{S}_1 = \frac{1}{2} \NO{(\phi_1)^2}\,, \quad \mca{S}_2 = \NO{\phi_1 \phi_2}\,, \qquad \mca{S}_3 = \NO{(\phi_2)^2}
\eeq
where $\NO{\phantom{ }}$ denotes normal ordering.
We leave it as an exercise to verify that $\mca{S}_a$ transforms correctly and that the three-point functions $\expec{\phi_a \phi_b \mca{S}_c}$ are consistent with conformal invariance. Summarizing, we expect the $\phi_a \times \phi_b$ OPE to contain:
\begin{itemize}
\item the unit operator $\unit$;
\item for every \emph{even} spin $\l$ and integer $n \geq 0$, a rank-3 multiplet of dimension $2\DD_\phi + \l + 2n$;
\item for every \emph{odd} spin $\l$ and integer $n \geq 0$, a rank-1 multiplet of dimension $2\DD_\phi + \l + 2n$\,.
\end{itemize}
In what follows, we will test this prediction by studying the $\phi_a$ four-point function and its conformal block decomposition. Let's first review the relevant four-point functions in the $\chi_{1,2}$ theory, which are given by:
\bsub
\begin{align}
  \expec{\chi_i(x_1) \chi_i(x_2) \chi_i(x_3) \chi_i(x_4)} &= \mbf{P}_{w_i w_i w_i w_i}(x) \, H_i(u,v) \qquad \qquad i=1,2\,, \\
  \expec{\chi_1(x_1) \chi_1(x_2) \chi_2(x_3) \chi_2(x_4)} &= \mbf{P}_{w_1 w_1 w_2 w_2}(x) \, H_{12}(u,v)\,,
\end{align}
\esub
writing $w_1 \ldef \DD_\phi + \eps$, $w_2 \ldef \DD_\phi$ and
\bsub
\begin{align}
  H_i(u,v) &=  (v/u^2)^{w_i/3} + (u/v^2)^{w_i/3} + (u v)^{w_i/3} \qquad i=1,2\,, \\
H_{12}(u,v) &= - (v/u^2)^{(w_1 + w_2)/6}\,.
\end{align}
\esub
The other four-point functions $\expec{\chi_1 \chi_2 \chi_2 \chi_2}$ and $\expec{\chi_1 \chi_1 \chi_1 \chi_2}$ vanish due to the $\mbb{Z}_2 \times \mbb{Z}_2$ symmetry of the $\chi_{1,2}$ theory. It's easy to see that the three functions $H_1(u,v)$, $H_2(u,v)$ and $H_{12}(u,v)$ obey the necessary crossing identities. 

The $\expec{\phi_a \phi_b \phi_c \phi_d}$ four-point functions can be computed either using Wick's theorem or using the limit construction. We find that they have the form predicted by Sec.~\ref{sec:ranktwo4pt}, with the functions $\mca{F}_a$ given by:
\bsub
\label{eq:gff4pt}
\begin{align}
  \mca{F}_4(u,v) &= \mca{F}_5(u,v) = 0 \,,\\
  \mca{F}_3(u,v) &= (u v)^{\DD_\phi /3} + (u/v^2 )^{\DD_\phi  /3} \,,\\
  \mca{F}_2(u,v) &= \frac{1}{2} \frac{\pd}{\pd \DD_\phi}  \mca{F}_3(u,v) +  (v/u^2)^{\DD_\phi/3}  \, \Phi(u,v) \,,\\ 
  \mca{F}_1(u,v) &= \frac{1}{4} \frac{\pd^2}{\pd \DD_\phi^2} \mca{F}_3(u,v)   +  (v/u^2)^{\DD_\phi/3}  \, \Phi^2(u,v) \,,
\end{align}
\esub
where we introduced the notation
\beq
\Phi(u,v) \ldef \frac{1}{6} \ln(v/u^2)\,.
\eeq
It may be checked that these $\mca{F}_i$ obey the crossing relations from~\reef{eq:cross4}.

In passing, we remark that it's not immediately obvious that the limit construction gives finite results, since correlators with insertions of $\phi_1$ come with poles in $1/\sqrt{\eps}$. The fact that the resulting expressions for the $\mca{F}_i(u,v)$ are finite follows from the following cancellations:
\bsub
\begin{align}
&\lim_{\eps \to 0} \left[ H_{12}(u,v) + H_{12}(v,u) + H_{12}(1/u,v/u) \right] = -H_2(u,v) \,,\\
&\lim_{\eps \to 0} \fracpt{}{\eps} \left[  \frac{ H_1(u,v) }{2} +  H_{12}(u,v) + H_{12}(v,u) + H_{12}(1/u,v/u) \right] = 0\,.
\end{align}
\esub

\subsubsection{Conformal block decompositions}

We now turn our attention to the conformal block decomposition of the $\phi_a$ four-point functions. This exercise will provide an extensive test of the formalism developed in Sec.~\ref{sec:CB}. As a starting point, we will recall the conformal block decompositions of the $\expec{\chi_i \chi_j \chi_k \chi_l}$ four-point functions in the $\chi_{1,2}$ theory. The four-point function $H_{12}(u,v)$ only contains the unit operator:
\beq
H_{12}(u,v) = -  G_{0}^{(0)}(u,v;w_1,w_1,w_2,w_2)\,,
\eeq
with the coefficient $-1$ arising from the sign of the $\expec{\chi_2 \chi_2}$ two-point function. In the cross- or $t$-channel, after exchanging $\chi_1(x_2) \lra \chi_2(x_4)$, we have the following conformal block decomposition:
\beq
\label{eq:h12vu}
H_{12}(v,u) = - \sum_{\l,n = 0}^\infty (-1)^\l \mfr{q}(\l,n;w_1,w_2) \, G_{w_1 + w_2 + \l + 2n}^{(\l)}(u,v;w_1,w_2,w_2,w_1)\,,
\eeq
with
\begin{multline}
  \mfr{q}(\l,n;\DD_1,\DD_2) \ldef \frac{2^{\l}}{\l!n!}\frac{(\DD_1 - \nu)_n(\DD_2 - \nu)_n(\DD_1)_{\l+n}(\DD_2)_{\l+n}}{(\l+d/2)_n(\DD_1 + \DD_2 + n -d+1)_n } \\
  \times \; \frac{1}{(\DD_1 + \DD_2 + \l+2n-1)_\l (\DD_1 + \DD_2 + \l + n -d/2)_n}\,.
  \label{eq:qDef}
\end{multline}
Here $(x)_n \ldef \Gamma(x+n)/\Gamma(x)$ is the Pochhammer symbol. These coefficients were first obtained in~\cite{Fitzpatrick:2011dm}. As expected, this is a sum over all double-trace operators $\Oo_{12}^{(\l,n)}$. Finally, we can analyze the $u$-channel with $\chi_1(x_2) \lra \chi_2(x_3)$, which yields
\beq
H_{12}(1/u,v/u) = - \sum_{\l,n = 0}^\infty \mfr{q}(\l,n;w_1,w_2) \, G_{w_1 + w_2 + \l + 2n}^{(\l)}(u,v;w_1,w_2,w_1,w_2)\,.
\eeq
For both $H_i(u,v)$ four-point functions, we rather have the conformal block decomposition
\beq
H_i(u,v) = G_0^{(0)}(u,v;w_i,\ldots,w_i) + \sum_{\text{even } \l} \sum_{n=0}^\infty 2\mfr{q}(\l,n;w_i,w_i)\,  G_{2\DD_i + \l + 2n}^{(\l)}(u,v;w_i,\ldots,w_i)\,.
\eeq
The first term clearly corresponds to the unit operator contribution, and the rest to even-spin double trace primaries $\Oo_{ii}^{(\l,n)}$.

Next, we will recycle these results to compute the conformal block decompositions of the functions $\mca{F}_{1 \dotsm 5}(u,v)$. This is a straightforward exercise: we express the four-point functions of $\phi_a$ as linear combinations of four-point functions of $\chi_{1,2}$ and take the limit $\eps \to 0$.  
To be precise, we stress that the $\chi_i$ four-point functions can depend on $\eps$ in four different ways: 1) through the scale factors $\mbf{P}$; 2) through the OPE coefficients $\mfr{q}(\l,n;w_i,w_j)$; 3) through the dimensions $w_i + w_j + \l + 2n$ of the exchanged operators, and finally 4) through the external dimensions appearing in the conformal blocks.
The $\eps$ dependence of the scale factors $\mbf{P}$ is trivial, since derivatives of $\mbf{P}$ simply give factors of $\zeta_1, \ldots, \zeta_4$ via Eq.~\reef{eq:zetadef2}.

This reduces the computation of the relevant conformal block decompositions to a bookkeeping exercise. For $\mca{F}_3$, we get in the $s$- and $t$-channels:
\begin{align}
\mca{F}_3(u,v) &= \sum_{\text{even } \l} \sum_{n=0}^\infty 2 \mfr{q}_{\l,n} \, G_{2\DD_\phi + \l + 2n}^{(\l)}(u,v;\DD_\phi) \label{eq:F3dec}\,,\\
\mca{F}_3(v,u) &= G_0^{(0)}(u,v;\DD_\phi) + \sum_{\l,n=0}^\infty \mfr{q}_{\l,n} \, G_{2\DD_\phi + \l + 2n}^{(\l)}(u,v;\DD_\phi)\,. \label{eq:F3cross}
\end{align}
Here we're writing $\mfr{q}_{\l,n} \equiv \mfr{q}(\l,n;\DD_\phi,\DD_\phi)$ for conciseness. It is easy to rewrite these expressions as differential operators $\mca{D}_{1122}$ resp.\@ $\mca{D}_{1221}$ acting on conformal blocks, following the logic of Sec.~\ref{sec:gooddiffop}. The fact that these operators are constant --- i.e.\@ there are no actual derivatives acting on the conformal block $G_\DD^{(\l)}$ --- means that several OPE coefficients vanish, as will be discussed later in more detail. For $\mca{F}_2(u,v)$ we find:
\beq
\label{eq:f2gff}
\mca{F}_2(u,v) =  (v/u^2)^{\DD_\phi/3}\Phi(u,v)  +  \sum_{\l,n = 0}^\infty \mca{D}^{(\l,n)}_{1112} \cdot G_{\DD_\Oo}^{(\l)}(u,v;\DD_1,\DD_2,\DD_3,\DD_\phi)\big|_{\DD_\Oo = 2\DD_\phi + \l + 2n,\, \DD_i = \DD_\phi}\,,
\eeq
with
\begin{align}
  \mca{D}^{(\l,n)}_{1112} &= \pd_{\DD_\phi} \mfr{q}_{\l,n}  +  \mfr{q}_{\l,n} \left( 2\pd_{\DD_\Oo} + \pd_{\DD_1} + \pd_{\DD_2} + 2\pd_{\DD_3} \right) &&\l \text{ even} \,,\\
  \mca{D}^{(\l,n)}_{1112} &= \mfr{q}_{\l,n} \, (\pd_{\DD_1} - \pd_{\DD_2})  &&\l \text{ odd}\,.
\end{align}
The difference between even- and odd-spin OPE coefficients comes from the factor $(-1)^\l$ in Eq.~\reef{eq:h12vu}. As a matter of fact, the odd $\l$ terms vanish due to the conformal block identity~\reef{eq:lima}. The first term in~\reef{eq:f2gff} corresponds to the unit operator contribution, namely
\beq
(v/u^2)^{\DD_\phi/3}\Phi(u,v) = \lim_{\DD \to 0} \left(\pd_{\DD_1} + \pd_{\DD_2} \right) G_\DD^{(0)}(u,v;\DD_1,\DD_2,\DD_\phi,\DD_\phi)\big|_{\DD_1 = \DD_2 = \DD_\phi}\,.
\eeq
Finally, for $\mca{F}_1$ we obtain
\beq
\mca{F}_1(u,v) =  (v/u^2)^{\DD_\phi/3}  \, \Phi^2(u,v)  +  \sum_{\l,n = 0}^\infty \mca{D}^{(\l,n)}_{1111} \cdot G_{\DD_\Oo}^{(\l)}(u,v;\DD_1,\DD_2,\DD_3,\DD_4)\big|_{\DD_\Oo = 2\DD_\phi + \l + 2n,\, \DD_i = \DD_\phi}\,,
\eeq
with
\begin{align}
  \mca{D}^{(\l,n)}_{1111} &= 2 \, \pd_{\DD_1} \pd_{\DD_2} \mfr{q}(\l,n;\DD_1,\DD_2) \big|_{\DD_1 = \DD_2 = \DD_\phi} 
  + \big[(\pd_{\DD_\phi} \mfr{q}_{\l,n}) +  2 \pd_{\DD_\Oo}\big] \sum_{i=1}^4 \pd_{\DD_i}   \nn 
  + 2(\pd_{\DD_\phi} \mfr{q}_{\l,n}) \pd_{\DD_\Oo} \\
  & \quad + \mfr{q}_{\l,n}  \Big[(\pd_{\DD_1} + \pd_{\DD_2}) (\pd_{\DD_3} + \pd_{\DD_4}) + 2(\pd_{\DD_1}\pd_{\DD_2} +  \pd_{\DD_3} \pd_{\DD_4}) \ + 2\pd_{\DD_\Oo}^2 \Big] &&\l \text{ even}\,,\\
  \mca{D}^{(\l,n)}_{1111} &= - \mfr{q}_{\l,n} \left(\pd_{\DD_1}\pd_{\DD_3} +\pd_{\DD_2}\pd_{\DD_4} -\pd_{\DD_1}\pd_{\DD_4} -\pd_{\DD_2}\pd_{\DD_3}\right) \nn \\
  &= -\mfr{q}_{\l,n} \, (\pd_{\DD_2} - \pd_{\DD_1})(\pd_{\DD_4} - \pd_{\DD_3}) &&\l \text{ odd}\,.
\end{align}
The first term is again the unit operator contribution:
\beq
 (v/u^2)^{\DD_\phi/3}\Phi(u,v)^2 = \lim_{\DD \to 0} \left(\pd_{\DD_1} + \pd_{\DD_2} \right)\left(\pd_{\DD_3} + \pd_{\DD_4} \right) G_\DD^{(0)}(u,v;\DD_1,\DD_2,\DD_3,\DD_4)\big|_{w_i = \DD_\phi}\,.
\eeq
These results provide several non-trivial checks on the formalism developed in the previous sections. In particular, by comparing the operators $\mca{D}_{ijkl}^{(\l,n)}$ with the results obtained in Sec.~\ref{sec:gooddiffop} this allows us (a) to verify that expressions for the differential operators are internally consistent and (b) to read off some of the OPE coefficients for the logarithmic GFF. Let us do this now.

In what follows we will denote the even-spin OPE coefficients $\displaystyle{a^{(\l,n)}_{p},\, b^{(\l,n)}_{p} \text{and } c^{(\l,n)}_{p}}$ in accordance with Sec.~\ref{sec:gooddiffop}. Likewise, we denote the odd-spin OPE coefficients $b^{(\l,n)}$.  Moreover, we will use the convention that all exchanged operators have a unit-normalized two-point function, at the expense of possibly getting imaginary OPE coefficients.
Let's first consider the odd-spin sector. From~\reef{eq:F3cross} we read off that 
\beq
\label{eq:oddres}
\big(b^{(\l,n)}\big)^2 = - \mfr{q}_{\l,n}
\eeq
and we find that the expressions for $\mca{D}^{(\l,n)}_{1111}$ and $\mca{D}^{(\l,n)}_{1112}$ are consistent with this result. 
Next, consider the even-spin sector. Because the even-spin operators transform as logarithmic triplets, reading off the OPE coefficients is more complicated than in the odd-spin case. For instance, conformal invariance dictates that the operator $\mca{D}_{2222}^{(\l,n)}$ is of the following form (cf.\@ Eq.~\reef{eq:d2222}):
\begin{align}
  \mca{D}_{2222}^{(\l,n)} &= \sum_{p,q=1}^3 c_p^{(\l,n)} c_q^{(\l,n)} V^{pq}(\pd_{\DD_\Oo}) \\
  &=   \, c_1^{(\l,n)} c_3^{(\l,n)} + \big(c_2^{(\l,n)} \big)^2 + 2 \, c_2^{(\l,n)} c_3^{(\l,n)} \pd_{\DD_\Oo} + \half  \big(c_3^{(\l,n)} \big)^2 \pd_{\DD_\Oo}^2 \nn
\end{align}
where we use the definition of the matrix $V^{pq}(\pd_{\DD_\Oo})$. Since $\mca{F}_5(u,v) = 0$ in the logarithmic GFF, we conclude that
\beq
\label{eq:firstres}
c_2^{(\l,n)} = c_3^{(\l,n)} = 0 \,,
\eeq
for all $\l,n$, but we cannot prove anything about the coefficients $c_1^{(\l,n)}$ from $\mca{F}_5$ alone.
By repeating this exercise for the other $\mca{F}_i$, we find:
\begin{multline}
  \label{eq:secondrest}
b_3^{(\l,n)} = 0\,, \quad
\big( b_2^{(\l,n)} \big)^2 = \mfr{q}_{\l,n}\,,  \quad
\thalf a_3^{(\l,n)} = c_1^{(\l,n)} = b_2^{(\l,n)}\,, \quad
a_2^{(\l,n)}b_2^{(\l,n)} + a_3^{(\l,n)}b_1^{(\l,n)} = \pd_{\DD_\phi} \mfr{q}_{\l,n}\,,\\
2  a_1^{(\l,n)}a_3^{(\l,n)} + \big(a_2^{(\l,n)} \big)^2 = 2  \, \pd_{\DD_1} \pd_{\DD_2} \mfr{q}(\l,n;\DD_1,\DD_2) \big|_{\DD_1 = \DD_2 = \DD_\phi} \,.
\end{multline}

Summarizing, we have obtained eight (quadratic) equations for nine unknown OPE coefficients, hence we can solve for nearly all OPE coefficients in the theory --- some, like $b_2^{(\l,n)}$, only up to a sign. It would be interesting to compare these predictions to a direct computation. Doing so would require two steps: first, to construct the exchanged primaries and unit-normalize them, and second to compute three-point functions of these primaries with $\phi_{a}$ using Wick's theorem.


\subsection{Self-avoiding walks and the $O(n\rightarrow 0)$ model}
The $O(n)$ model with order parameter $\phi_i$ is described by the Landau-Ginzburg action
\beq
\mca{L} = \frac{z}{2} (\pd \phi_i)^2 + \frac{r}{2} \, \NO{\phi_i^2} + \frac{g}{8} \NO{(\phi_i^2)^2}\,.
\label{eq:lambda4action}
\eeq
The normalization $z$ is chosen such that $\expec{\phi_i(x) \phi_j(0)} = \dd_{ij} |x|^{2-d}$ at the free massless point $r = g = 0$. 
In $d=4-\eps$ dimensions this theory has a weakly coupled critical point at $g_* = \sO(\eps)$, although we will consider $d$ to lie in the entire range $2 \leq d < 4$. For fixed $d$, we can analytically continue the $O(n)$ fixed point to fractional $n$, at the expense of breaking unitarity~\cite{Maldacena:2011jn}. In particular we can take the limit $n \to 0$, which describes a special type of polymer statistics, called self-avoiding walks~\cite{Duplantier:1987sh,Cardy:1996xt,sawlectures}. LogCFT aspects of this limit have been studied in Refs.~\cite{Cardy:1999zp,Cardy:2013rqg,Movahed:2004nr}. In the rest of this section, we will review the evidence that the $n=0$ theory is a logCFT and how it fits into our formalism.

Because we will be interested in the $\phi_i$ four-point function, our discussion is restricted to the $\phi_i \times \phi_j$ OPE. The latter can only contain operators in the singlet $(S)$, rank-two traceless symmetric tensor $(T)$ and rank-two antisymmetric (A) irreps of $O(n)$. Correlation functions of these operators will have a tensor structure that is fixed by $O(n)$ invariance, e.g.\@ for a tensor primary operator $T_{ij}$ we have
\beq
\label{eq:T2pt}
\expec{T_{ij}(x)T_{kl}(0)} = c_T(n)\left[\dd_{ik}\dd_{jl} + \dd_{il}\dd_{jk} - \frac{2}{n} \dd_{ij}\dd_{kl}\right] \frac{1}{|x|^{2\DD_T(n)}}\,.
\eeq
We allow both the normalization $c_T(n)$ and the scaling dimension $\DD_{T}(n)$ to depend on $n$. The coefficient $c_T(n)$ is of course arbitrary, although we can imagine fixing it by specifying a renormalization scheme, say minimal subtraction. We will assume that $c_T(n)$ has a finite limit as $n \to 0$, in which case the correlator~\reef{eq:T2pt} is ill-defined at $n=0$ due to the $1/n$ pole. Without loss of generality, we will set $c_T(n) = 1$ in what follows, which is justified in a neighborhood around $n=0$.

There is a simple mechanism to obtain a finite limit in~\reef{eq:T2pt}. Suppose that there exists a scalar $S$ in the singlet channel of $O(n)$ whose dimension $\DD_{S}(n)$ satisfies $\DD_{T}(0) = \DD_{S}(0)$ i.e.\@ there is a degeneracy in the spectrum of the theory at $n=0$. The two-point function of $S$ can be written as
\beq
\label{eq:SS2pt}
\expec{S(x)S(0)} =  \frac{2n\, c_S(n)}{|x|^{2\DD_S(n)}}\,,
\eeq
where we have extracted a factor of $2n$ in the normalization of $S$ for convenience.\footnote{In minimal subtraction, many operators of interest, such as $\no{\phi_i^2}$, have a two-point function proportional to $n$, at least to leading order in epsilon. For such operators, we have $c_{S}(n) \sim 1$ in the normalization of~\reef{eq:SS2pt}.
} We will assume that $c_S(n)$ also has a finite limit as $n \to 0$, and as above we will set $c_S(n)  = 1$ for simplicity.  Now define the operator
\beq
\overline{T}_{ij} \ldef T_{ij} + \frac{1}{n} \, \mu^{\DD_T - \DD_S} \, \dd_{ij}\, S\,,
\eeq
where we were forced to introduce a scale $\mu$. By construction, $\overline{T}$ has a finite two-point function in the limit $n \to 0$:
\beq
  \expec{\overline{T}_{ij}(x)\overline{T}_{kl}(0)} 
 \limu{n \to 0}  \frac{1}{|x|^{2\DD_T(0)}} \left[ (\dd_{ik}\dd_{jl}+\dd_{il}\dd_{jk}) - 2 \a \,  \dd_{ij}\dd_{kl} \, \ln \mu^2 x^2 \right]
\eeq
where $\a$ is defined as
\beq
\label{eq:diffDelta}
\a \ldef \lim_{n \to 0} \, \frac{1}{n} \left[\DD_{S}(n) - \DD_{T}(n) \right].
\eeq
We see that resolving the $1/n$ pole in~\reef{eq:T2pt} has given rise to a logarithm. 
Moreover, we have
\beq
\expec{\overline{T}_{ij}(x) S(0)} \; \limu{n \to 0} \; 2\dd_{ij}  \, |x|^{-2\DD_{T}(0)} \qquad \text{and} \qquad
\expec{S(x) S(0)} \; \limu{n \to 0} \; 0\,.
\eeq
This implies that the operators $\overline{T}_{ij}$ and $\overline{S} \ldef \a S$ form a logarithmic rank-two multiplet of dimension $\DD_{T}(0)$ in the $n=0$ theory. 

However, if $T$ and $S$ combine into a logarithmic multiplet, their three-point functions with $\phi_i$ must also be related. For symmetry reasons, at finite $n$ the latter are of the form
\bsub
\label{eq:TS3pt}
\begin{align}
\expec{\phi_i(x_1)\phi_j(x_2) T_{kl}(x_3)} &= \la_T(n) \left[\dd_{ik}\dd_{jl} + \dd_{il}\dd_{jk} - \frac{2}{n} \dd_{ij}\dd_{kl}\right] \mbf{P}_{\DD_\phi \DD_\phi \DD_T(n)\,,} \\
\expec{\phi_i(x_1)\phi_j(x_2) S(x_3)}  &= 2n \la_S(n) \, \dd_{ij} \, \mbf{P}_{\DD_\phi \DD_\phi \DD_S(n)}\,,
\end{align}
\esub
where we have extracted a factor of $2n$ for convenience. We will assume that $\la_T(0)$ is finite. But then the only way to cancel the $1/n$ pole in $\expec{\phi \phi T}$ is if
\beq
\la_S(n) \; \limu{n \to 0} \frac{\la_T(0)}{n} + \wt{\la} + \sO(n)\,. \label{eq:condope}
\eeq
In that case, we obtain
\bsub
\label{eq:better3pt}
\begin{align}
  \expec{\phi_i(x_1)\phi_j(x_2) \overline{T}_{kl}(x_3)} &\limu{n \to 0}  \Big[\la_T(0) \left(\dd_{ik}\dd_{jl} + \dd_{il}\dd_{jk} \right) \nn \\
    &\qquad\qquad + 2 \dd_{ij} \dd_{kl} \left(\wt{\la} - \la_T'(0) + \a \la_T(0) \tau_3 \right)\! \Big] \mbf{P}_{\DD_\phi \DD_\phi \DD_T(0)} \,,\\
\expec{\phi_i(x_1)\phi_j(x_2) \overline{S}(x_3)}   &\limu{n \to 0} 2\a \la_T(0)  \, \dd_{ij} \, \mbf{P}_{\DD_\phi \DD_\phi \DD_T(0)}\,,
\end{align}
\esub
consistent with conformal invariance. In~\reef{eq:better3pt}, we have set $\mu = 1$ for simplicity.

Finally, let's consider the contributions of $T_{ij}$ and $S$ to the $\phi_i$ four-point function. By writing down the $\phi_i \times \phi_j \sim T_{ij} + S$ OPEs, it follows that at finite $n$ these are given by
\begin{multline}
  \prod_{i < j} |x_{ij}|^{2\DD_\phi/3} \, \expec{\phi_i \phi_j \phi_k \phi_l} \, \supset \, \big(\la_{T}(n)\big)^2 \left[\dd_{ik}\dd_{jl} + \dd_{il}\dd_{jk} - \frac{2}{n} \dd_{ij}\dd_{kl}\right] G_{\DD_T(n)}^{(0)}(u,v;\DD_\phi,\ldots,\DD_\phi) \\
  +  2n \big(\la_{S}(n)\big)^2 \, \dd_{ij} \dd_{kl} \, G_{\DD_S(n)}^{(0)}(u,v;\DD_\phi,\ldots,\DD_\phi)\,.
\end{multline}
Taking the limit $n \to 0$, this becomes
\begin{multline}
  \label{eq:n0cb}
   \big(\la_{T}(0)\big)^2  \left[\dd_{ik}\dd_{jl} + \dd_{il}\dd_{jk} \right]  G_{\DD_T(0)}^{(0)}(u,v;\DD_\phi,\ldots,\DD_\phi) \\
  + 2\dd_{ij}\dd_{kl} \left[2 \la_T(0) \big(\wt{\la} - \la_T(0)\big) +  \a \big(\la_T(0)\big)^2 \pd_\DD \right] G_{\DD}^{(0)}(u,v;\DD_\phi,\ldots,\DD_\phi)\big|_{\DD = \DD_T(0)}\,.
\end{multline}
Eq.~\reef{eq:n0cb} shows that the contribution of $S$ and $T_{ij}$ to the four-point function at $n=0$ is governed by a logarithmic conformal block: a linear combination of a normal block $G_{\DD}^{(0)}(u,v)$ together with its derivative $\pd_\DD G_{\DD}^{(0)}(u,v)$, in agreement with the fact that $\overline{T}_{ij}$ and $\overline{S}$ form a rank-two multiplet. In passing, we notice that Eq.~\reef{eq:n0cb} could have been obtained by simply reading off the OPE coefficients from~\reef{eq:better3pt}.

\subsubsection{$O(n)$ in perturbation theory}

Let us now specify the general discussion of the previous section to a concrete example. Here we will study the $O(n)$ model at leading order in the quartic coupling $\lambda$. Let us consider the model defined in Eq.~\ref{eq:lambda4action} with $r=0$.
We are interested in the leading order correction to the four point function in the neighbourhood of $d=4$, where there is a weakly coupled fixed point:
\be
\langle\phi_i \phi_j \phi_k \phi_l \rangle_{g} =  \langle\phi_i \phi_j \phi_k \phi_l \rangle_{g=0} - \frac{g}8 \int d^d y  \langle\phi_i \phi_j \phi_k \phi_l \, \NO{ (\phi_a(y)^2)^2} \rangle_{g=0} + O(g^2)\,,
\ee
and we have suppressed the arguments of the external fields $x_1,x_2,x_3,x_4$. The first piece is the free, disconnected four point function:
\begin{multline}
 |x_{12}|^{d-2}|x_{34}|^{d-2}  \; \expec{\phi_i \phi_j \phi_k \phi_l}_{\lambda=0} =  \delta_{ij}\delta_{kl} + u^{\frac{d-2}2} \,\delta_{ik}\delta_{jl}+ \left(u/v\right)^{\frac{d-2}2} \delta_{il}\delta_{jl}\\
\phantom{----} \equiv 2n \mathcal \delta_{ij}\delta_{kl}\,  \mca{S}(u,v) +\left( \delta_{il}\delta_{jk}+\delta_{ik}\delta_{jl}-\frac2n\delta_{ij}\delta_{kl} \right)\mathcal T(u,v)+ \left( \delta_{il}\delta_{jk}-\delta_{ik}\delta_{jl}\right)\mathcal A(u,v)\,. 
\end{multline}
The functions $\mca{S}(u,v)$ and $\mca{T}(u,v)$ admit the following conformal block decomposition:
\be
\mathcal S(u,v) = \sum_{\ell \text{ even}}  (\lambda^S_{\ell})^2 \, \wh{G}_{\ell+d-2}^{(\l)}(u,v)\,,
\quad
\mathcal T(u,v) = \sum_{\ell \text{ even}}  (\lambda^T_{\ell})^2 \, \wh{G}_{\ell+d-2}^{(\l)}(u,v)\,,
\ee
where
\bea
(\lambda^T_\ell)^2 = \frac{p_{\ell}}2\,,\qquad (\lambda^S_\ell)^2 = \frac{p_{\ell}}{2n^2} \,.
\eea
For convenience we are working with the conventional conformal blocks $\wh{G}_{\DD}^{(\l)}$ in this section. The coefficients $p_\l$ can be expressed in terms of the coefficients $\mfr{q}$ from Eq.~\reef{eq:qDef}, however their value will not matter for this discussion. Notice that the coefficients $\la_\l^S$ and $\la_\l^T$ are in agreement with the asymptotics of Eq.~\reef{eq:condope}, with $\wt{\la} = 0$.

Next we compute the leading correction to the four point function. First, notice that
\bea
\frac18\int d^d y  \langle\phi_i \phi_j \phi_k \phi_l \, \NO{ (\phi_a(y)^2)^2} \rangle_{g=0}  = \left(  \delta_{ij}\delta_{kl} + \delta_{ik}\delta_{jl}+\delta_{il}\delta_{kj}  \right)\int d^d y  \prod_{i=1}^4 \frac{1}{|y-x_i|^{d-2}}\,.
\eea
At one-loop order in the epsilon expansion, this integral must be evaluated at $d=4$. Although the resulting integral is finite, it will be useful to rewrite it as follows:
\beq
\int d^d y  \prod_{i=1,2} \frac{1}{|y-x_i|^{\Delta}}  \prod_{i=3,4}  \frac{1}{|y-x_i|^{d-\Delta}}
\eeq
where we have in mind that in the end the limit $d \to 4$, $\DD \to 2$ must be taken.
Following~\cite{Dolan:2011dv}, this yields:
\begin{multline}
   \frac{\pi^{d/2}}{|x_{12}|^\Delta |x_{34}|^{d-\Delta}}
    \left[
      \frac{\Gamma(d/2-\Delta)}{\Gamma(\Delta)}\frac{\Gamma(\Delta/2)^2}{\Gamma((d-\Delta)/2)^2} \wh{G}_{\DD}^{(0)}(u,v)
      + \frac{\Gamma(\Delta-d/2)}{\Gamma(d-\Delta)}\frac{\Gamma((d-\Delta)/2)^2}{\Gamma(\Delta/2)^2} \wh{G}^{(0)}_{d-\DD}(u,v)
      \right].
\end{multline}
Near $\Delta=d/2$  the above formula reduces to a linear combination of $\wh{G}_{d/2}^{(0)}$ and $\partial_\Delta \wh{G}_{\DD}^{(0)}|_{\DD = d/2}$. Finally, after taking the limit $d \to 4$ we obtain 
\be
\label{eq:identity}
-\frac18\int d^4 y  \langle\phi_i \phi_j \phi_k \phi_l \, \NO{ (\phi_a(y)^2)^2} \rangle_{g=0} = - 2\pi^{2}\frac{\left(  \delta_{ij}\delta_{kl} + \delta_{ik}\delta_{jl}+\delta_{il}\delta_{jl}  \right)}{|x_{12}|^2 |x_{34}|^2}
\left(\wh{G}_{2}^{(0)}(u,v) - \frac{\pd}{\pd \DD} \wh{G}_{\DD}^{(0)}(u,v)\big|_{\DD = 2} \right).
\ee
We conclude that the only effect of this leading-order contribution to the four-point function is the following:  it modifies the OPE coefficients of the leading tensor and singlet scalar operators, and it gives rise to two anomalous dimensions:
\bsub
\begin{align}
\mathcal S(u,v)_{g\neq0} &= (\lambda^S_{0})^2 \,  \wh{G}^{(0)}_{2+\gamma_S}(u,v) + \sum_{\ell \geq 2 \text{ even}}  (\lambda^S_{\ell})^2\, \wh{G}_{\ell+2}^{(\ell)}(u,v) + O(g^2)\,, \\
\mathcal T(u,v)_{g\neq0} & =(\lambda^T_{0})^2 \, \wh{G}_{2+\gamma_T}^{(0)}(u,v) + \sum_{\ell \geq 2 \text{ even}}  (\lambda^T_{\ell})^2 \, \wh{G}_{\ell+2}^{(\l)}(u,v)+ O(g^2)
\end{align}
\esub
where
\beq
(\lambda^T_{0})^2  =  \frac{p_{0}}2 - 2 \pi^2 g\,,
\quad
(\lambda^S_{0})^2 = \frac{p_{0}}{2n^2} - \pi^2 g\frac{n+2}{n^2} \,,
\quad
\gamma_T =  \frac{4 \pi^2 g}{p_0}\,,
\quad
\gamma_S =  \frac{2 \pi^2 g(n+2)}{p_0}\,.
\eeq
Here $g$ must be tuned to its critical value $g_* = \sO(\eps)$ and we use that
\beq
\wh{G}_{2+\gamma}^{(0)}(u,v) = \wh{G}_{2}^{(0)}(u,v) + \gamma \, \fracpt{}{\DD} \wh{G}_{\DD}^{(0)}(u,v)\big|_{\DD=2} + \sO(\gamma^2)\,.
\eeq
All OPE coefficients $\la^{S,T}_{\l}$ with $\l \geq 2$ are not modified compared to their values in the free theory.
Let us now focus on the scalar primaries with dimension $\simeq 2$ in the $\mca{S}$ and $\mca{T}$ channel.  Notice that $\gamma_S/\gamma_T=(n+2)/2$, consistent with the standard $\epsilon$-expansion prediction:
\be
\gamma_S = \frac{n+2}{n+8}\epsilon \,,
\qquad
\gamma_T = \frac{2}{n+8}\epsilon\,.
\ee
In the $n\to 0$ limit we see that the conditions \reef{eq:diffDelta} and \reef{eq:condope} are satisfied, with%
\bea
\alpha=\frac{2\pi^2 g}{p_0}\,,
\qquad
\lambda_T(0)^2=\frac{p_0}2-2\pi^2 g\,,
\qquad
\lambda'_T(0)=0\,,
\qquad
\wt{\lambda} =\frac{2\pi^2 g}{p_0^2}\,.
\eea
which can now be plugged back into \reef{eq:n0cb}. We stress that the appearance of the derivative $\pd_\DD \wh{G}_\DD$ of a conformal block in~\reef{eq:identity} is an artifact of working in perturbation theory. At finite $n$, resumming all terms in the epsilon expansion would eliminate such derivatives.

Summarizing, we have confirmed that to leading order in perturbation theory, the $\phi_i$ four-point function of the $O(n \to 0)$ model behaves as it would in a logarithmic CFT. In particular, the scaling dimensions of all operators in the $\mca{S}$ and $\mca{T}$ channels of $O(n)$ collide as $n \to 0$, i.e.\@ we have ${\lim_{n \to 0} \Delta_S(n)-\Delta_T(n) = 0}$. Provided that this persists to all orders in perturbation theory, the $O(n \to 0)$ model is a logCFT.

Note that at leading order in epsilon, only two operators obtained a nonzero anomalous dimension. At the 3$d$ critical point --- which is nonperturbative --- all operators in the $\phi_i \times \phi_j$ OPE (except for the stress tensor) are expected to have a finite anomalous dimension. The fact that all of them must collide pairwise in the limit $n \to 0$ is clearly a strong constraint on the spectrum of the $O(n)$ model.

\subsection{Percolation and the $Q \to 1$ Potts model}

The $Q$-state Potts model can be thought of as the theory of an order parameter field $\phi_a(x)$ with $a = 1,\ldots,Q$, with interactions invariant under a global symmetry group $S_Q$, the permutation group acting on $Q$ elements.  The order parameter satisfies $\sum_{a=1}^Q \phi_a = 0$, and it forms an irreducible representation of $S_Q$.  The model is described by the Landau-Ginzburg action
\beq
\mca{L} = \frac{1}{2} \sum_a (\pd_\mu \phi_a)^2 + \frac{r}{2} \sum_a \phi_a^2 + \frac{g}{3!} \sum_a \phi_a^3 \label{eq:Pottsaction}\,,
\eeq
which has a weakly coupled fixed point in $6-\eps$ dimensions for sufficiently small $Q$.

We will be interested in the $Q \to 1$ limit, which is known to describe the theory of percolation in $2 \leq d < 6$ dimensions. As with the $O(n)$ model, logarithmic behavior in the $Q \to 1$ Potts model arises due to group-theoretical considerations. The irreducible representations appearing in the OPE of the field $\phi_a$ with itself were worked out a long time ago \cite{Wallace1978}, with a more detailed analysis appearing in \cite{Vasseur:2012tz,Vasseur:2013baa}. 
We can schematically write the OPE as
\bea
\phi_a \times \phi_b=P_{ab} \, S+(V_{a}+V_b)+F_{[ab]}+T_{(ab)}\,,
\eea
with the projector $P_{ab}$ being defined by
\bea
P_{ab} \ldef \delta_{ab}-\frac 1Q\,\qquad \sum_{a=1}^Q P_{ab}=0.
\eea
One should understand the above as a sum of scalar, vector, antisymmetric, and tensorial representations, with dimensions $1, (Q-1), \frac{(Q-1)(Q-2)}2$ and $\frac{Q(Q-3)}2$ respectively. Two point functions of the fields appearing in these components must take the form
\bea
\langle S S\rangle \propto 1 \qquad \langle V_a V_b\rangle \propto  P_{ab} \qquad \langle F_{ab} F_{cd}\rangle \propto P_{a[c}P_{d]b} \qquad \langle T_{ab} T_{cd}\rangle \propto P_{abcd} \,,
\eea
with
\bea
P_{abcd} \ldef \delta_{a\neq b} \delta_{c \neq d}\left[\delta_{ac}\delta_{bd}+\delta_{ad}\delta_{bc}-\frac{1}{Q-2}(\delta_{ac}+\delta_{ad}+\delta_{bc}+\delta_{bd})+\frac{2}{(Q-1)(Q-2)}\right]\,.
\eea
The field $\phi_a$ itself is not logarithmic: its two-point function is of the form
\beq
\expec{\phi_a(x) \phi_b(0)} \propto  \frac{P_{ab}}{|x|^{2\DD_\phi(Q)}}\,,
\eeq
hence it is finite at $Q = 1$. However, in the OPE $\phi_a \times \phi_b$ there will be logarithmic operators, with the simplest logarithmic doublet built out of $\phi^2$ and $\phi^2_{ab}$, the ``watermelon'' or two-leg operators\footnote{In~\cite{Vasseur:2013baa}, these are written as $\phi^2\equiv t^{(0,2)}$ and $\phi^2_{ab}\equiv t^{(2,2)}$. In \cite{Cardy:2013rqg}, $\phi^2_{ab}$ is written $\hat \phi^{(2)}_{ab}$, and $\phi_{ab}^{(2)}$ is the finite $Q\to 1$ limit combination.} which are the leading operators in $S$ and $T_{ab}$.
The fact that various operators organize themselves in logarithmic multiplets at $Q=1$ also follows from the four point function of $\phi_a$:
\begin{multline}
\langle \phi_a(x_1) \phi_b(x_2) \phi_c(x_3) \phi_d(x_4)\rangle= \prod_{i < j} \frac{1}{|x_{ij}|^{2\DD_\phi/3}} \, \bigg(P_{ab}P_{cd} \, \mca{S}(u,v)+P_{a(c}P_{d)b} \, \mca{V}(u,v)  \\
+P_{a[c}P_{d]b}\,\mca{F}(u,v)+P_{abcd} \, \mca{T}(u,v)\bigg)\,.
\end{multline}
The divergence of the projector $P_{abcd}$ as $Q\to 1$ implies that the scalar and tensor components in the OPE must combine as 
\bea
\overline{T}_{ab}:=T_{ab}+\frac{1}{Q-1}\, \mu^{\Delta_T-\Delta_S}\, S
\eea
Thus the story here is almost exactly the same as for the previous section, and we will not repeat it. Let us note however that in particular, a smooth $Q\to 1$ limit requires that the dimensions of the operators $\phi^2, \phi^2_{ab}$ agree when $Q\to 1$. Wallace and Young~\cite{Wallace1978} have shown this is true to all orders in perturbation theory.
For instance, to leading order we have~\cite{Amit1976,Theumann1984}:
\bea
\Delta_{\phi^2}-\Delta_{\phi^2}^{\mbox{\tiny free}}&=&-\frac{5 (Q-2)}{3(10-3Q)} \epsilon \limu{Q\to 1}\frac{5}{21} \epsilon\,,\\
\Delta_{\phi^2_{ab}}-\Delta_{\phi^2_{ab}}^{\mbox{\tiny free}}&=&\frac{Q+4}{3(10-3Q)} \epsilon \limu{Q\to 1} \frac{5}{21}\epsilon \,.
\eea
To finish this section, let us mention that the $Q\to 2$ limit reproduces the logarithmic extension of the Ising model~\cite{winter2008geometric,Cardy:2013rqg}.
In that case, logarithms appear in OPEs of higher-dimensional operators, for instance in $\phi^2_{ab}\times \phi^2_{cd}$.  Another case of interest is the $Q \to 0$ limit of the Potts model, which has a geometrical interpretation in terms of \emph{spanning forests}~\cite{Jacobsen:2003qp,Caracciolo:2004hz,Deng:2006ur}.


\section{Discussion}\label{sec:disc}

In this paper we pursued a systematic discussion of logarithmic CFTs in $d$ dimensions, exploiting constraints imposed by the global conformal group $SO(d+1,1)$. Our work is complementary to most of the existing literature on 2$d$ logCFTs, which uses Virasoro and $\mca{W}$ algebra techniques.  We obtained the most general form of correlation functions consistent with logarithmic conformal invariance in a number of cases. Special attention was paid to four-point functions. In particular we examined the consequences of Bose symmetry for these correlators and we showed that logarithmic multiplets contribute to them via ``logarithmic'' conformal blocks that can be computed in terms of derivatives of ordinary conformal blocks with respect to scaling dimensions. Along the way, we made explicit how to reconcile scale invariance with the presence of logarithms in correlation functions. As discussed in detail in Sec.~\ref{sec:scale}, the running of coefficients appearing in two- and three-point functions compensates for the non-trivial scale transformations of logarithms, giving rise to RG-invariant correlation functions. 

The formalism developed in this paper applies to any spacetime dimension, in particular to two dimensions. In this context we want to address some remarks made in the literature which we believe are incorrect. Let us consider a rank-two scalar primary $\phi_a$ of dimension $\DD_\phi > 0$. The two-point function $\expec{\phi_2 \phi_2}$ of its bottom component $\phi_2$ vanishes, as shown in Sec.~\ref{sec:twopt}. It was argued in~\cite{Flohr:2000mc} that for cluster decomposition to hold, all $n$-point functions of $\phi_2$ must vanish as well:
\beq
\label{eq:claim0}
\expec{\phi_2(z_1) \dotsm \phi_2(z_n)} = 0\,.
\eeq
This is certainly not required by global conformal invariance. The argument that this is required by cluster decomposition cannot be correct, since in any conformally invariant theory (including logarithmic ones), cluster decomposition is automatically satisfied thanks to the OPE as long as all operator dimensions are positive. Indeed, the fact that the two point function vanishes 
does not prevent $\phi_2$ from having a non-zero OPE. Section~\ref{sec:holoscalar} provides a specific holographic example of this fact, see for instance Eq.~\reef{eq:ads3pt}.

Nevertheless, the $\phi_2$ four-point function vanishes in many 2$d$ examples. This can be understood from the following argument. What is definitively true is that the identity operator does not contribute to the $\phi_2\times \phi_2$ OPE. This holds both when the identity operator is of rank one and when it's part of a larger logarithmic multiplet, and it is a simple consequence of the results derived in this paper. In 2$d$ CFTs, this means that the Virasoro conformal block of the identity operator is absent from the $\phi_2$ four-point function. This does \emph{not} necessarily imply that the full four-point function vanishes, since other Virasoro blocks may appear in principle. One should only expect the $\phi_2$ four-point function to vanish if all primaries in the theory belong to an extended bosonic chiral algebra, as is the case for the logarithmic minimal models --- since in that case, all possible contributions to the four-point function are related to that of the identity. 

In Sec.~\ref{sec:ads} we discussed some holographic models of logarithmic CFTs. Two results stand out. First, we have shown how some of these models can be derived by coupling an ordinary bulk theory to disorder. It would be  interesting to understand this in more detail, including interactions, and to extend these results to spin two at the non-linear level. We also derived for the first time models of logarithmic spin-1 fields, including conserved currents. Altogether, there now exists a logarithmic generalization of Einstein-Maxwell-scalar theory in the bulk, variations of which have been extensively used in holographic models of condensed matter phenomena. Using our new models, it seems likely that a range of similar applications can now be made for strongly coupled, disordered, boundary theories. For instance, it would be very interesting to study a disordered analog of a holographic superconductor.

Finally, our results lay the groundwork for any future bootstrap applications. An interesting question for the immediate future is whether a logarithmic bootstrap analysis of the $O(n \to 0)$ model can reproduce or even improve known critical exponents~\cite{Pelissetto:2000ek,Blote,denganiso} for self-avoiding walks in 3$d$. The 3$d$ $O(n)$ model at finite $n$ has already been studied in great detail using numerical bootstrap methods~\cite{Rattazzi:2010yc,Kos:2013tga,Kos:2015mba,Kos:2016ysd}.  Likewise, it would be interesting to study the Potts model in $2 \leq d < 6$ in the limit $Q \to  1$ using bootstrap techniques. Any results could be compared to predictions coming from the $6-\eps$ expansion or Monte Carlo methods~\cite{Gracey:2015tta,deAlcantaraBonfim:1981sy,Gori:2015rta}. 

Unfortunately, since logCFTs are non-unitary, bootstrapping them is not quite straightforward. One possibility is to study the logarithmic theories directly using the determinant method developed by Gliozzi~\cite{Gliozzi:2013ysa,Gliozzi:2014jsa}. 
A different workaround comes from the fact that many logCFTs arise as limits of ordinary CFTs. This was exploited in an interesting recent paper~\cite{Shimada:2015gda} where the $O(n \to 0)$ critical point was studied by computing bootstrap constraints at fractional $n < 1$ and extrapolating these results to $n=0$. In the paper in question unitarity was assumed for fractional $n$, although in principle unitarity violations occur~\cite{Maldacena:2011jn}. This does not necessarily lead to large errors: for instance, the Wilson-Fisher fixed point in $4-\eps$ dimensions is known to be nonunitary~\cite{Hogervorst:2015akt} although a bootstrap analysis of the same model gave results consistent with RG predictions~\cite{El-Showk:2013nia}. The reason for this was that unitarity violations only affected high-dimension operators in the CFT spectrum. It may be an interesting problem to quantify the unitarity violations in the $O(n)$ model at fractional $n$ and to see to which extent they affect low-lying operators, and mutatis mutandis for the $Q$-state critical Potts model.

In this paper we have scratched the surface of the theory of logarithmic CFTs in $d$ dimensions. There are many obvious extensions in this direction, for instance by considering supersymmetric theories or defects and boundaries. We are optimistic that the near future will bring breakthroughs both in understanding formal properties of these theories and in cornering concrete examples through the conformal bootstrap.


\subsection*{Acknowledgements}

We thank Sheer El-Showk, Kristan Jensen, Nikita Nekrasov, Leonardo Rastelli, Slava Rychkov and Balt van Rees for useful conversations and in particular Paolo De Los Rios for spurring the work which led to the this paper.
MH and AV thank the Simons Center for Geometry and Physics at Stony Brook University for its hospitality during the Simons Summer Workshop 2015. This research was supported in part by Perimeter Institute for Theoretical Physics. Research at Perimeter Institute is supported by the Government of Canada through Industry Canada and by the Province of Ontario through the Ministry of Research and Innovation.
MFP was supported by a Marie Curie Intra-European Fellowship of the European Community's 7th Framework Programme under contract number PIEF-GA-2013-623606.

\appendix

\section{Simplifications for the two-point function}\label{sec:twopts}

In Sec.~\ref{sec:twopt}, two statements about two-point functions in logCFTs were made but not proven: first, that two-point functions of operators in different multiplets vanish, and second that two-point functions can be brought to a canonical form. Here we will discuss these claims in more detail. For simplicity we consider the scalar ($\l = 0$) case, but for higher spins the analysis is similar.

Let $\Oo_a$ be a primary of rank $\rnk$ and $\wt{\Oo}_a$ a primary of rank $\rnk'$. Then we will argue that after a suitable change of basis the two-point function $\expec{\Oo_a(x) \wt{\Oo}_b(x)}$ vanishes. The argument goes as follows. Without loss of generality we can assume that $\rnk \leq \rnk'$ and the two primaries have equal scaling dimension $\DD$. Conformal invariance requires that the different two-point functions depend on $2\rnk + \rnk'$ parameters $k_{1\dotsm \rnk}$, $\rho_{1 \dotsm \rnk}$ and $\wt{k}_{1 \dotsm \rnk'}$ as follows:
\bsub
\begin{align}
  \expec{\Oo_a(x) \Oo_b(0)} &= \frac{1}{|x|^{2\DD}} \sum_{m=0}^{\rnk-n} k_{n+m} \frac{(-1)^m}{m!}(\ln x^2)^m \qquad [n = a+b-\rnk-1]\,,\\
  \expec{\Oo_a(x) \wt{\Oo}_b(0)} &= \frac{1}{|x|^{2\DD}} \sum_{m=0}^{\rnk-n} \rho_{n+m} \frac{(-1)^m}{m!}(\ln x^2)^m \qquad [n = a+b-\rnk-1]\,,\\
  \expec{\wt{\Oo}_a(x) \wt{\Oo}_b(0)} &= \frac{1}{|x|^{2\DD}} \sum_{m=0}^{\rnk'-n} \wt{k}_{n+m} \frac{(-1)^m}{m!}(\ln x^2)^m \qquad [n = a+b-\rnk'-1]\,.
\end{align}
\esub
It's convenient to combine the two multiplets into one vector $\Oo_I = ( \Oo_a,\wt{\Oo}_b )$, with $I = 1,\ldots,\rnk+\rnk'$. The dilatation operator acts on $\Oo_I$ as follows:
\beq
\label{eq:Ddef}
D \ket{\Oo_I} = - i \msc{D}\du{I}{J} \ket{\Oo_J}, \qquad \msc{D}\du{I}{J} = 
\begin{pmatrix}
  \mbf{\DD}_{\rnk \times \rnk} & \mbf{0}_{\rnk \times \rnk'} \\
  \mbf{0}_{\rnk' \times \rnk} & \mbf{\DD}_{\rnk' \times \rnk'}
\end{pmatrix}\,,
\eeq
where the matrix $\mbf{\DD}$ is defined in Eq.~\reef{eq:Jblock}.

Next, we will need to consider for which values of the parameters $k_m$, $\rho_m$ and $\wt{k}_m$ the Hilbert space is non-degenerate. This will require the Gram matrix, which will be discussed in more detail in Appendix~\ref{sec:gram}. We will consider the cases $\rnk < \rnk'$ and $\rnk = \rnk'$ separately.

If $\rnk < \rnk'$, the determinant of the Gram matrix equals $\pm (k_\rnk)^{\rnk} (\wt{k}_{\rnk'})^{\rnk'}$, so requiring that there are no null states imposes that $k_\rnk \neq 0$ and $\wt{k}_{\rnk'} \neq 0$. Then consider the following change of basis:
\beq
\label{eq:Cmatdef}
\ket{\Oo_I} \; \to \; \ket{\Oo'_I} = \mbf{C}\du{I}{J} \ket{\Oo_J}, \qquad \mbf{C}\du{I}{J} =
\left(\begin{array}{c c c c | c} 
  &  & \unit_{\rnk \times \rnk} &  & \mbf{0}_{\rnk \times \rnk'} \\
  \hline
   \a_1 & \a_2 & \dotsm & \a_{\rnk} & \\
   0 & \a_1 & \dotsm & \a_{\rnk - 1} & \\
   \vdots &  & \ddots & \vdots  &  \unit_{\rnk' \times \rnk'} \\   
   0 & 0 & \dotsm & \a_1 &  \\ 
    &   & \mbf{0}_{(\rnk'- \rnk) \times \rnk}  & &
\end{array}\right)\,,
\eeq
which depends on $\rnk$ parameters $\a_1, \ldots, \a_{\rnk}$. The matrix $\mbf{C}$ commutes with $\msc{D}$ defined in Eq.~\reef{eq:Ddef}, hence the matrix $\expec{\Oo'_I(x) \Oo'_J(0)}$ is of the same form as $\expec{\Oo_I(x) \Oo_J(0)}$, only with coefficients that depend on the choice of $\a_{1 \dotsm \rnk}$. 
Using the assumption that $k_\rnk \neq 0$, there exists a suitable choice of parameters $\a_1, \ldots, \a_{\rnk}$ such that
\beq
\expec{\Oo'_I(x) \Oo'_J(0)} = \begin{pmatrix} * & \mbf{0}_{\rnk \times \rnk'} \\ \mbf{0}_{\rnk' \times \rnk} & * \end{pmatrix}\,,
\eeq
implying that the off-diagonal two-point functions in the new basis are vanishing. We have not found a compact expression for the parameters $\a_i$, although it is straightforward to find the explicit change of basis using computer algebra. This is consistent with a counting of parameters: there are $\rnk$ adjustable parameters $\a_i$ that we use to set $\rnk$ coefficients $\rho_i$ to zero.

Next, we consider the case $\rnk = \rnk'$. In that case, the absence of null states requires that $k_\rnk \, \wt{k}_{\rnk} \neq (\rho_\rnk)^2$. We can isolate three subcases: (a) $k_\rnk \neq 0$, (b) $k'_\rnk \neq 0$ and (c) $k_\rnk = k'_{\rnk} = 0$ but $\rho_{\rnk} \neq 0$. (When both $k_{\rnk}$ and $\wt{k}_{\rnk}$ are non-zero, both cases (a) and (b) apply.) In case (a) we can recycle the previous argument. In case (b) the same holds, after swapping $\Oo_a \lra \wt{\Oo}_a$. Finally, for case (c) we consider a different change of basis: 
\beq
\label{eq:worseCoB}
\ket{\Oo_I} \; \to \; \ket{\Oo'_I} = \mbf{E}\du{I}{J} \ket{\Oo_J}, \qquad \mbf{E}\du{I}{J} =
\left(\begin{array}{c c c c | c} 
  &  & \unit_{\rnk \times \rnk} &  & \unit_{\rnk \times \rnk} \\
  \hline
   \b_1 & \b_2 & \dotsm & \b_{\rnk} & \\
   0 & \b_1 & \dotsm & \b_{\rnk - 1} & \\
   \vdots &  & \ddots & \vdots  &  \unit_{\rnk \times \rnk} \\   
   0 & 0 & \dotsm & \b_1 &  \\ 
  \end{array}\right)\,,
\eeq
which depends on $\rnk$ parameters $\b_{1 \dotsm \rnk}$. Notice that the upper right corner in~\reef{eq:worseCoB} is non-zero, hence $\mbf{E}$ is of a different form than $\mbf{C}$ from Eq.~\reef{eq:Cmatdef}. Again one can find a suitable choice of parameters $\b_i$ that makes all off-diagonal matrix elements in $\expec{\Oo'_I \Oo_J'}$ vanish. 
The fact that such a change of basis exists is consistent with the counting of degrees of freedom: there are $\rnk$ coefficients $\rho_i$ that need to be set to zero and $\rnk$ adjustable parameters $\b_i$.

Next, consider a single primary multiplet $\Oo_a$ of rank $\rnk$. By conformal invariance, its two-point function is of the following form:
\beq
\label{eq:O2ptorig}
\expec{\Oo_a(x) \Oo_b(0)} = \frac{1}{|x|^{2\DD}} \sum_{m=0}^{\rnk+1-a-b} k_{m+a+b-1} \, \frac{(-1)^m}{m!}(\ln x^2)^m\,.
\eeq
The determinant of the Gram matrix is equal to $\pm (k_\rnk)^\rnk$, hence we will assume that $k_\rnk \neq 0$. We want to prove that after a suitable change of basis $\ket{\Oo_a} \to \ket{\Oo'_a} = \mbf{R}\du{a}{b} \ket{\Oo_b}$, we have 
\beq
\label{eq:O2desired}
\expec{\Oo'_a(x) \Oo'_b(0)} = \frac{(-1)^{n}}{n!} \frac{ k_{\rnk} }{|x|^{2\DD}}(\ln x^2)^{n} \qquad \text{if } \;  n = \rnk + 1 - a-b \geq 0
\eeq
and $\expec{\Oo'_a \Oo'_b} = 0$ if $a+b > \rnk+1$, corresponding to the ``canonical form'' shown in Eq.~\reef{eq:2ptscal}. To achieve this, consider the following change of basis:
\beq
\mbf{R}\du{a}{b} =   \begin{pmatrix}
1 & R_1       & R_2       & \cdots  & R_{\rnk-1} \\
0       & 1 & R_1       & \cdots  & R_{\rnk-2} \\
\vdots  & \vdots  & \vdots& \ddots  & \vdots \\
0       & 0       & 0        & 1 & R_1       \\
0       & 0       & 0       & 0       & 1 \end{pmatrix}\,,
\eeq
which depends on $\rnk-1$ parameters $R_1, \ldots, R_{\rnk-1}$. Since $\mbf{R}\du{a}{b}$ commutes with the matrix $\mbf{\DD}$ from~\reef{eq:Jblock}, it follows that the two-point function~$\expec{\Oo'_a(x)\Oo'_b(0)}$ is of the same form as~\reef{eq:O2ptorig} but with shifted coefficients $k_m \to k'_m(R)$ for $m=1,\ldots,\rnk-1$. The coefficient $k_\rnk$ does not change, since the diagonal elements of the matrix $\mbf{R}\du{a}{b}$ are equal to one. Again, it is possible to adjust the parameters $R_i$ to achieve the desired form~\reef{eq:O2desired}, consistent with the counting of parameters.

\section{Logarithmic OPE}
\label{appendix:OPE}

In this appendix we use the formal replacement introduced in Eq.~\reef{eq:formal-id} to compute the OPE $\phi_1 \times \chi_1$ of the upper components of two distinct rank-2 scalar multiplets $\phi_a,\chi_a$. We only focus on the contribution of a spin-1 non-logarithmic operator, since this result is used in the main text. The analysis presented in this section can be easily generalized to higher rank cases. In the following we assume that the exchanged operator is unit-normalized i.e.:
\beq
\expec{\Oo_{\mu}(x) \Oo_{\nu}(0)} = \frac{1}{|x|^{2\DD}} {\mca{I}_{\mu \nu}(x) }\,.
\eeq
where $\Delta$ is the dimension of $\Oo_{\mu}(x) $. We will also denote $\gamma = \DD_\phi - \DD_\chi$ and $\nu = (d-2)/2$. 

Recalling the notation introduced in Eq.~\ref{eq:3ptgensol}, the general structure of the three point function can be parametrized as:
\beq
\label{eq:3pfappendix}
\expec{\phi_1(x_1) \chi_1(x_2) \Oo_\mu(0)} = \left[ \lambda^{\phi\chi\Oo}_{11} + \frac{\lambda^{\phi\chi\Oo}_{+}}{2} \, (\tau_1 + \tau_2) + \frac{\lambda^{\phi\chi\Oo}_{-}}{2} \, (\tau_1 - \tau_2) + \lambda^{\phi\chi\Oo}_{22} \, \tau_1 \tau_2 \right] \, \mbf{P}_{\DD_\phi\DD_\chi\DD}\,\, X_\mu\,,
\eeq
where $ \lambda^{\phi\chi\Oo}_{\pm}=\lambda^{\phi\chi\Oo}_{12}\pm\lambda^{\phi\chi\Oo}_{21}$, while $\mbf{P}_{\DD_\phi\DD_\chi\DD}$ and $X_\mu$ are defined in Eq.~\reef{eq:Polyakov} and Eq.~\reef{eq:capitalX}.

First, let us recall that the OPE of the bottom components has the standard form:
\beq
 \phi_2(x)\chi_2(0) = \frac{1}{|x|^{\DD_1 + \DD_2 - \DD_\Oo + 1}} \sum_{n,j} \left[
  A_{n,j} \, |x|^{2n}  x^\mu (x\cdot \pd)^j  \Box^n
  + B_{n,j} \, |x|^{2n+2} \pd^\mu (x \cdot \pd)^j  \Box^n
  \right]\!\Oo_\mu(0)\,.
\eeq
We report the first terms of the above series for the sake of completeness:
\begin{align}
&  A_{0,0} = 1, \quad\qquad\qquad\quad A_{1,0} = -\frac{(\DD + \l + \gamma )(\DD + \l -\gamma)}{16(\DD+\l)(\DD+\l+1)(\DD - \nu)}, \nn \\
 &  A_{0,1} = \frac{\DD+\gamma+\l}{2(\DD+\l)}, \qquad
  A_{0,2} = \frac{(\DD+\ga+\l)(\DD+\ga+\l+2)}{8(\DD+\l)(\DD+\l+1)}\,,\\
 & B_{0,0} = \frac{\gamma}{2(\DD+1)(\DD-d+1)}, \quad B_{0,1} = \frac{(\gamma +\Delta +1) (-2 \gamma  \Delta -3 \gamma +\gamma  d+d-\Delta -1)}{4 (\Delta +1) (\Delta +2) (d-2 \Delta -2) (d-\Delta -1)}\,. \nn
\end{align}
Using the derivative trick of Eq.~\reef{eq:formal-id} to express both $\phi_1$ and $\chi_1$ as derivatives of  $\phi_2$ and $\chi_2$ and finally matching with the three point function~\reef{eq:3pfappendix} we obtain:
\beq
\phi_1(x)\chi_1(0) = \frac{1}{|x|^{\DD_1 + \DD_2 - \DD + 1}} \, \mca{D}^\mu(x,\pd) \Oo_\mu(0)\,,
\eeq
where
\begin{align}
  \mca{D}^\mu(x,\pd) &= \left( \lambda^{\phi\chi\Oo}_{11} -\lambda^{\phi\chi\Oo}_{+} \ln x +\lambda^{\phi\chi\Oo}_{22} (\ln x)^2 \right) \, \left[ x^\mu + \frac{1}{2} x^\mu \, x \cdot \pd + \ldots \right] \nonumber\\
  &+ \lambda^{\phi\chi\Oo}_{-} \left[ \frac{1}{2(\DD+1)} x^\mu(x \cdot \pd) + \frac{1}{2(\DD+1)(\DD - d + 1)} x^2 \pd^\mu + \ldots \right] \nonumber\\
  &+\lambda^{\phi\chi\Oo}_{22}\left[ \text{subleading non-logarithmic terms} \right]\,.
\end{align}


\section{Partial wave decomposition}\label{sec:gram}
In this appendix we will prove two claims that are needed in Sec.~\ref{sec:general-cb}.

\addtocontents{toc}{\protect\setcounter{tocdepth}{1}}
\subsection{Gram matrix and proof of Eq.~\reef{eq:logPW}}
\addtocontents{toc}{\protect\setcounter{tocdepth}{2}}

We have in mind a primary operator $\Oo_p$ of rank $\rnk$, hence the label $p$ runs over $1,\ldots,\rnk$. For simplicity we take $\Oo_p$ to be a scalar primary, although the result generalizes to spin $\l \geq 1$ without difficulty. Recall that a primary state is defined by inserting the operator $\Oo_p$ at the origin, i.e.
\beq
\ket{\Oo_p} \ldef \Oo_p(0) \ket{0}\,.
\eeq
All descendants within the multiplet of $\Oo_p$ are obtained by acting with $P_\mu$, the generator of translations. We will denote them as follows:
\beq
\ket{\Oo_p;\a} \ldef P_\a \ket{\Oo_p}\,,
\eeq
where $\a$ is a multi-index, and $P_\a$ is actually $P_{\alpha_1}P_{\alpha_2}\ldots P_{\alpha_k}$ with $k=|\alpha|$ . We can always decompose descendants into irreps of rotations, and so we may choose $\a$ symmetric and traceless. 
Out-states are defined by means of the inversion $x^\mu \, \to \, x^\mu/|x|^2$. For a logarithmic operator, we have
\beq
\label{eq:bra}
\bra{\Oo_p} = \lim_{w \to \infty}  |w|^{2\DD} \sum_{m=0}^{\rnk -p} \frac{(\ln w^2)^m}{m!}  \bra{0} \Oo_{p+m}(w)\,,
\eeq
assuming that $\Oo_p$ is hermitian. We will denote descendant out-states as
\beq
\bra{\Oo_p;\a} \ldef \bra{\Oo_p} K_\a\,.
\eeq
Let's assume that the two-point function of $\Oo_p$ has been brought to its canonical form~\reef{eq:2ptscal}, with coefficient $k_\Oo$. Then the Gram matrix restricted to the primaries is given by
\beq
\brakket{\Oo_p}{\Oo_q} = k_\Oo 
 \begin{pmatrix}
   0       & 0 & \dotsm   & 0       & 1 \\
   0       & 0 & \dotsm  & 1       & 0 \\
   \vdots  & \vdots  &  \Ddots  &   \vdots & \vdots \\
   0       &  1 & \dotsm & 0       & 0 \\
   1       &  0 &\dotsm  & 0       & 0
 \end{pmatrix} = k_\Oo \, \dd_{p+q,\rnk+1}\,.
 \label{eq:curlyI}
 \eeq
 It is easy to see that $\brakket{\Oo_p}{\Oo_q}$ has $\floor{\rnk/2}$ eigenvalues $-k_\Oo$ and $\ceil{\rnk/2}$ eigenvalues $+k_\Oo$. Assuming that $\rnk \geq 2$, this proves that there is at least one negative-norm state in any logarithmic multiplet, proving that logCFTs are non-unitary. 
  
 Next, we need to compute the Gram matrix for the descendants, namely
 \beq
 \label{eq:genGram}
 \mbf{G}_{p\a;q\b} \ldef \brakket{\Oo_p;\a}{\Oo_q;\b} = \bra{\Oo_p}K_\a P_\b \ket{\Oo_q}\,.
 \eeq
 It is easy to see that this vanishes unless $\a = \b$. However, the Gram matrix $\mbf{G}_{p\a;q\b}$ does not completely factorize. In order to compute~\reef{eq:genGram}, we can use the commutation relations~\reef{eq:commu} to eliminate $P_\mu$ and $K_\mu$ in favor of $M_{\mu \nu}$ and $D$. Since $M_{\mu \nu} \ket{\Oo_p} = 0$ in the scalar case, we can express the result as follows:
 \beq
 \bra{\Oo_p}K_\a P_\b \ket{\Oo_q} = \dd_{\a\b} \, \braket{\Oo_p}{g_{\a}(iD)}{\Oo_q}
 \eeq
 cf.\@ Eq.~\reef{eq:normgab}.
 The functions $g_{\a}(iD)$ are all polynomials in $D$, and in general transform as $SO(d)$ tensors. If $\Oo_p$ were a non-logarithmic operator, then the above would evaluate to
 \beq
 \braket{\Oo}{g_{\a}(iD)}{\Oo} = k_\Oo \, g_{\a}(\DD)  \qquad [\Oo \text{ non-logarithmic}]\,.
 \eeq
  In order to generalize this to logarithmic multiplets, we first compute that
 \beq
 \braket{\Oo_p}{(iD)^n}{\Oo_q} = k_\Oo\, (\mbf{\DD}^n \cdot \mca{I})_{pq} = k_\Oo \times \begin{cases}
 
  \displaystyle{\binom{n}{m}} \, \DD^{n-m} & \text{if} \quad m \equiv \rnk + 1 - p -q \geq 0 \\
   0 & \text{otherwise}
 \end{cases} \; .
 \eeq
 This can be rewritten using the following identity:
 \beq
 \binom{n}{m} \DD^{n-m} = \frac{1}{m!} \frac{\pd^m}{\pd \DD^m} \DD^n \,.
 \eeq
 The latter allows us to write
 \beq
 \label{eq:normGram}
 \mbf{G}_{p\a;q\b}= k_\Oo  \, \dd_{\a\b} \, V_{pq}\!\left(\frac{\pd}{\pd \DD} \right) g_{\a}(\DD), \quad  V_{pq}(\pd_{\DD}) = \begin{cases} 
  1/m! \, (\pd_{\DD})^m  & \text{if} \quad m \equiv \rnk + 1 - p -q \geq 0 \\
   0 & \text{otherwise}
 \end{cases} \; .
 \eeq
 For the partial wave decomposition, we need the resolution of the identity, restricted to the multiplet of $\Oo$. This is given by
 \beq
 \sum_{p,q=1}^\rnk \sum_{\a,\b} \mbf{G}^{p\a;q\b} \, \ket{\Oo_p;\a}\bra{\Oo_q;\b}\,,
 \eeq
 where $\mbf{G}^{p\a;q\b}$ is the inverse Gram matrix:
\beq
\sum_{s,\ga} \mbf{G}^{p\a,s\gamma}\,\mbf{G}_{s \gamma,q \beta} = \dd^p_q \, \dd^{\a}_\b\,.
\eeq
Using Eq.~\reef{eq:normGram}, the inverse Gram matrix evaluates to
\beq
\mbf{G}^{p\a, q\b} = \dd^{\a \b}\, k_\Oo^{-1} \, V^{pq}(\pd_{\DD}) \cdot g_{\a}(\DD)^{-1}\,,
\eeq
introducing the matrix $V^{pq}$ defined as follows:
\beq
\label{eq:Vdef}
V^{pq}(\pd_\DD) =   \begin{cases}
   1/n!\, \pd_{\DD}^n    & \text{if} \quad n \equiv p + q - \rnk -1 \geq 0  \\
  0 & \text{if} \quad n < 0
\end{cases}\,.
\eeq
This proves Eq.~\reef{eq:logPW}.

\addtocontents{toc}{\protect\setcounter{tocdepth}{1}}
\subsection{Proof of Eq.~\reef{eq:fullPW}}
\addtocontents{toc}{\protect\setcounter{tocdepth}{3}}

Next, we want to show that Eq.~\reef{eq:logPW} can be written as~\reef{eq:fullPW}. Let us restate the problem here. We are given two differential operators
\bsub
\begin{align}
K_{abp} &= \sum_{k=0}^{\rnk_1 - a}\sum_{l=0}^{\rnk_2 - b}\sum_{m=0}^{\rnk - p} \la^{12\Oo}_{(a+k)(b+l)(p+m)} \frac{1}{k!l!m!} \frac{\pd^k}{\pd \DD_1^k}  \frac{\pd^l}{\pd \DD_2^l}  \frac{\pd^m}{\pd \DD^m}\,, \\
K'_{cdq} &= \sum_{k=0}^{\rnk_3 - c}\sum_{l=0}^{\rnk_4 - d}\sum_{m=0}^{\rnk - q} \la^{34\Oo}_{(c+k)(d+l)(q+m)}  \frac{1}{k!l!m!} \frac{\pd^k}{\pd \DD_3^k}  \frac{\pd^l}{\pd \DD_4^l}  \frac{\pd^m}{\pd \DD^m}  \,,
\end{align}
\esub
that depend on $\rnk_1 \times \rnk_2 \times \rnk$ OPE coefficients $\la^{12\Oo}_{abp}$ and $\rnk_3 \times \rnk_4 \times \rnk$ OPE coefficients $\la^{34\Oo}_{cdq}$. Then we need to show that
\begin{multline}
 \sum_{p,q=1}^\rnk   K_{abp} \cdot \mca{M}[\a,\DD,\DD_1,\DD_2] \left[ V^{pq}\!\left( \fracpt{}{\DD} \right) \cdot  g_\a(\DD)^{-1} \right]  K'_{cdq} \cdot \mca{M}'[\a,\DD,\DD_3,\DD_4]   \\
   = \;   \sum_{l_1=0}^{\rnk_1-a}   \sum_{l_2=0}^{\rnk_2-b}\sum_{l_3=0}^{\rnk_3-c} \sum_{l_4 = 0}^{\rnk_4-d}
  \sum_{p,q=1}^{\rnk}  \la^{12\Oo}_{(a+l_1)(b+l_2)p} \, \la^{34\Oo}_{(c+l_3)(d+l_4)q}   \prod_{i=1}^4 \frac{1}{l_i!} \frac{\pd^{l_i}}{\pd \DD_i^{l_i}} \; V^{pq}\!\left(\fracpt{}{\DD}\right)  \\
   \qquad \qquad \qquad \times \Big[ \mca{M}[\a,\DD,\DD_1,\DD_2] \; g_\a(\DD)^{-1} \; \mca{M}'[\a,\DD,\DD_3,\DD_4] \, \Big] \,.
  \label{eq:toProve}
\end{multline}
In order to obtain~\reef{eq:fullPW}, it is sufficient to sum the above expression over all descendant states $\a$.

Let us now prove the above identity. First notice that we can trivially rewrite the LHS of~\reef{eq:toProve} as
\beq
\label{eq:secondStep}
  \sum_{l_1=0}^{\rnk_1-a}   \sum_{l_2=0}^{\rnk_2-b}\sum_{l_3=0}^{\rnk_3-c} \sum_{l_4 = 0}^{\rnk_4-d}  \prod_{i=1}^4 \frac{1}{l_i!} \frac{\pd^{l_i}}{\pd \DD_i^{l_i}}
  \sum_{p,q=1}^\rnk   \mca{F}^{(l_1 l_2)}_p \cdot \mca{M} \, \left[ V^{pq}\!\left( \fracpt{}{\DD} \right) \cdot g_\a(\DD)^{-1} \right]  \mca{H}^{(l_3 l_4)}_q \cdot  \mca{M}' \,.
\eeq
with
\beq
  \mca{F}^{(l_1 l_2)}_p = \sum_{m=1}^{\rnk-p} \la^{12\Oo}_{(a+l_1)(b+l_2)(p+m)} \frac{1}{m!} \frac{\pd^m}{\pd \DD^m}\,, \qquad
   \mca{H}^{(l_3 l_4)}_q = \sum_{n=1}^{\rnk-q} \la^{34\Oo}_{(c+l_3)(d+l_4)(q+n)} \frac{1}{n!} \frac{\pd^n}{\pd \DD^n}\,.
\eeq
Next, we will rewrite the sum over $p,q$. To do so, we appeal to the following lemma:

\noindent {\bf Lemma}. Suppose that we are given two sets of constants $\mu_i$, $\mu'_i$ with $1 \leq i \leq \rnk$. Let's define the following families of differential operators:
\beq
C_p \ldef \sum_{m=0}^{\rnk-p} \frac{1}{m!} \mu_{p+m} \, \frac{\pd^m}{\pd \DD^m}\,,
\qquad
C'_q \ldef \sum_{n=0}^{\rnk-q} \frac{1}{n!} \mu'_{q+n} \, \frac{\pd^n}{\pd \DD^n}\,, \qquad p,q = 1,\ldots,\rnk.
\eeq
Then for any functions $h_1,h_2,h_3$ of $\DD$ we have
\beq
\sum_{p,q=1}^\rnk  \left( C_p \cdot h_1 \right) \;  \left( V^{pq}(\pd_\DD)\cdot h_2 \right) \;  \left( C'_q \cdot h_3 \right)  = \sum_{p,q=1}^\rnk \mu_p \, \mu'_q  \; V^{pq}(\pd_\DD) \cdot h_1 h_2 h_3 \,. 
\eeq
\noindent \emph{Proof}: a direct computation using the explicit form of the matrix $V^{pq}$, see Eq.~\reef{eq:Vdef}. \qed

To conclude, we apply this lemma to Eq.~\reef{eq:secondStep} with $\mu_p = \la^{12\Oo}_{(a+l_1)(b+l_2)p}$, $\mu'_q = \la^{34\Oo}_{(c+l_3)(d+l_4)q}$ and
\beq
h_1 = \mca{M}[\a,\DD_1,\DD_2,\DD], \quad
h_2 = g_a(\DD)^{-1}, \quad
h_3 = \mca{M}'[\a,\DD_3,\DD_4,\DD]\,.
\eeq
The result evaluates to the RHS of~\reef{eq:toProve}, so we are done.


\section{Conformal block identities}
The conformal blocks $G_\DD^{(\l)}(u,v;\DD_1,\DD_2,\DD_3,\DD_4)$ obey various identities which follow from the properties of the $\widehat{G}_\DD^{(\l)}(u,v;\rho_1,\rho_2)$~\cite{DO1,DO2,DO3}. These identities are used to simplify various equations in this work. 
 First, since the conventional blocks only depend on $\rho_1$ and $\rho_2$, different partial derivatives with respect of the external dimensions $\DD_i$ are related:
 \beq
\label{eq:del12}
\left(\fracpt{}{\DD_1} + \fracpt{}{\DD_2} \right)\!G_\DD^{(\l)}(u,v;\DD_i)
= \left( \fracpt{}{\DD_3} + \fracpt{}{\DD_4} \right)\!G_\DD^{(\l)}(u,v;\DD_i) =
\frac{1}{6} \ln\!\left(v/u^2\right) G_\DD^{(\l)}(u,v;\DD_i).
\eeq
Moreover, under the coordinate change $(u,v) \to (u/v,1/v)$ the blocks transform as
\beq
\label{eq:uvRels}
G_\DD^{(\l)}(u/v,1/v;\DD_i)
= (-1)^\l \, G_\DD^{(\l)}(u,v;\DD_i)\big|_{\DD_1 \lra \DD_2}
= (-1)^\l \, G_\DD^{(\l)}(u,v;\DD_i)\big|_{\DD_3 \lra \DD_4}\,,
\eeq
which implies that
\beq
\label{eq:ext1}
G_{\DD}^{(\l)}(u,v;\DD_1,\DD_2,\DD_3,\DD_4) = G_{\DD}^{(\l)}(u,v;\DD_2,\DD_1,\DD_4,\DD_3)\,.
\eeq
The invariance of the conventional blocks $\widehat{G}$ under $\rho_1 \lra \rho_2$ translates to
\beq
\label{eq:ext2}
G_{\DD}^{(\l)}(u,v;\DD_1,\DD_2,\DD_3,\DD_4) = G_{\DD}^{(\l)}(u,v;\DD_4,\DD_3,\DD_2,\DD_1)\,.
\eeq
Finally, the conventional blocks satisfy
\beq
\fracpt{}{\rho_1} \widehat{G}_\DD^{(\l)}(u,v;\rho_1,0)\big|_{\rho_1 = 0} = \fracpt{}{\rho_2} \widehat{G}_\DD^{(\l)}(u,v;0,\rho_2)\big|_{\rho_2 = 0} = - \frac{\ln v}{2}\,  \widehat{G}_\DD^{(\l)}(u,v;0,0)\,,
\eeq
which implies that for arbitrary $\eta$ we have
\bea
&&  \fracpt{}{\DD_1} G_{\DD}^{(\l)}(u,v;\DD_1,\eta,\eta,\eta)\big|_{\DD_1 = \eta} = \; \ldots \; = \fracpt{}{\DD_4} G_{\DD}^{(\l)}(u,v;\eta,\eta,\eta,\DD_4)\big|_{\DD_4 = \eta}   \nn\\
 &&\phantom{ \fracpt{}{\DD_1} G_{\DD}^{(\l)}(u,v;\DD_1,\eta,\eta,\eta)\big|_{\DD_1 = \eta}} = \frac{1}{12} \ln(v/u^2) \, G_{\DD}^{(\l)}(u,v;\eta,\eta,\eta,\eta)\,.
\label{eq:lima}
\eea
This can also be proved by combining Eqs.~\reef{eq:ext1} and~\reef{eq:del12}.


\section{Free field limit of logarithmic GFF}

In this Appendix we consider the logarithmic GFF in the limit $\Delta_\phi \to  (d-2)/2$, which is the scaling dimension of a free field, describing various interesting features that arise in this limit.

First, recall what happens to a normal  GFF $\chi$ of dimension $\DD_\chi$ in this limit. The conformal block decomposition of the four-point function $\expec{\chi \chi \chi \chi}$ is controlled by the coefficients $\mfr{q}(\l,n;\DD_\chi,\DD_\chi)$ from Eq.~\reef{eq:qDef}. In the free field limit $\DD_\chi \to (d-2)/2$, the coefficients with $n=0$ remain finite, but all coefficients $\mfr{q}(\l,n \geq 1,\DD_\chi,\DD_\chi)$ vanish. The reason is that nearly all double-trace operators decouple, due to the equation of motion $\pd^2 \chi = 0$ which holds iff $\DD_\chi = (d-2)/2$.

Naively, one may expect that the same decoupling persists in the logarithmic GFF. To be precise, suppose that we set $\mfr{q}(\l,n \geq 1) \equiv 0$ in Eqs.~\reef{eq:oddres} and~\reef{eq:secondrest}, which define the relevant OPE coefficients. This does not give rise to the correct four-point function $\expec{\phi_a \phi_b \phi_c \phi_d}$, as can be traced back to an order-of-limits issue. The reason is the following. Remark that several OPE coefficients in~\reef{eq:secondrest} feature \emph{derivatives} of the coefficients $\mfr{q}(\l,n,\DD_1,\DD_2)$. It is no longer true that all of these derivatives vanish if $n \geq 1$. For instance, 
\beq
\fracpt{}{\DD_\phi} \mfr{q}(\l,n=1,\DD_\phi,\DD_\phi) \; \limu{\DD_\phi \to \nu} \text{finite}\,, \qquad \nu \equiv (d-2)/2\,,
\eeq
although the above coefficient with $n \geq 2$ still vanishes in the free field limit. Such coefficients appear in the CB decomposition of $\mca{F}_2(u,v)$. The conformal block decomposition of $\mca{F}_1(u,v)$ is even more subtle. Let us parametrize the free field limit as $\DD_\phi \equiv \nu + \dd$, hence we are interested in the limit $\dd \to 0$. We remark that the following coefficient diverges as $\dd \to 0$:
\beq
2  \, \pd_{\DD_1} \pd_{\DD_2} \mfr{q}(\l,1;\DD_1,\DD_2) \big|_{\DD_1 = \DD_2 = \nu + \dd} \; \limu{\dd \to 0} \; \frac{\rho_\l}{\dd}  \; + \; \sO(1)\,,
\quad
\rho_\l = \frac{2^{\l-2}\,(\nu)_\l^2}{\l!\, (\l+2\nu)_\l} \frac{\l+2\nu}{\l+\nu+1}\,.
\eeq
This coefficient multiplies a conformal block of dimension $2\DD_\phi + \l+2$ and spin $\l$. We conclude that there is an infinite tower of divergent contribution to $\mca{F}_1(u,v)$, namely:
\beq
\label{eq:evenDiv}
\mca{F}_1(u,v) \; \supset \; \frac{1}{\dd} \sum_{\text{even } \l}  \rho_\l \, G_{\l + d}^{(\l)}(u,v;\nu,\nu,\nu,\nu) \; + \;  \sO(1)  \; + \; \text{odd spins} \,.
\eeq
But this is paradoxical: the function $\mca{F}_1(u,v)$ has a finite, well-defined free field limit. To resolve this paradox, we will look for any divergences in the odd-spin sector. The odd-spin OPE coefficients $b^{(\l,n)}$ --- see Eq.~\reef{eq:oddres} --- all remain finite as $\dd \to 0$; as we remarked before, only the coefficients with $n=0$ survive. However, the odd-spin conformal blocks appearing in $\mca{F}_1(u,v)$ will diverge.  Concretely, we have
\beq
\label{eq:div0}
\lim_{\dd \to 0} \; (\pd_{\DD_2} - \pd_{\DD_1})(\pd_{\DD_4} - \pd_{\DD_3}) \, G_{\l + d-2+\dd}^{(\l)}(u,v;\DD_1,\DD_2,\DD_3,\DD_4) \big|_{\DD_i = \nu} = \infty\,,
\eeq
for all odd $\l$.  
By subtracting finite terms, the divergence in Eq.~\reef{eq:div0} can be traced back to a divergence in the conventional conformal blocks $\widehat{G}_{\DD}^{(\l)}$ as follows:
\begin{multline}
  (\pd_{\DD_2} - \pd_{\DD_1})(\pd_{\DD_4} - \pd_{\DD_3}) \, G_{\l + d-2+\dd}^{(\l)}(u,v;\DD_1,\DD_2,\DD_3,\DD_4) \big|_{\DD_i = \nu}\\
  \limu{\dd \to 0} \; - \frac{1}{2} (v/u^2)^{\nu/3} \, \fracpt{}{\rho_1}\fracpt{}{\rho_2} \widehat{G}^{(\l)}_{\l + d-2 + \dd}(u,v;\rho_1,\rho_2)\big|_{\rho_1 = \rho_2 = 0} + \sO(1)\,,
\end{multline}
which diverges as $1/\dd$. This divergence can be understood by noting that in the limit $\dd \to 0$ a level-one descendant of spin $\l-1$ becomes null. A standard argument of conformal representation theory~\cite{zamolodchikov1984,Zamolodchikov:1987,Kos:2013tga,Penedones:2015aga} then predicts that
\beq
- \frac{1}{2} \fracpt{}{\rho_1}\fracpt{}{\rho_2} \widehat{G}^{(\l)}_{\l + d-2 + \dd}(u,v;\rho_1,\rho_2)\big|_{\rho_1 = \rho_2 = 0} \; \limu{\dd \to 0} \; \frac{ \kappa_\l }{\dd} \, \widehat{G}^{(\l-1)}_{\l + d-1}(u,v;0,0) + \sO(1)\,,
\eeq
for some constant $\kappa_\l$. A short computation shows that this is indeed the case, and the constant of proportionality is
\beq
\kappa_\l = \frac{\l(\l+2\nu-1)}{4(\l+\nu-1)(\l+\nu)}\,.
\eeq
In conclusion, we have a tower of divergent contributions to $\mca{F}_1$ given by
\beq
\label{eq:oddDiv}
\mca{F}_1(u,v) \; \supset \;  \frac{1}{\dd} \sum_{\text{odd } \l} (-1)\, \kappa_\l  \, \mfr{q}(\l,0;\nu,\nu) \, G_{\l + d-1}^{(\l-1)}(u,v;\nu,\nu,\nu,\nu) \; + \; \sO(1) \;  + \; \text{even spins}.
\eeq
The minus sign comes from Eq.~\reef{eq:oddres}.

Finally, we need to confirm that the divergences coming from even~\reef{eq:evenDiv} and odd~\reef{eq:oddDiv} operators cancel, such that $\mca{F}_1(u,v)$ is finite. We claim that this cancellation happens term by term, i.e.\@ the contributions of spin $\l=2k$ and $\l=2k+1$ cancel out. This easy to see --- it's an immediate consequence of the identity
\beq
\rho_{2k} = \kappa_{2k+1}\, \mfr{q}(2k+1,0;\nu,\nu)\,, \quad k =0,1,2,\ldots.
\eeq

\small
\parskip=-10pt
\bibliography{biblio}
\bibliographystyle{utphys}
\end{document}